\documentclass[fleqn,usenatbib]{mnras}

\usepackage{times} 
\usepackage{listings}
\usepackage{graphics}
\usepackage{graphicx,xspace}
\usepackage{amsmath}
\usepackage{amssymb}
\usepackage{hyperref}
\usepackage{lscape}
\usepackage{rotating}
\usepackage{placeins}
\usepackage{natbib}
\usepackage{color}
\usepackage{txfonts}
\usepackage{mathptmx}
\usepackage{newtxtext,newtxmath}

\usepackage{subcaption}
\usepackage[dvipsnames]{xcolor}
\usepackage{caption}

\usepackage{grffile}
\usepackage{tabularx}
\usepackage{adjustbox}
\usepackage{array, makecell, cellspace}
\usepackage{multirow}
\usepackage{siunitx} %allow angles like arcseconds
\usepackage{tikz}

\newcommand{\hi}{H{\sc\,i}\xspace}

\title[The $M_{\rm HI}-M_{\star}$ relation of massive galaxies]{MIGHTEE-\hi: The $M_{\rm HI}-M_{\star}$ relation of massive galaxies and the \hi mass function at $0.25 < z < 0.5$}

\author[H. Pan et al.]{Hengxing Pan$^{1,2,3}$\thanks{E-mail:panhengxing@nao.cas.cn},
Matt J.~Jarvis$^{3,4}$, Ian Heywood$^{3,5,6}$, Tariq Yasin$^{3}$, 
Natasha Maddox$^{7}$, 
Mario G. Santos$^{4,6}$,
\newauthor
Maarten Baes$^{8}$,
Anastasia A.~Ponomareva$^{9,3}$ and Sambatriniaina H. A. Rajohnson$^{10}$\\
$^{1}$National Astronomical Observatories, Chinese Academy of Sciences, Beijing 100101, People’s Republic of China\\
$^{2}$Guizhou Radio Astronomical Observatory, Guizhou University, Guiyang 550000, China\\
$^{3}$Astrophysics, University of Oxford, Denys Wilkinson Building, Keble Road, Oxford OX1 3RH, UK\\
$^{4}$Department of Physics and Astronomy, University of the Western Cape, Cape Town 7535, South Africa\\
$^{5}$Centre for Radio Astronomy Techniques and Technologies, Department of Physics and Electronics, Rhodes University, PO Box 94, Makhanda, 6140, South Africa\\
$^{6}$South African Radio Astronomy Observatory (SARAO), 2 Fir Street, Observatory, 7925, South Africa\\
$^{7}$School of Physics, H.H. Wills Physics Laboratory, Tyndall Avenue, University of Bristol, Bristol, BS8 1TL, UK\\
$^{8}$Sterrenkundig Observatorium, Universiteit Gent, Krijgslaan 281 S9, 9000 Gent, Belgium\\
$^{9}$Centre for Astrophysics Research, School of Physics, Astronomy and Mathematics, University of Hertfordshire, College Lane, Hatfield, AL10 9AB, UK\\
$^{10}$INAF - Osservatorio Astronomico di Cagliari, Via della Scienza 5, I-09047 Selargius (CA), Italy\\
}

\date{Accepted XXX. Received YYY; in original form ZZZ}

\pubyear{2022}

\begin{document}
\label{firstpage}
\pagerange{\pageref{firstpage}--\pageref{lastpage}}
\maketitle

\begin{abstract}
The relationship between the already formed stellar mass in a galaxy and the gas reservoir of neutral atomic hydrogen, is a key element in our understanding of how gas is turned into stars in galaxy haloes. In this paper, we measure the $M_{\rm HI}-M_{\star}$ relation based on a stellar-mass selected sample at $0.25 < z < 0.5$ and the MIGHTEE-\hi DR1 spectral data. Using a powerful Bayesian stacking technique, for the first time we are also able to measure the underlying bivariate distribution of \hi mass and stellar mass of  galaxies with $M_\star > 10^{9.5}$\,M$_{\odot}$, finding that an asymmetric underlying \hi distribution is strongly preferred by our complete samples. We define the concepts of the average of the logarithmic \hi mass, $\langle\log_{10}(M_{\rm HI})\rangle$, and the logarithmic average of the \hi mass, $\log_{10}(\langle M_{\rm HI}\rangle)$, and find that the difference between $\langle\log_{10}(M_{\rm HI})\rangle$ and $\log_{10}(\langle M_{\rm HI}\rangle)$ can be as large as $\sim$0.5 dex for the preferred asymmetric \hi distribution. We observe shallow slopes in the underlying $M_{\rm HI}-M_{\star}$ scaling relations, suggesting the presence of an upper \hi mass limit beyond which a galaxy can no longer retain further \hi gas. From our bivariate distribution we also infer the \hi mass function at this redshift and find tentative evidence for a decrease of 2-10 times in the co-moving space density of the most H{\sc i} massive galaxies up to $z\sim 0.5$. 
\end{abstract}


%%%%%%%%%%%%%%%%%%%%%%%%%%%%%%%%%%%%%%%%%%%%%%%%%%

%%%%%%%%%%%%%%%%% BODY OF PAPER %%%%%%%%%%%%%%%%%%

\section{Introduction}

Neutral atomic hydrogen (\hi) gas serves as the raw fuel for star formation in galaxies. The relationship between neutral atomic hydrogen gas and stars provides insight into a galaxy's evolutionary stage. However, the relationship between \hi and stellar mass ($M_{\rm HI}-M_{\star}$) is complex because of the intricate physical processes involved in galaxy evolution. Consequently, a simple correlation describing the $M_{\rm HI}-M_{\star}$ relation is insufficient to capture the full complexity of this relation. Gaining a deeper understanding of this relation is crucial for uncovering the mechanisms that drive galaxy evolution from youth to maturity \citep[e.g.][]{Rodr_guez_Puebla_2020, Saintonge2022}.

In particular, \cite{Maddox2015} and \cite{Pan2024} explored the upper envelope of the \hi and stellar mass ($M_{\rm HI}-M_{\star}$) relation based on \hi selected samples from ALFALFA, MIGHTEE and FAST COSMOS \hi surveys, enlightening the processes of gas consumption and star formation. In complementary studies, \cite{Catinella_2010} and \cite{Parkash2018} explored the underlying $M_{\rm HI}-M_{\star}$ relation based on stellar mass-selected samples, highlighting the role of gas poor early-type galaxies in regulating the intrinsic $M_{\rm HI}-M_{\star}$ relation.

The direct detection of emission lines from the neutral hydrogen component of galaxies has been limited to the local Universe or massive \hi\ systems, constrained by the sensitivity of modern radio instruments such as the Parkes and Arecibo telescopes in the past few decades. Despite these limitations, several \hi studies have been carried out from the \hi\ Parkes All-Sky Survey (HIPASS; \citealt{barnes2001h}), the Arecibo Legacy Fast ALFA (ALFALFA) survey (\citealt{giovanelli2005arecibo}), to the more recent, MeerKAT International GHz Tiered Extragalactic Exploration \citep[MIGHTEE;][]{Jarvis2016} survey, Looking at the Distant Universe with the MeerKAT Array \citep[LADUMA;][]{Blyth2016} and the FAST All Sky HI survey \citep[FASHI;][]{zhang2024} using MeerKAT \citep{Jonas2016mksJ} and the Five-hundred-meter Aperture Spherical Radio Telescope \citep[FAST;][]{NAN_2011}, respectively.

At higher redshift (z$>$0.1), the number of direct detections from \hi surveys is limited due to the faintness of the 21-cm line, with only recent surveys gradually building up larger samples \citep[e.g.][]{Xi2024,Jarvis2025}, thus \hi studies tend to rely on statistical approaches. The straightforward way is to average the \hi fluxes at known locations of sources detected in other wavelengths, thus increasing the signal to noise ratio using the stacking technique \citep[e.g.][]{Delhaize_2013,Healy_2019, Sinigaglia_2022}. However, a simple averaging in the flux space risks losing information on the intrinsic scatter of \hi distribution \citep[or shape of \hi mass function, see e.g.][]{Bera_2022,Sinigaglia2025}, limiting the accessible knowledge of the $M_{\rm HI}-M_{\star}$ relation from the noisy \hi data at high redshifts. The underlying \hi mass distribution as a function of the stellar mass reveals how strong the \hi mass correlates with the stellar mass and hence traces the conversion of gas into stars, which is crucial for constraining the evolutionary stage of galaxies, and inferring the total amount of \hi gas by linking to the \hi mass function \citep[e.g.][]{Ponomareva2023,2024arXiv241211426K}.

In this paper, we implement a Bayesian stacking technique, based on our previous work \citep{Pan_2021,Pan_2023}, for measuring the underlying $M_{\rm HI}-M_{\star}$ scaling relation by combining the MIGHTEE DR1 \hi images and an optical catalogue at $0.25 < z < 0.5$, while taking into account the intrinsic \hi scatter as a function of the stellar mass. This technique employs fluxes of \hi emission line as measurables while modelling the underlying \hi mass distribution that can naturally account for the thermal noise scatter from the radio receiver on the linear flux scale, and the intrinsic \hi scatter on the logarithmic mass scale. 

This paper is organized as follows. We describe our MIGHTEE-\hi and the ancillary data in Section \ref{sec:data}, and the Bayesian technique in Section \ref{sec:method}. We present our main results in Section \ref{sec:results}, and conclude in Section \ref{sec:conclusions}. We use the standard $\Lambda$CDM cosmology with a Hubble constant $H_{0}=70$ km$\cdot$s$^{-1}\cdot {\rm Mpc}^{-1}$, total matter density $\Omega_{\rm m}=0.3$ and dark energy density $\Omega_{\Lambda}=0.7$, and AB magnitudes throughout when magnitudes are used.

\section{Data}
\label{sec:data}

\subsection{MIGHTEE-\hi}
\label{sec:HIcubes}

\begin{figure}
    \centering
    \includegraphics[width=\columnwidth]{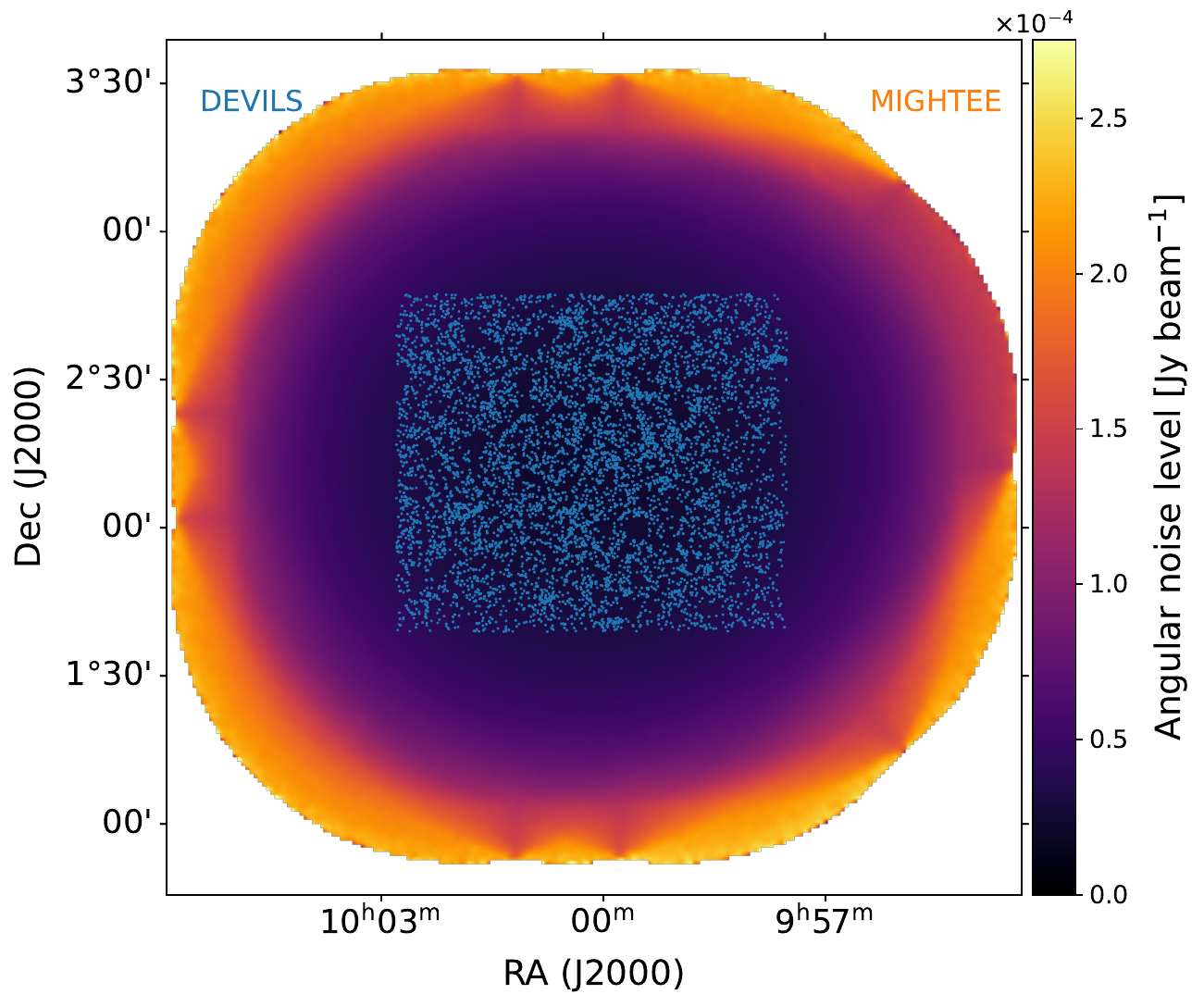}
  \caption{DEVILS spectroscopic samples overlaid on the MIGHTEE field. The colour scheme indicates the noise level of MIGHTEE DR1 L1 spectral data on a projected 2-dimensional sky.}
  \label{fig:mightee}
\end{figure}

MIGHTEE-\hi is the \hi emission project within the MIGHTEE survey. The MIGHTEE Early Science (E.S.) data were collected between mid-2018
and mid-2019 and are described by \citet{maddox2021} in detail. For this study, we use the latest MIGHTEE Data Release 1, observed with the 32k channel mode of the MeerKAT telescope \citep{Heywood_2024} across the L-band. The MeerKAT 32k correlator mode provides 32,768 channels with a spectral resolution of 26.1 kHz, corresponding to 5.5 km s$^{-1}$ at 1420 MHz. The L-band has been split into two sub-bands (L1 and L2) with the frequency ranges of these sub-bands are 960–
1150 MHz (0.23 $< z_{\rm HI} <$ 0.48) and 1290–1520 MHz (0 $< z_{\rm HI} <$0.1) respectively. Note that the L1 band cube has a velocity resolution of  27.6 km s$^{-1}$ for H{\sc i} at $z=0.25$, and we round the \hi redshift range of L1 band as 0.25 $< z <$ 0.5 hereafter. The COSMOS field was observed using this mode with  15$\times$8h tracks, arranged in a tightly-dithered mosaic covering approximately $\sim$6 deg$^2$ for L1 band. Each pointing was imaged using robustness parameters of 0.0 and 0.5 after two rounds of RFI flagging, self-calibration, and visibility-domain continuum subtraction. Subsequently, all pointings were combined for homogenization, mosaicking, and an additional image-domain continuum subtraction \citep[][]{Heywood_2024}. We show the averaged noise level of the L1 data across the sub-band in Figure~\ref{fig:mightee}, where the galaxies with optical spectroscopy sit in the deepest region of our survey field. We list a few key parameters of the MIGHTEE Data Release 1 (DR1) L1 spectral data in Table~\ref{tab1} which are used in this paper.

\begin{table}
        \small
        \centering
        \caption{Key parameters of the MIGHTEE DR1 L1 spectral data.}
        \label{tab1}
        \begin{tabular}{l l}
                \hline
                \hline
                Survey area & 4.94-6.49 deg$^{2}$\\
                Integration time  & 94.2 hrs \\
                Frequency range  & 960–1150 MHz\\
                Channel width   & 104.5 kHz         \\
                Synthesized Beam & 14.8$^{\prime\prime}$, 19.6$^{\prime\prime}$ (r=0, 0.5)\\
                Sensitivity (Median 1$\sigma$ RMS) & 40 $\mu$Jy beam$^{-1}$ (r=0.5)\\
                \hline
        \end{tabular}
\end{table}

\subsection{Ancillary data}
\label{sec:stmass}

\begin{figure}
    \centering
    \includegraphics[width=0.97\columnwidth]{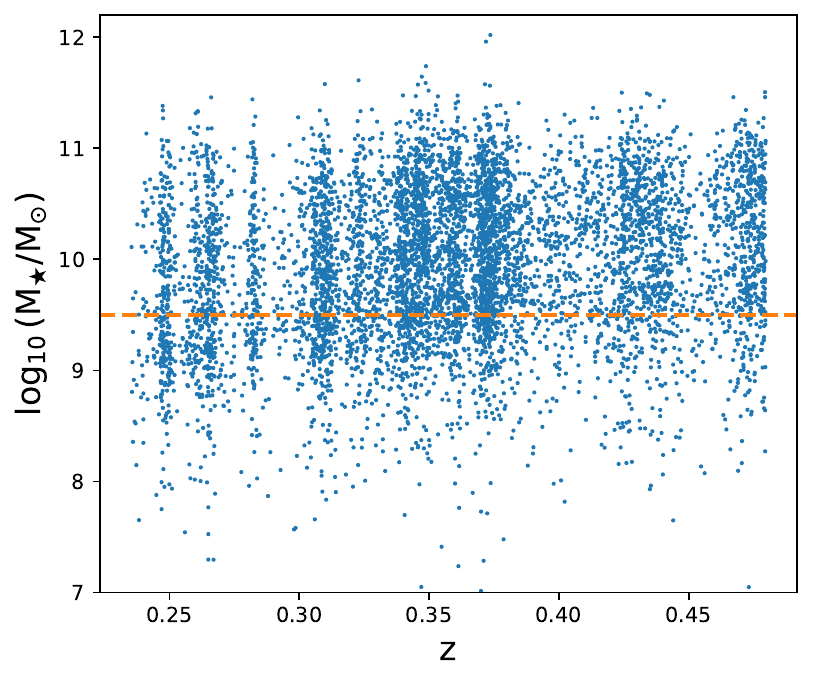}
  \caption{Stellar mass as a function of redshift. The dashed line is the limit of the sample completeness of $\log_{10}(M_\star/M_\odot$) = 9.5.}
  \label{fig:mstar_vs_z}
\end{figure}

The MIGHTEE COSMOS field is covered by multi-wavelength photometric and spectroscopic surveys ranging from X-ray to far-infrared bands \citep[e.g.][]{Cuillandre2012, McCracken2012, Aihara_2017, Davies_2018, Aihara2019}. We use the Deep Extragalactic VIsible Legacy Survey \citep[DEVILS;][]{Davies_2018} for stellar properties, which are constructed in \cite{Hashemizadeh_2022}. DEVILS is a magnitude-limited spectroscopic and multi-wavelength survey, with the spectroscopic observations taken with the Anglo-Australian Telescope (AAT), providing spectroscopic redshift completeness of $>$ 95\% to Y-mag $<$ 21.2~mag. This spectroscopic sample has derived physical properties, such as colour, stellar mass and star formation rate based on fitting the full spectral energy distribution with the modelling code PROSPECT which has well-motivated parametrizations for dust attenuation, star formation histories, and metallicity evolution \citep{Thorne2021}. In this paper, we limit the stellar mass to $\log_{10}(M_\star/M_\odot$) $\geq 9.5$ to reduce the effects of incompleteness (see Figure~\ref{fig:mstar_vs_z}) and to include the morphological classifications following the same practice in \cite{Hashemizadeh_2022}.

\begin{figure}
    \centering
    \includegraphics[width=\columnwidth]{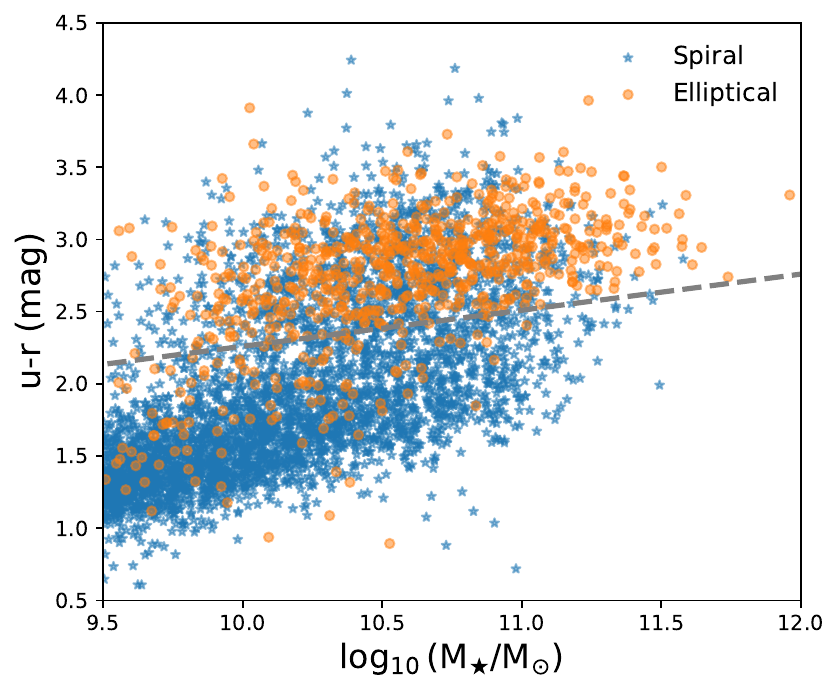}
  \caption{Colour (u-r) against the stellar mass. The blue stars are spirals while the orange dots are ellipticals. The dashed line is from \protect\cite{Schawinski2014} for seperating the full sample into blue and red sub-samples.}
  \label{fig:ur_vs_mstar}
\end{figure}

\begin{figure}
    \centering
    \includegraphics[width=\columnwidth]{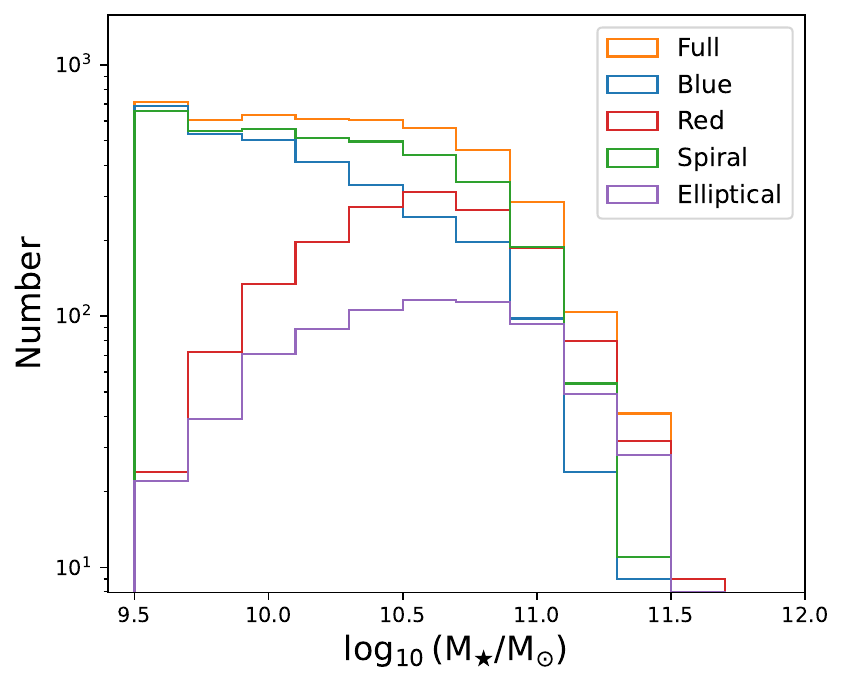}
  \caption{Histogram of the stellar mass for all galaxy samples. The orange, blue, red, green and purple are the full sample, blue, red, spiral and elliptical sub-samples.}
  \label{fig:gsmf}
\end{figure}

\subsection{Morphology and colour}
\label{sec:morpho}

The optical morphologies of the cataloged galaxies are classified into pure disk (D), pesudo/diffuse bulge + disk (pBD), classical bulge + disk (cBD), elliptical (E), compact (C) and hard (H). We group the D, pBP and cBD as 'Spirals (or disk galaxies)', and also combine the C and E as 'Ellipticals (or Spheroids)' since the C subcategory is dominated by unresolved and most likely compact spheroidal systems \citep{Hashemizadeh_2022}. The sample for visual inspection is selected to have large enough surface brightness and angular size, to be visually classified in the COSMOS Hubble Space Telescope (HST) data, based on a number of quality checks outlined in \cite{Hashemizadeh2021}. The process of morphological classification initially used a number of automatic pre-classification methods followed by a full visual inspection by multiple classifiers. If the checker felt that the source was erroneously classified it was removed. All erroneous sources were then re-classified by each classifier independently. The final morphology classification is taken where the majority of classifiers agree. This process is described in more detail in \cite{Hashemizadeh2021}.

The galaxies are also well separated into blue and red colours (u-r) in Figure~\ref{fig:ur_vs_mstar} using the upper limit of green valley galaxies in \cite{Schawinski2014}. The u-band images for COSMOS field were obtained as part of the Canada-France-Hawaii Telescope Legacy Survey (CFHT-LS1)\footnote{\url{http://www.cfht.hawaii.edu/Science/CFHLS}}, supplemented by independent CFHT observations outlined in \cite{Capak_2007}, and the r-band images in this region were from the new second public data release \citep[PDR2;][]{Aihara2019} of the Subaru Telescope’s Hyper Suprime-Cam Subaru Strategic Program \citep[HSC-SSP;][]{Aihara2018}. The Planck $E(B-V)$ map was used to correct the measured magnitudes for Galactic extinction, and the attenuation correction for each band was determined in the traditional manner (i.e. $A_x = [A_x/E(B-V)]\times E(B-V)$) by using the extinction coefficients listed in Table 2 of \cite{Davies_2021}. Overall, the blue and red galaxies correlate with the spirals and ellipticals, respectively. However, we note that a large fraction of red spirals (i.e. blue stars in the upper panel of Figure~\ref{fig:ur_vs_mstar}) exist, indicating a significantly reduced star formation rate in these galaxies compared to typical blue spirals, often caused by a depletion of gas needed for new stars to form, potentially due to environmental factors, such as ram pressure stripping in galaxy clusters or internal processes such as strong stellar feedback.

We show the histograms of the full galaxy sample, and the subsamples based on the morphology and colour classifications in Figure~\ref{fig:gsmf}. Clearly, the blue or spiral galaxies trace the majority of the full sample and their numbers drop as the stellar mass increases. The spirals have even larger counts at the massive end compared to the blue galaxies as a fair fraction of red spirals exist \citep[e.g.][]{Masters_2010,Mahajan_2019}, although a different color (e.g. NUV-r) selection may reduce the fraction of red spirals to a certain degree \citep{Zhou_2021}. In contrast, the red or elliptical galaxies are concentrated in the intermediate mass range, and we can see that the number of red spirals is on the same order of magnitude as that of the red ellipticals. Overall, the red galaxies and ellipticals  constitute only a small fraction of the total galaxy population, except for the high mass end of $\log_{10}(M_\star/M_\odot$)$>$10.5.

\subsection{Flux extraction}
\label{sec:hiflux}

The \hi fluxes were measured from the MIGHTEE L1 spectral cube based on the angular positions and spectroscopic redshifts of objects in the ancillary galaxy catalog. We first extracted cubelets centered on all cataloged sources with an angular aperture size of 16 arcsec and a spectral window of 1500 km s$^{-1}$. We further restrict the spectral window to 600 km s$^{-1}$ at the center of the full spectra for measuring the flux density of the \hi line to minimize the level of source confusion \citep{Pan_2019}, while balancing the need to include all flux without significantly increasing the noise given that the maximum \hi line width is $\sim$500 km s$^{-1}$ \citep{Ponomareva_2021}. The rest of the spectral windows serve as a measurement of the spectral baseline that is subsequently removed from the central spectra. We then measure the integrated flux, $S_{\rm m}$, from the extracted flux densities, while correcting for the total beam area. The total area under a 2D Gaussian is exactly twice the area integrated within its FWHM ellipse. The correcting factor for our extraction window is slightly lower than the typical beam size ($\sim$20 arcsec) at the frequency range of 960-1150 MHz. Therefore, we correct the integral flux numerically based on the integral to peak flux for a bright maser discovered by \cite{Jarvis2024} in Appendix~\ref{sec:mhi_m}. The minor evolution of the beam size within this frequency range is also taken into account \citep{Heywood_2024}. We quantify the flux uncertainties by moving the flux extraction box to 100 random positions surrounding the source in a distance range of 5-10 beams and computing their standard deviation. We do not require a formal \hi detection in order to measure the flux since we extract the \hi flux based on the source location from an optical catalog. We also note that we are extracting fluxes below the nominal noise threshold of the data, as such the fluxes that we are measuring are not cleaned, but are beam-corrected fluxes from the synthesised beam. However, the sidelobes within the MIGHTEE H{\sc i} sythesised beam are very low level for the robust=0.5 spectral cube we use in this paper\footnote{In Figure~\ref{fig:extraction} we show the ratio of the integrated to peak flux of the unresolved OH Megamaser from \citep{Jarvis2024} which shows that the flux extracted at 16~arcsec is consistent with the synthesised beam}, as such we do not expect significant offsets in the measured flux densities. We also carry out the same experiment with the robust=0.0 cube and find consistent results for all our analyses, from which we can infer that the flux extraction is robust.

\begin{figure}
    \centering
    \includegraphics[width=\columnwidth]{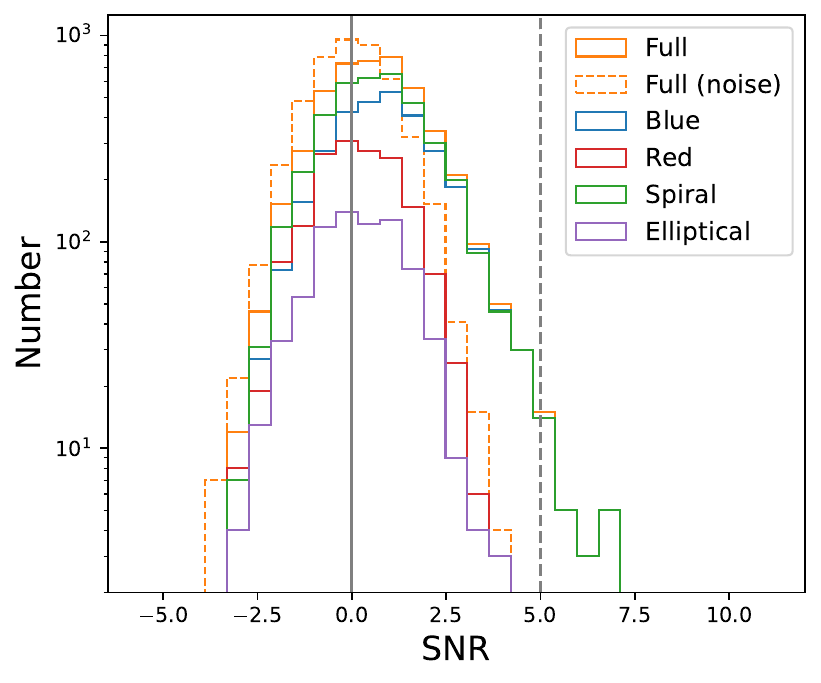}
  \caption{Distribution of SNR of the measured flux for all galaxy samples. The orange, blue, red, green and purple are the full sample, blue, red, spiral and elliptical sub-samples. The dashed grey line indicates a nominal detection threshold of 5$\sigma$. The dashed line is for the measured noise at positions offset from the optical galaxy locations.}
  \label{fig:snr_distribution}
\end{figure}

We show the distribution of the signal to noise ratio ($\frac{S_{\rm m}}{\sigma_{\rm n}}$) of the measured flux in solid lines in Figure~\ref{fig:snr_distribution}. Although there are some luminous \hi sources present in our full sample, the majority of the extracted fluxes are below the 5$\sigma_{\rm n}$ detection threshold. As expected, the blue (or spiral) galaxies dominate the high SNR end since they represent the most \hi-rich population of galaxies. We also show the histogram of the measured noise with the dashed orange line for the full sample at offset from the spectroscopic source position by five times the beam size, demonstrating quasi-Gaussian noise behavior in comparison to the obvious "long tail" feature for the full galaxy sample due to the \hi signal at the source location. We are essentially modelling the difference between the distribution of measurements from the offset position and that of the measurement at the position of the galaxies, with each measurement weighted by its noise. The weighting scheme is similar to the approach employed in traditional stacking investigations \citep[e.g.][]{Healy_2019,Sinigaglia_2022,Bianchetti_2025}.

At the redshift range of 0.25 $< z <$ 0.5, a large number of our measurements are below the nominal detection threshold, therefore we only use the measured fluxes to constrain the underlying \hi mass distribution rather than to calculate the individual \hi masses. However, we do need to establish a relationship between the \hi mass and the intrinsic flux to model the underlying \hi mass distribution (see the next section for details). Under the optically thin gas assumption, we convert the intrinsic flux to an \hi mass (or vice versa) via
\begin{equation}
    M_{\rm HI} = 2.356 \times 10^5 D^2_L(1+z)^{-1} S,
	\label{eq:factor}
\end{equation}
where $M_{\rm HI}$ is the \hi mass in solar masses, $D_L$ is the luminosity distance in Mpc, and $S$ is the integrated flux in Jy\,km\,s$^{-1}$ \citep{meyer2017tracing}. 

\section{Bayesian analysis}
\label{sec:method}

\subsection{Bayesian framework}

Our analysis is based on Bayes’ theorem:
\begin{equation}
\centering
\mathcal{P}(\Theta|D,H) = \frac{\mathcal{L}(D|\Theta,H) \Pi(\Theta|H)}{\mathcal{Z}(D|H)},
\label{eqn:bayes}
\end{equation}
where $\mathcal{P}$ represents the posterior distribution of the model parameters $\Theta$, given the data $D$ and a model $H$. The likelihood $\mathcal{L}$ quantifies the probability of the data $D$ given the parameters and the model. The prior $\Pi$ reflects our initial knowledge or assumptions about the parameter values. The Bayesian evidence $\mathcal{Z}$ acts as a normalization factor and is expressed as an integral over the $n$-dimensional parameter space $\Theta$:
\begin{equation}
\mathcal{Z}(D|H) = \int \mathcal{L}(D|\Theta,H) \Pi(\Theta|H) \textrm{d}^n\Theta.
\label{eqn:Z}
\end{equation}
The evidence $\mathcal{Z}$ is essential for model selection, as it represents the probability of the data given a model, marginalized over all free parameters. By comparing evidences, we can quantitatively evaluate the relative performance of different models. The difference in the logarithm of the evidences, $\Delta\ln(\mathcal{Z}) = \ln(\mathcal{Z_B}) - \ln(\mathcal{Z_A})$, is known as the Bayes factor and indicates how much better Model $B$ is compared to Model $A$. This method naturally penalizes overly complex models by accounting for their flexibility.

We adopt the criteria from \cite{Pan_2023}, where $\Delta\ln(\mathcal{Z}) < 1$ is categorized as "not significant," $1 < \Delta\ln(\mathcal{Z}) < 2.5$ as "significant," $2.5 < \Delta\ln(\mathcal{Z}) < 5$ as "strong," and $\Delta\ln(\mathcal{Z}) > 5$ as "decisive." For our analysis, we use {\sc Multinest} \citep{Feroz_2009, Buchner_2014}, an efficient and robust Bayesian inference tool commonly applied in cosmology and particle physics. {\sc Multinest} allows us to sample the parameter space, explore the posterior distribution for parameter estimation, and calculate the Bayesian evidence for model comparison.

\begin{table}
    \centering
    \caption{Parameters and priors for three Models used to measure the \hi and stellar mass relation of the Full, Blue and Spiral subsamples.}
    \label{tab:priors}
    \begin{tabular}{cll}
        \hline
        Model & Parameter & Prior Probability Distribution \\
        \hline
        \multirow{3}{*}{A} 
            & $a$ & uniform $\in [-2.5, 2.5] $  \\
            & $b$ & uniform $\in [7, 12] $ \\
            & $\sigma_{\rm HI}$ & uniform $\in [0, 2] $ \\
        \hline
        \multirow{3}{*}{B} 
            & $a$ & uniform $\in [-2.5, 2.5] $  \\
            & $b$ & uniform $\in [7, 12] $ \\
            & $\alpha$ & uniform $\in [-2.5, 2.5] $ \\
        \hline
        \multirow{4}{*}{C} 
            & $a$ & uniform $\in [-2.5, 2.5] $ \\
            & $b$ & uniform $\in [8, 13] $ \\
            & $\alpha$ & uniform $\in [-47.5, 2.5] $  \\
            & $\sigma$ & uniform $\in [0, 3] $ \\
        \hline
    \end{tabular}
\end{table}

\renewcommand{\arraystretch}{1.05}
\begin{table*}
	\centering
	\caption{Best fitting parameters of the measured conditional \hi mass distribution with Models A, B and C (corresponding to Normal, Schechter and Skew normal models, respectively) for the full galaxy sample, blue, red, spiral, and elliptical sub-samples at $0.25<z<0.5$. Fits are omitted for subsamples with sizes too small to produce robust results.}
	\label{tab:post}
 	\begin{adjustbox}{max width=\textwidth}
        \begin{tabular}{lcl@{\hskip 0.2cm}r@{\hskip 0.8cm}r@{\hskip 0.8cm}r@{\hskip 0.8cm}r@{\hskip 0.8cm}r@{\hskip 0.6cm}}
        \hline
	\hline
        % Top row with adjusted spacing
        \multicolumn{1}{@{\extracolsep{\fill}}l}{Sample} & 
        \multicolumn{1}{c}{Model} & 
        \multicolumn{1}{l@{\hskip -0.8cm}}{Parameter} & 
        \multicolumn{1}{c@{\hskip 0.1cm}}{$9.5<\log_{10}(M_{\star})<10$} & 
        \multicolumn{1}{c@{\hskip 0.1cm}}{$10<\log_{10}(M_{\star})<10.5$} & 
        \multicolumn{1}{c@{\hskip 0.1cm}}{$10.5<\log_{10}(M_{\star})<11$} & 
        \multicolumn{1}{c@{\hskip 0.1cm}}{$11<\log_{10}(M_{\star})<11.5$} & 
        \multicolumn{1}{c}{$9.5<\log_{10}(M_{\star})$} \\
	\hline
        & & $a$ & 0.05 ± 0.13 & -0.17 ± 0.15 & 0.46 ± 0.19 & -0.48 ± 0.65 & -0.0 ± 0.03 \\
        & A & $b$ & 9.29 ± 0.04 & 9.28 ± 0.04 & 8.88 ± 0.15 & 9.55 ± 0.74 & 9.23 ± 0.02 \\
        & & $\sigma_{\rm HI}$ & 0.36 ± 0.02 & 0.39 ± 0.02 & 0.44 ± 0.03 & 0.56 ± 0.09 & 0.41 ± 0.01 \\
        \cline{2-8}
        & & $a$ & 0.06 ± 0.13 & -0.07 ± 0.14 & 0.49 ± 0.17 & -0.47 ± 0.59 & 0.03 ± 0.03 \\
        & B & $b$ & 9.44 ± 0.06 & 9.53 ± 0.06 & 9.26 ± 0.14 & 10.37 ± 0.7 & 9.53 ± 0.03 \\
        & & $\alpha$ & 0.0 ± 0.13 & -0.21 ± 0.11 & -0.37 ± 0.09 & -0.69 ± 0.12 & -0.25 ± 0.06 \\
        Full & & $\Delta\ln(\mathcal{Z})$ & 6.28 ± 0.09 & 5.62 ± 0.03 & 9.26 ± 0.17 & 1.47 ± 0.22 & 19.89 ± 0.45 \\
        \cline{2-8}
        & & $a$ & 0.06 ± 0.12 & -0.04 ± 0.13 & 0.51 ± 0.14 & -0.41 ± 0.53 & 0.04 ± 0.02 \\
        & & $b$ & 9.91 ± 0.06 & 9.94 ± 0.06 & 9.68 ± 0.12 & 10.38 ± 0.69 & 9.94 ± 0.03 \\
        & C & $\alpha$ & -5.74 ± 9.27 & -6.46 ± 3.48 & -11.48 ± 10.11 & -4.78 ± 4.7 & -6.8 ± 1.44 \\
        & & $\sigma$ & 0.97 ± 0.13 & 1.11 ± 0.14 & 1.42 ± 0.12 & 1.45 ± 0.36 & 1.17 ± 0.09 \\
        & & $\Delta\ln(\mathcal{Z})$ & 5.23 ± 0.09 & 4.77 ± 0.11 & 11.85 ± 0.34 & -0.21 ± 0.19 & 19.03 ± 0.09 \\
        \hline
        & & $a$ & 0.23 ± 0.14 & 0.09 ± 0.13 & 0.7 ± 0.18 & -0.11 ± 0.63 & 0.22 ± 0.03 \\
        & A & $b$ & 9.4 ± 0.04 & 9.44 ± 0.04 & 9.06 ± 0.14 & 9.76 ± 0.72 & 9.4 ± 0.02 \\
        & & $\sigma_{\rm HI}$ & 0.33 ± 0.02 & 0.31 ± 0.02 & 0.32 ± 0.03 & 0.34 ± 0.09 & 0.32 ± 0.01 \\
        \cline{2-8}
        & & $a$ & 0.21 ± 0.12 & 0.14 ± 0.13 & 0.71 ± 0.18 & -0.3 ± 0.56 & 0.22 ± 0.03 \\
        & B & $b$ & 9.43 ± 0.06 & 9.35 ± 0.07 & 9.0 ± 0.15 & 10.04 ± 0.65 & 9.39 ± 0.03 \\
        & & $\alpha$ & 0.23 ± 0.17 & 0.52 ± 0.21 & 0.43 ± 0.25 & 0.18 ± 0.49 & 0.35 ± 0.11 \\
        Blue & & $\Delta\ln(\mathcal{Z})$ & 4.23 ± 0.05 & 5.71 ± 0.43 & 7.16 ± 0.28 & 1.2 ± 0.05 & 13.58 ± 0.03 \\
        \cline{2-8}
        & & $a$ & 0.2 ± 0.12 & 0.17 ± 0.12 & 0.6 ± 0.16 & -0.49 ± 0.63 & 0.21 ± 0.02 \\
        & & $b$ & 9.91 ± 0.07 & 9.93 ± 0.06 & 9.69 ± 0.12 & 10.76 ± 0.74 & 9.94 ± 0.03 \\
        & C & $\alpha$ & -4.26 ± 9.79 & -5.61 ± 3.92 & -10.69 ± 13.98 & -5.34 ± 15.67 & -5.69 ± 1.77 \\
        & & $\sigma$ & 0.77 ± 0.13 & 0.75 ± 0.11 & 0.83 ± 0.09 & 0.88 ± 0.29 & 0.8 ± 0.07 \\
        & & $\Delta\ln(\mathcal{Z})$ & 2.66 ± 0.18 & 3.07 ± 0.03 & 8.83 ± 0.06 & 0.67 ± 0.07 & 14.4 ± 0.76 \\
        \hline
        & & $a$ & \multicolumn{1}{c}{$-$}  & 2.45 ± 2.27 & -0.41 ± 0.58 & -1.17 ± 0.98 & 0.25 ± 0.13 \\
        & A & $b$ & \multicolumn{1}{c}{$-$}  & 7.46 ± 1.0 & 9.21 ± 0.43 & 10.06 ± 1.17 & 8.47 ± 0.15 \\
        & & $\sigma_{\rm HI}$ & \multicolumn{1}{c}{$-$} & 0.62 ± 0.15 & 0.29 ± 0.15 & 0.5 ± 0.26 & 0.49 ± 0.08 \\
        \cline{2-8}
        & & $a$ & \multicolumn{1}{c}{$-$}  & 5.12 ± 3.36 & -0.37 ± 0.55 & -0.66 ± 1.04 & 0.24 ± 0.13 \\
        & B & $b$ & \multicolumn{1}{c}{$-$}  & 7.4 ± 1.38 & 9.12 ± 0.47 & 10.25 ± 1.16 & 9.01 ± 0.33 \\
        & & $\alpha$ & \multicolumn{1}{c}{$-$}  & -0.84 ± 0.48 & 0.46 ± 0.96 & -0.7 ± 1.21 & -0.45 ± 0.82 \\
        Red & & $\Delta\ln(\mathcal{Z})$ & \multicolumn{1}{c}{$-$}  & -1.0 ± 0.1 & 1.05 ± 0.09 & 0.17 ± 0.06 & -0.21 ± 0.12 \\
        \cline{2-8}
        & & $a$ & \multicolumn{1}{c}{$-$}  & 1.3 ± 1.29 & -0.53 ± 0.57 & -0.75 ± 1.24 & 0.22 ± 0.13 \\
        & & $b$ & \multicolumn{1}{c}{$-$}  & 8.58 ± 0.33 & 9.87 ± 0.44 & 9.86 ± 1.32 & 8.92 ± 0.17 \\
        & C & $\alpha$ & \multicolumn{1}{c}{$-$}  & -1.82 ± 5.15 & -10.52 ± 4.14 & -0.91 ± 4.78 & -1.38 ± 2.12 \\
        & & $\sigma$ & \multicolumn{1}{c}{$-$}  & 1.04 ± 0.38 & 0.92 ± 0.41 & 0.66 ± 0.51 & 0.74 ± 0.25 \\
        & & $\Delta\ln(\mathcal{Z})$ & \multicolumn{1}{c}{$-$}  & -3.32 ± 0.1 & -0.16 ± 0.11 & -0.84 ± 0.06 & -1.47 ± 0.11 \\
        \hline
        & & $a$ & 0.14 ± 0.13 & -0.06 ± 0.15 & 0.74 ± 0.19 & 0.34 ± 0.68 & 0.1 ± 0.03 \\
        & A & $b$ & 9.34 ± 0.04 & 9.31 ± 0.05 & 8.79 ± 0.15 & 8.91 ± 0.81 & 9.29 ± 0.02 \\
        & & $\sigma_{\rm HI}$ & 0.35 ± 0.02 & 0.38 ± 0.02 & 0.4 ± 0.03 & 0.46 ± 0.1 & 0.38 ± 0.01 \\
        \cline{2-8}
        & & $a$ & 0.14 ± 0.13 & 0.01 ± 0.14 & 0.68 ± 0.17 & 0.23 ± 0.66 & 0.11 ± 0.03 \\
        & B & $b$ & 9.44 ± 0.07 & 9.49 ± 0.06 & 9.08 ± 0.14 & 9.47 ± 0.76 & 9.48 ± 0.03 \\
        & & $\alpha$ & 0.06 ± 0.15 & -0.08 ± 0.12 & -0.14 ± 0.13 & -0.39 ± 0.23 & -0.09 ± 0.07 \\
        Spiral & & $\Delta\ln(\mathcal{Z})$ & 5.64 ± 0.36 & 6.07 ± 0.57 & 8.12 ± 0.1 & 0.87 ± 0.18 & 18.15 ± 0.03 \\
        \cline{2-8}
        & & $a$ & 0.13 ± 0.12 & 0.07 ± 0.13 & 0.64 ± 0.15 & -0.07 ± 0.61 & 0.11 ± 0.02 \\
        & & $b$ & 9.91 ± 0.07 & 9.94 ± 0.06 & 9.62 ± 0.12 & 10.22 ± 0.72 & 9.96 ± 0.03 \\
        & C & $\alpha$ & -5.37 ± 8.08 & -6.56 ± 4.32 & -13.32 ± 12.4 & -9.56 ± 14.95 & -7.38 ± 2.06 \\
        & & $\sigma$ & 0.91 ± 0.14 & 1.04 ± 0.15 & 1.23 ± 0.12 & 1.45 ± 0.29 & 1.08 ± 0.08 \\
        & & $\Delta\ln(\mathcal{Z})$ & 4.06 ± 0.27 & 4.23 ± 0.09 & 10.7 ± 0.17 & 0.68 ± 0.1 & 19.26 ± 0.09 \\
        \hline
        & & $a$ & -1.16 ± 0.91 & -0.79 ± 1.16 & \multicolumn{1}{c}{$-$}  & \multicolumn{1}{c}{$-$}  & -0.29 ± 0.18 \\
        & A & $b$ & 8.7 ± 0.4 & 8.92 ± 0.43 & \multicolumn{1}{c}{$-$}  & \multicolumn{1}{c}{$-$}  & 8.88 ± 0.15 \\
        & & $\sigma_{\rm HI}$ & 0.41 ± 0.23 & 0.46 ± 0.25 & \multicolumn{1}{c}{$-$}  & \multicolumn{1}{c}{$-$}  & 0.43 ± 0.12 \\
        \cline{2-8}
        & & $a$ & -1.27 ± 0.79 & -0.48 ± 1.08 & \multicolumn{1}{c}{$-$}  & \multicolumn{1}{c}{$-$}  & -0.31 ± 0.17 \\
        & B & $b$ & 9.05 ± 0.39 & 9.23 ± 0.45 & \multicolumn{1}{c}{$-$}  & \multicolumn{1}{c}{$-$}  & 9.23 ± 0.3 \\
        & & $\alpha$ & -0.37 ± 1.0 & -0.3 ± 1.11 & \multicolumn{1}{c}{$-$}  & \multicolumn{1}{c}{$-$}  & -0.26 ± 0.96 \\
        Elliptical & & $\Delta\ln(\mathcal{Z})$ & 0.97 ± 0.04 & 0.79 ± 0.05 & \multicolumn{1}{c}{$-$}  & \multicolumn{1}{c}{$-$}  & 1.1 ± 0.07 \\
        \cline{2-8}
        & & $a$ & -1.23 ± 0.76 & -0.01 ± 1.06 & \multicolumn{1}{c}{$-$}  & \multicolumn{1}{c}{$-$}  & -0.32 ± 0.17 \\
        & & $b$ & 9.5 ± 0.28 & 9.1 ± 0.37 & \multicolumn{1}{c}{$-$}  & \multicolumn{1}{c}{$-$}  & 9.64 ± 0.15 \\
        & C & $\alpha$ & -38.33 ± 15.43 & -25.28 ± 15.78 & \multicolumn{1}{c}{$-$}  & \multicolumn{1}{c}{$-$}  & -9.86 ± 15.99 \\
        & & $\sigma$ & 1.38 ± 0.47 & 0.01 ± 0.46 & \multicolumn{1}{c}{$-$}  & \multicolumn{1}{c}{$-$}  & 1.09 ± 0.28 \\
        & & $\Delta\ln(\mathcal{Z})$ & 0.93 ± 0.04 & 0.42 ± 0.05 & \multicolumn{1}{c}{$-$}  & \multicolumn{1}{c}{$-$}  & 0.6 ± 0.24 \\
        \hline
	\end{tabular}
	\end{adjustbox}
\end{table*}

\subsection{Likelihood}
The relationship between the \hi mass and stellar mass of galaxies can be fully described by a bivariate distribution function of \hi mass and stellar mass, i.e. the conditional \hi mass function on the stellar mass. 

We first assume the probability of having a \hi mass $(M_{\rm HI})$ at a given stellar mass $(M_{\star})$ follows the normal function,
\begin{equation}
    P(M_{\rm HI}|M_\star) = \frac{1}{\sqrt{2\pi}\sigma_{\rm HI}} e^{-\frac{1}{2}\left(\frac{\log_{10}(M_{\rm HI})-\mu}{\sigma_{\rm HI}}\right)^2},
    \label{eq:norm}
\end{equation}
where $\mu = a[\log_{10}(M_{\star})-10]+b$, and $\sigma_{\rm HI}$ is the intrinsic scatter in \hi mass at a given stellar mass. We define this normal function as our base ``\textbf{Model A}".

A priori we do not know the underlying relation between stellar mass and \hi mass. We therefore also define two additional models where the underlying distribution of \hi mass is allowed to be asymmetric.

The first asymmetric model is a conditional probability density  with the form of a Schechter function:
\begin{equation}
    P(M_{\rm HI}|M_\star) = \ln(10) \space \phi_\ast\left(\frac{M_{\rm HI}}{M_\ast}\right)^{\alpha + 1} e^{-\frac{M_{\rm HI}}{M_\ast}},
    \label{eq:schechter}
\end{equation}
where $\log_{10}(M_\ast)=a [\log_{10}(M_{\star})-10]+b$ and $\alpha$ correspond to the characteristic mass and faint-end slope respectively, and $\phi_\ast$ is the normalization such that the integration of Eq.~\ref{eq:schechter} over the logarithmic \hi mass is equal to 1, therefore not a free parameter. We note this Schechter function form as ``\textbf{Model B}".

The second asymmetric model that we propose is a skew-normal probability density distribution given by:
\begin{equation}
P(M_{\rm HI}|M_\star) = \frac{2}{\sigma \sqrt{2\pi}} \, e^{-\frac{(\log_{10}(M_{\rm HI}) - \mu)^2}{2\sigma^2}} \, \Phi\left(\alpha \cdot \frac{\log_{10}(M_{\rm HI}) - \mu}{\sigma}\right),
\label{eq:skewnorm}
\end{equation}
where $\alpha, \sigma, \mu=a [\log_{10}(M_{\star})-10]+b$ are free parameters, and $\Phi(x) = \frac{1}{2} \left[ 1 + \mathrm{erf}\left(\frac{x}{\sqrt{2}}\right) \right]$. We note this skew-normal form as "\textbf{Model C}".

With an assumed conditional probability distribution, the probability of measuring a flux, $S_{\rm m}$, for a single source, can be expressed as
\begin{equation}
P(S_{\rm m}|M_\star) = \int d M_{\rm HI}P(M_{\rm HI}|M_\star) P_{\rm n}(S_{\rm m}-S(M_{\rm HI})),
\label{eq:probability}
\end{equation}
where $S(M_{\rm HI})$ is given by Eq.~\eqref{eq:factor}, and $P_{\rm n}$ is the noise distribution which we assume to be consistent with a normal distribution centered on zero with standard deviation $\sigma_{\rm n}$. Figure B2 in \cite{Heywood_2024} shows that the mean Pearson kurtosis are close to the ideal Gaussian value of 3, and a slight deviation from Gaussianity for the noise behavior does not affect our main results. (We tested this using a model distribution for the $M_{\star} - M_{\rm HI}$ relation and injected the simulated sources in the real data cube and performed the same analaysis. Full details can be found in Appendix~\ref{sec:injection_and_recovery}). We emphasise that this assumption is the same for the more traditional stacking technique, where the signal is weighted by the noise measured in nearby offset positions or either side of the emission line, also under the assumption of Gaussian noise \citep[e.g.][]{Healy_2019}.

The likelihood of all the sources having the measured fluxes, given the model and known stellar masses, is given by
\begin{equation}
    \begin{aligned}
    \mathcal{L} \propto  \prod_{\rm source} P(S_{\rm m} &|M_\star, {\rm A} (a, b, \sigma_{\rm HI}),\\[-1.0em]
    \rm or\; &|M_\star, {\rm B}(a, b, \alpha),\\
    \rm or\; &|M_\star, {\rm C}(a, b, \alpha, \sigma)).
    \label{eq:likeli}
    \end{aligned}
\end{equation}
By maximizing Eq.~\eqref{eq:likeli}, we obtain the best fitting $M_{\rm HI} - M_{\star}$ relation described by Models A, B and C for a given galaxy sample. 

At this point, we would like to note again that our Eq.~\ref{eq:probability} does not require a formal \hi detection to calculate the probability of measuring the flux, $S_{\rm m}$,  since this flux can always be extracted based on the source location from an optical catalog regardless of its signal to noise ratio (SNR) in the radio image. The $\sigma_{\rm n}$ is measured on a per source basis from propagating the per channel noise measurement over the 600 km s$^{-1}$ spectral window and $\sigma_{\rm n}$ is not assumed to be constant. A higher noise level for a low SNR source will result in a wider $P_{\rm n}$, which will reduce the contribution of the $S_{\rm m}$ for this source to the total likelihood of the Eq.~\ref{eq:likeli} for constraining the assumed model of $P(M_{\rm HI}|M_\star)$ in the Eqs.~\ref{eq:norm},~\ref{eq:schechter} and~\ref{eq:skewnorm}.

The average of logarithmic \hi mass for a conditional distribution of $\log_{10}(M_{\rm HI})$ on the stellar mass is
\begin{equation}
    \langle{\log_{10}(M_{\rm HI})}\rangle=\int \log_{10}(M_{\rm HI})P(M_{\rm HI}|M_\star)d\log_{10}(M_{\rm HI}).
	\label{eq:meanlog}
\end{equation}

In comparison, the logarithmic of average \hi mass is given from our modelling by
\begin{equation}
    \log_{10}(\langle{M_{\rm HI}}\rangle)=\log_{10}\left(\int M_{\rm HI}P(M_{\rm HI}|M_\star)d\log_{10}(M_{\rm HI})\right).
	\label{eq:logmean}
\end{equation}
We note that the $\log_{10}(\langle{M_{\rm HI}}\rangle)$ could also be measured directly from the standard co-adding stacking approach, i.e. $\log_{10}\left(\frac{\sum w{M_{\rm HI}^{\rm m}}}{\sum w}\right)$, where $w$ is the weight of each galaxy taken as the reciprocal of the noise variance, and $M_{\rm HI}^{\rm m}$ is the measured \hi mass which is the intrinsic $M_{\rm HI}$ added to the noise. To differentiate these two measurements, we denote the latter one as "Co-adding" measurement.

\begin{figure*}
    \centering
  \begin{subfigure}[b]{0.33\textwidth}
    \includegraphics[height=0.94\columnwidth, width=1\columnwidth]{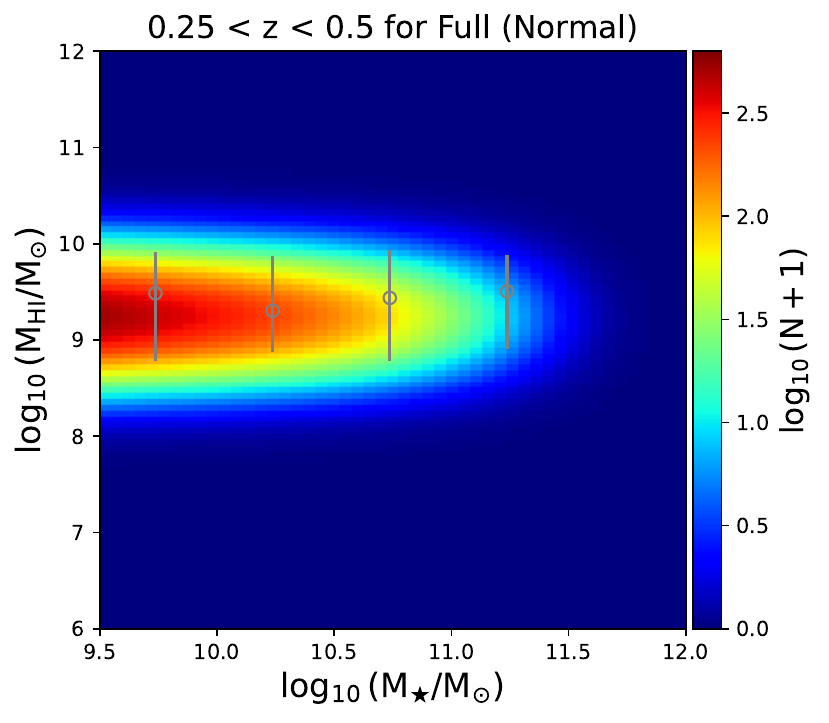}
    \includegraphics[height=0.94\columnwidth, width=1\columnwidth]{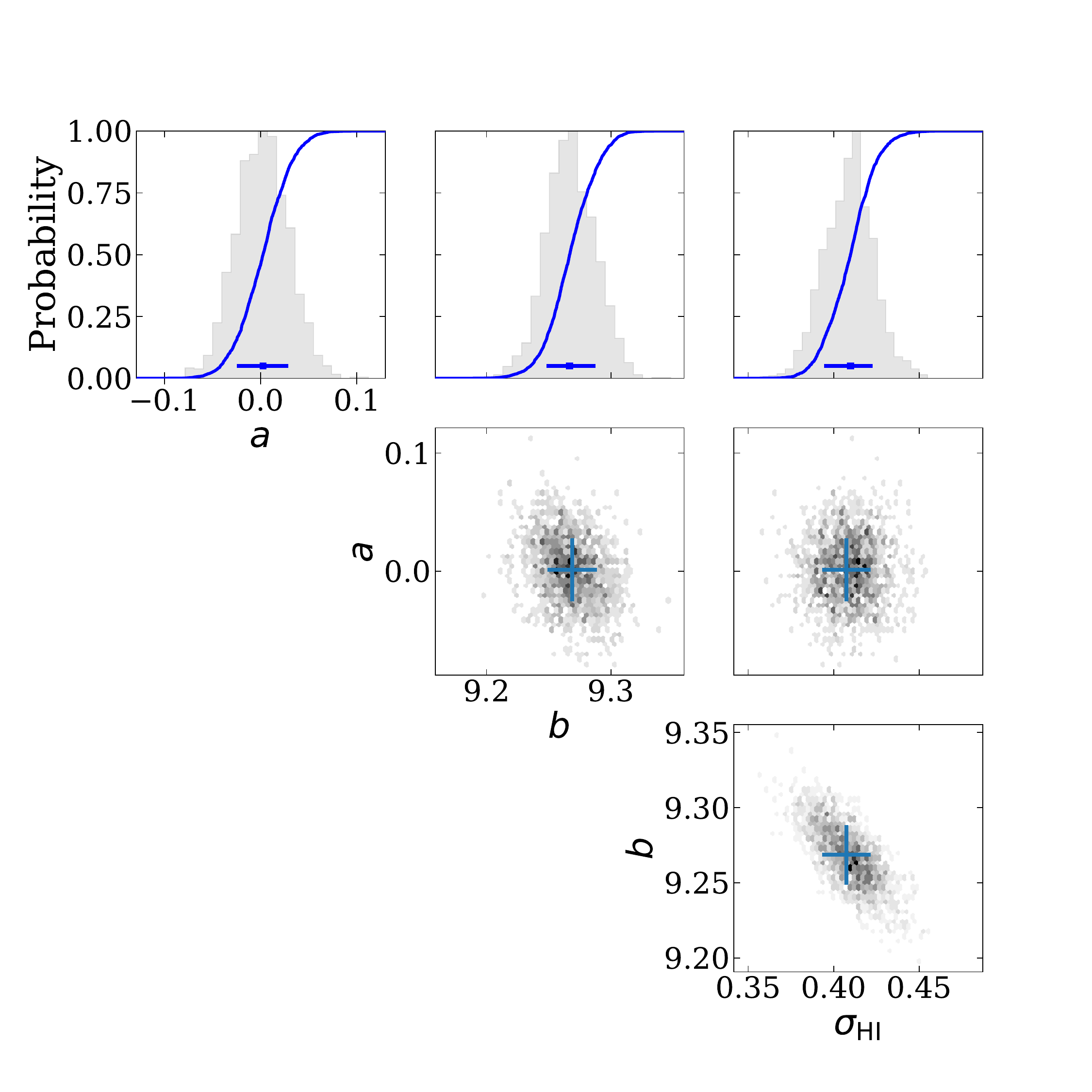}
  \end{subfigure}%
    \hfill
  \begin{subfigure}[b]{0.33\textwidth}
    \includegraphics[height=0.94\columnwidth, width=1.\columnwidth]{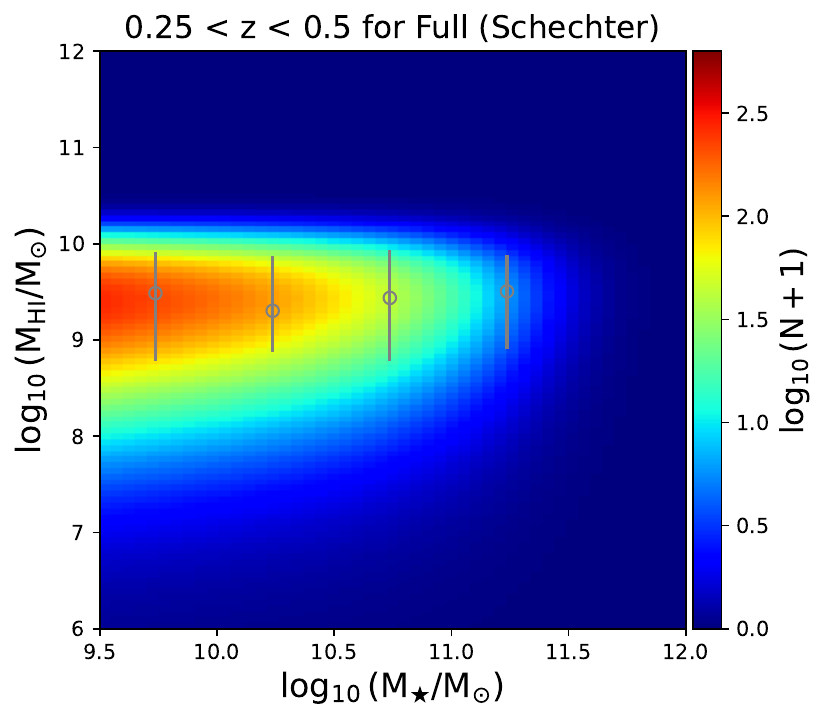}
    \includegraphics[height=0.94\columnwidth, width=1\columnwidth]{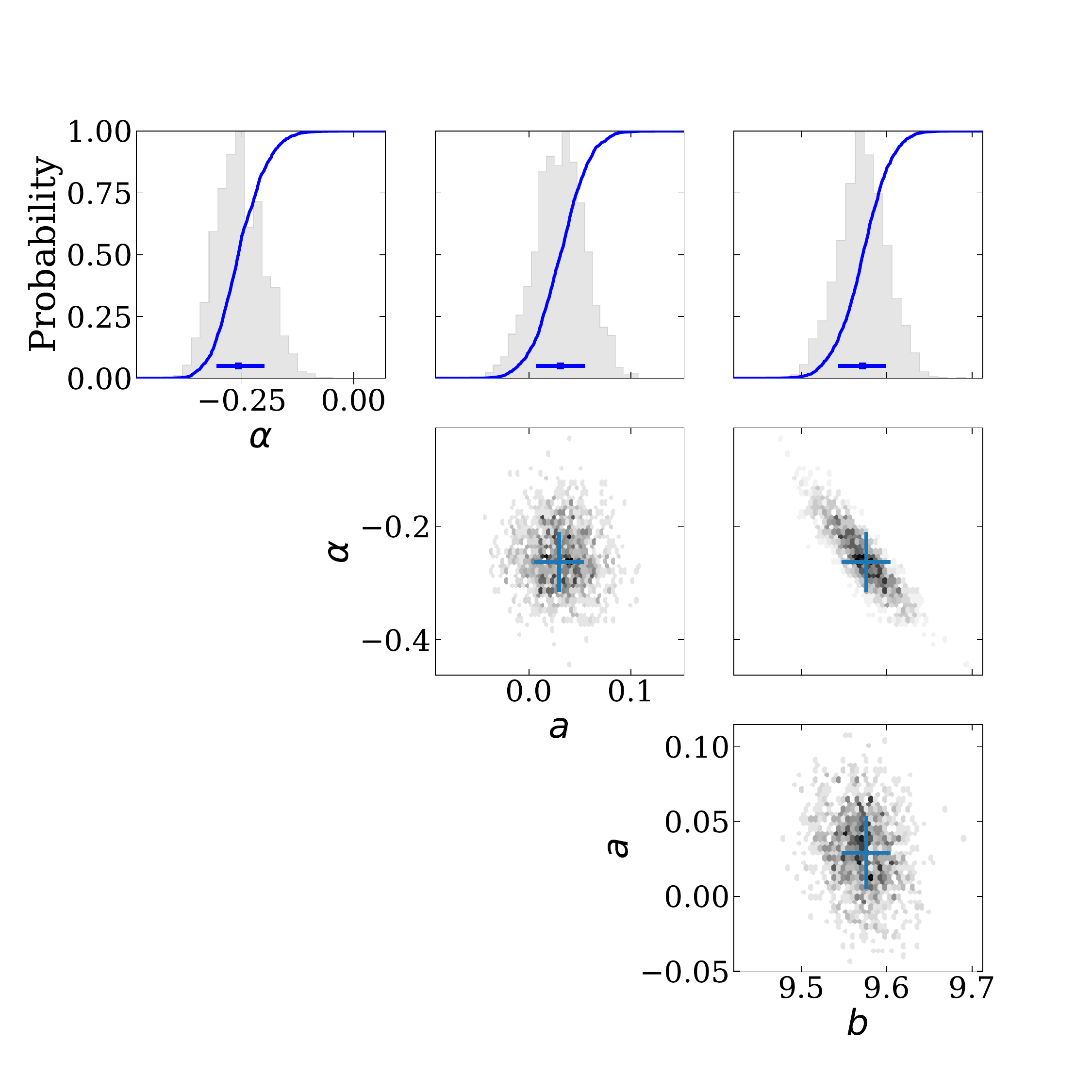}
  \end{subfigure}%
    \hfill
  \begin{subfigure}[b]{0.33\textwidth}
    \includegraphics[height=0.94\columnwidth, width=1.\columnwidth]{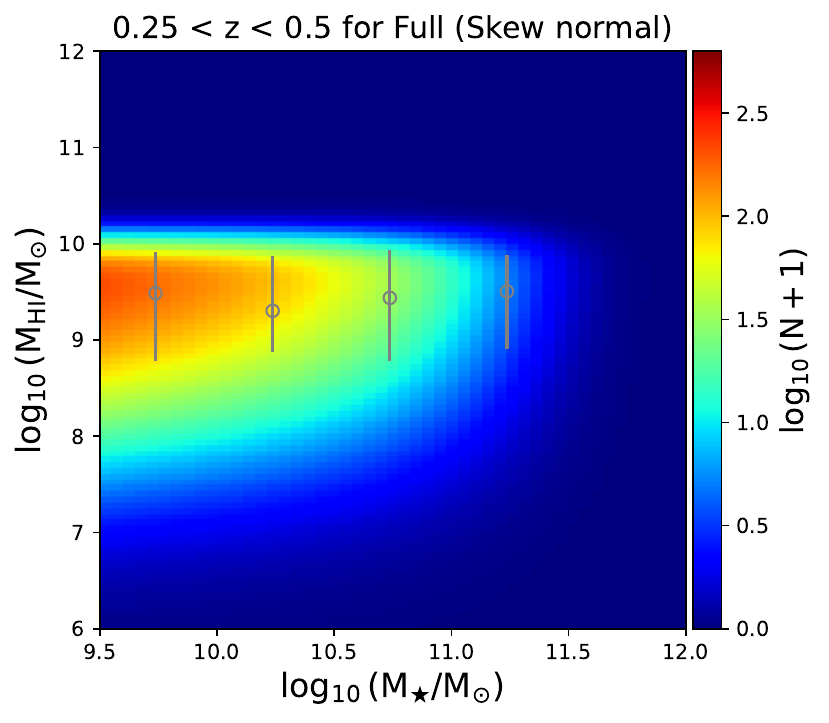}
    \includegraphics[height=0.94\columnwidth, width=1.\columnwidth]{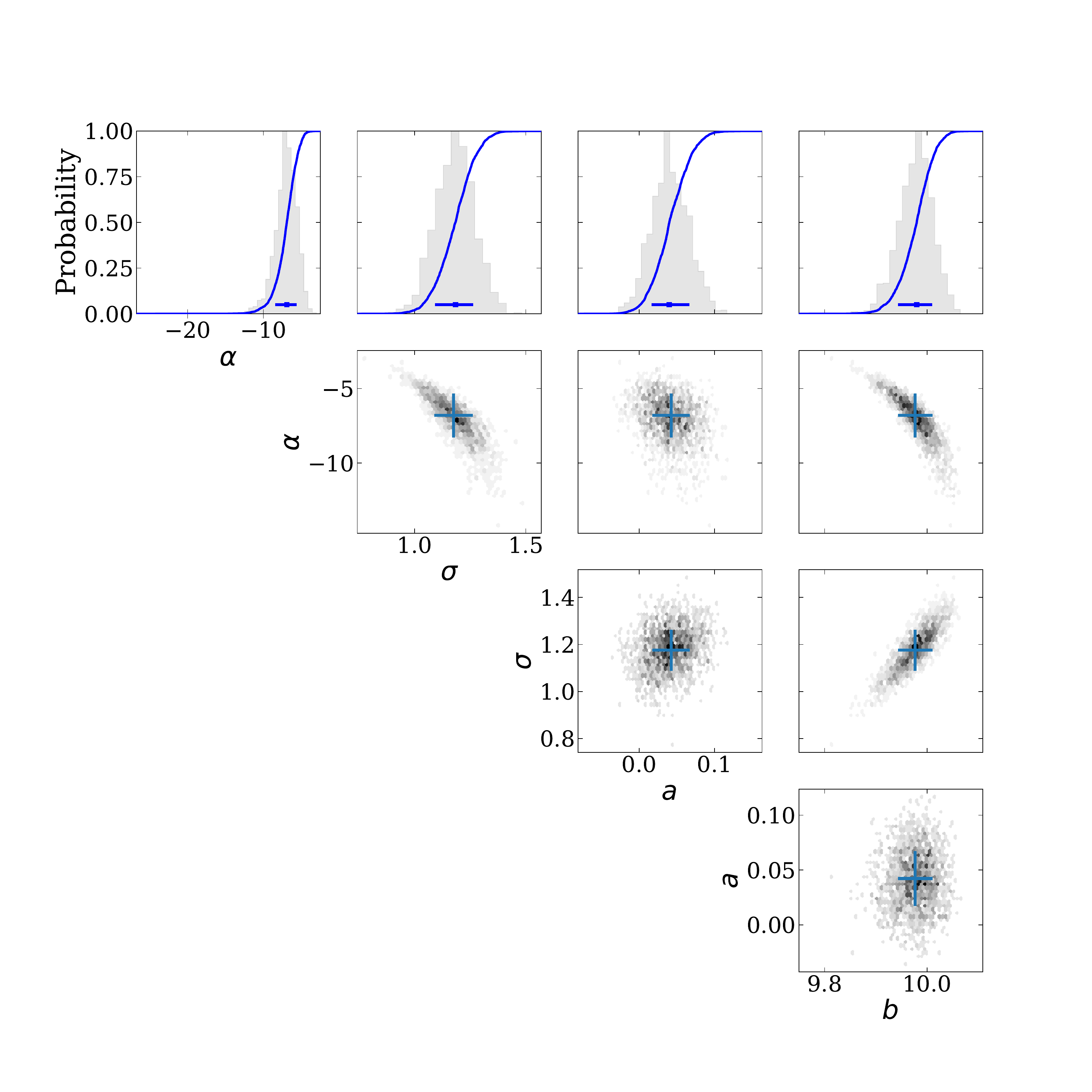}
  \end{subfigure}%
  \caption{Top row: Bivariate distribution of \hi mass and stellar mass for the full sample at $0.25 < z < 0.5$. Three models (Normal, Schechter, Skew normal) are fit to the data from left to right panels. The grey dots with the error bars corresponds to the 16,50 and 84th percentiles for the \hi mass in four stellar mass bins from the SIMBA simulation \protect\citep{Dav__2019}. Bottom row: Posterior probability distributions for the three fitting models correspondingly from from left to right panels. The gray histograms represent the marginal posterior probabilities (1 or 2 dimensional), while the blue curves indicate the cumulative distributions. In the 2 dimensional (2D) posterior plots, the blue crosses mark the parameter set with the maximum likelihood, and the 1$\sigma$ error bars are estimated from the 1D marginal posterior distributions.}
  \label{fig:full_hist}
\end{figure*}

\begin{figure}
    \centering
    \includegraphics[width=\columnwidth]{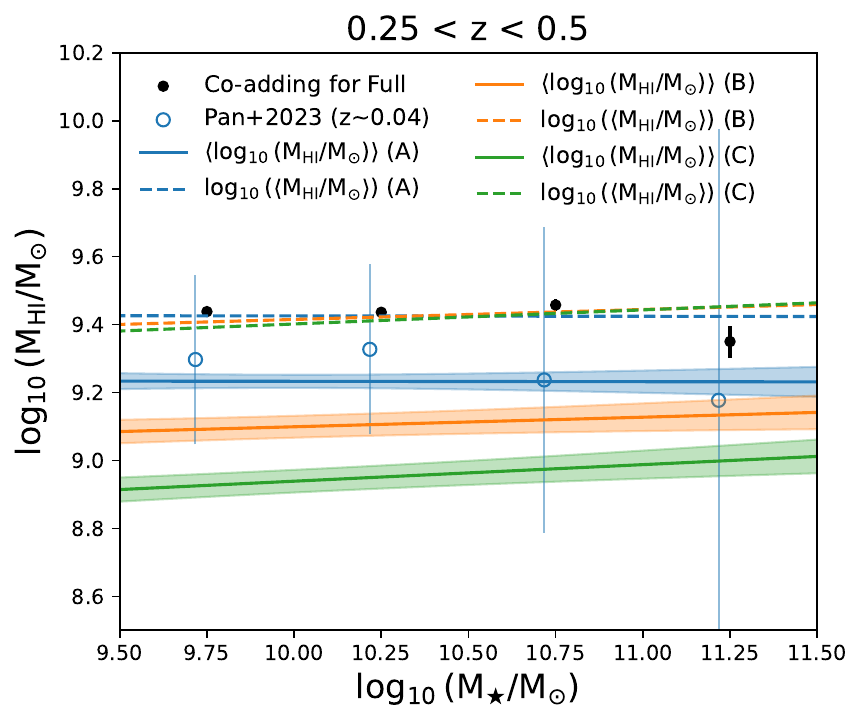}
  \caption{Best fitting $M_{\rm HI} - M_{\star}$ scaling relation for the full sample. The solid and dashed lines are the $\langle\log_{10}(M_{\rm HI})\rangle$ and $\log_{10}(\langle M_{\rm HI}\rangle)$, respectively. The blue, orange and green lines are our Model A, B and C corresponding to the Normal, Schechter and Skew normal distributions. The solid dots are the logarithmic of directly averaging $M_{\rm HI}$, labelled as ``Co-adding" to be differentiated from the $\log_{10}(\langle M_{\rm HI}\rangle)$ driven from our Bayesian modelling. The circles are the average of logarithmic \hi mass from \protect\cite{Pan_2023}. The shaded areas are 1$\sigma$ statistical uncertainties on the $\langle\log_{10}(M_{\rm HI})\rangle$.}
  \label{fig:full_ave}
\end{figure}

The conditional \hi mass function or bivariate distribution  can then be written as
\begin{equation}
    \Phi(M_{\rm HI}|M_\star) = P(M_{\rm HI}|M_\star) \Phi(M_\star),
	\label{eq:chimf}
\end{equation}
where $\Phi(M_\star)$ is the galaxy stellar mass function (GSMF). Therefore, the marginalisation of $\Phi(M_{\rm HI}|M_\star)$ over $M_\star$ results in the \hi mass function. In this paper, we present the \hi mass function for a stellar mass-selected sample with $\log_{10}(M_{\star}/M_\odot)>9.5$ as
\begin{equation}
    \begin{aligned}
    \Phi(M_{\rm HI})_{9.5} & = \int_{9.5}^{+\infty} \Phi(M_{\rm HI}|M_\star)d\log_{10}(M_{\star}) \\ 
    & =  \int_{9.5}^{+\infty} P(M_{\rm HI}|M_\star) \Phi(M_\star)d\log_{10}(M_{\star}).
    \end{aligned}
	\label{eq:himf}
\end{equation}
We use the GSMF measurements for the full sample, spirals, and ellipticals from \cite{Hashemizadeh_2022}, and also split the GSMF of the full sample into GSMFs for the blue and red galaxies.

\subsection{Priors}
The priors, which encode our knowledge of each parameter before incorporating new experimental data, are listed in Table~\ref{tab:priors} for the  Full, Blue and Spiral subsamples. Minor adjustments are made for the red and elliptical subsamples due to their small sample sizes.

\section{Results}
\label{sec:results}

\subsection{The $M_{\rm HI} - M_{\star}$ relation for the full sample}
\label{sec:full}

We present the best fitting bivariate distribution of \hi mass and stellar mass for the full sample in Figure~\ref{fig:full_hist} (top row), with the three best fitting models of Normal, Schechter, and Skew normal from the left to right panels, respectively (the measured bivariate distribution is shown in Appendix~\ref{sec:mhi_m} for completeness). Compared to the Normal model where the \hi scatter is symmetric at a given stellar mass, the Schechter and Skew normal models have the capability of capturing asymmetry in the underlying \hi mass distribution as a function of the stellar mass. The posterior probability distributions for all parameters in the three models are well converged as shown in the bottom row, and our data decisively favour the asymmetric \hi mass distribution for the full galaxy sample, given that the Bayes factors of the Schechter (Model B) and Skew normal (Model C) models compared to the Normal distribution (Model A) are much larger than the critical threshold for most stellar mass bins as reported in Table~\ref{tab:post}. The Skew normal has no better performance than the Schechter model, despite the fact that the Skew normal has four free parameters which add a certain level of flexibility of fitting for the data in comparison with the Schechter form. The physics behind this stems from the fact that the \hi gas mass fraction is non-linear with stellar mass and although there appears to be a strong upper envelope in \hi mass as a function of stellar mass \citep[e.g.][]{Pan_2023}, the distribution of \hi mass is not likely to have such a strong low-\hi mass envelope and that this is likely to be different for spheroids and disk galaxies. We also show the band of \hi mass for the 16, 50 and 84th percentiles in four stellar mass bins at $0.25 < z < 0.5$ from the SIMBA simulation \citep{Dav__2019}. Overall, the bivariate distribution of the \hi mass and stellar mass traces the SIMBA percentiles, especially for the Schechter and Skew normal models where the high density area just above the mass limit is more consistent between our model and the simulated galaxies.

\subsection{Log of the average versus average of the log}
The average of the logarithmic \hi mass, i.e. $\langle\log_{10}(M_{\rm HI})\rangle$, and the logarithmic average of the \hi mass, i.e. $\log_{10}(\langle M_{\rm HI}\rangle)$, as a function of the stellar mass, are shown in solid and dashed lines respectively for the full sample in Figure~\ref{fig:full_ave}. The three models of Normal, Schechter, and Skew normal are colour-coded in blue, orange and green. The difference between $\langle\log_{10}(M_{\rm HI})\rangle$ and $\log_{10}(\langle M_{\rm HI}\rangle)$ for all models is considerable with 0.2-0.3 dex for Normal and Schechter models, and $\sim$0.5 dex for the Skew normal. The solid dots are the logarithmic of direct averaging the measured $M_{\rm HI}$, labelled as ``Co-adding" to be differentiated from the $\log_{10}(\langle M_{\rm HI}\rangle)$ driven from our Bayesian modelling. The $\log_{10}(\langle M_{\rm HI}\rangle)$ would be equal to the Co-adding measurement if our models could perfectly describe the conditional \hi mass distribution on the stellar mass. As shown, the dashed lines are indeed in excellent agreement with the solid dots overall, except for the most massive end of $\log_{10}(M_\star/M_\odot$)$>$11 where the number of sources is only on a scale of a few hundreds, hence contributing low-number statistical uncertainties to our modelling. The overall minimal offsets (less than 0.1 dex) indicate that the modeling of the \hi distribution is  not in perfect agreement with the bivariate distribution of \hi mass and stellar mass from traditional stacking. This may be because the relationship is more complex and is not fully captured by our simple analytic models with three or four parameters. On the other hand, the traditional stacking method is sensitive to outliers, which could result in a biased estimate of the average \hi mass, whereas our modelling of the underlying distribution provides more information on this.

On the other hand, the $\langle\log_{10}(M_{\rm HI})\rangle$ for our base Model A (i.e. the Normal model) also broadly agrees with that from \cite{Pan_2023} where their average of the logarithmic \hi mass is derived from the \hi detections at $z<0.08$. Such an agreement could indicate minimal evolution on the \hi distribution in massive galaxies at $z<0.5$. However, the $\langle\log_{10}(M_{\rm HI})\rangle$ is noticeably lower than the measurement from \cite{Pan_2023} at $\log_{10}(M_\star/M_\odot$)$<$10.5 but slightly higher than that at $\log_{10}(M_\star/M_\odot$)$>$11. We return to this when discussing the conditional \hi mass function in Section~\ref{sec:chimf}.

The finding is also in accord with \cite{Sinigaglia_2022}, suggesting a scenario where the massive galaxies have undergone a significant \hi replenishment through some accretion mechanism, such as minor mergers or gas accretion from the nearby cosmic web \citep{Kleiner_2016}, to maintain their neutral gas storage during the last $\sim$4 Gyr.  However, we note that the approach taken here using a large number of non-\hi detections is quite different from \cite{Pan_2023} using a limited number of \hi detections with large statistical uncertainties of $\sim0.3$ dex. The shallow slopes across all models suggest the existence of an upper \hi mass limit beyond which a galaxy can no longer retain more \hi gas, independent of the amount of their stellar component. This trend is also in line with the findings from \cite{huang2012}, \cite{Maddox2015} and \cite{Pan_2023} on the non-linear $M_{\rm HI} - M_{\star}$ scaling relation with a transition stellar mass between $10^9$ and $10^{9.5} M_\odot$.

\begin{figure*}
  \centering
  \begin{subfigure}[s]{0.45\textwidth}
    \includegraphics[height=0.94\columnwidth, width=1\columnwidth]{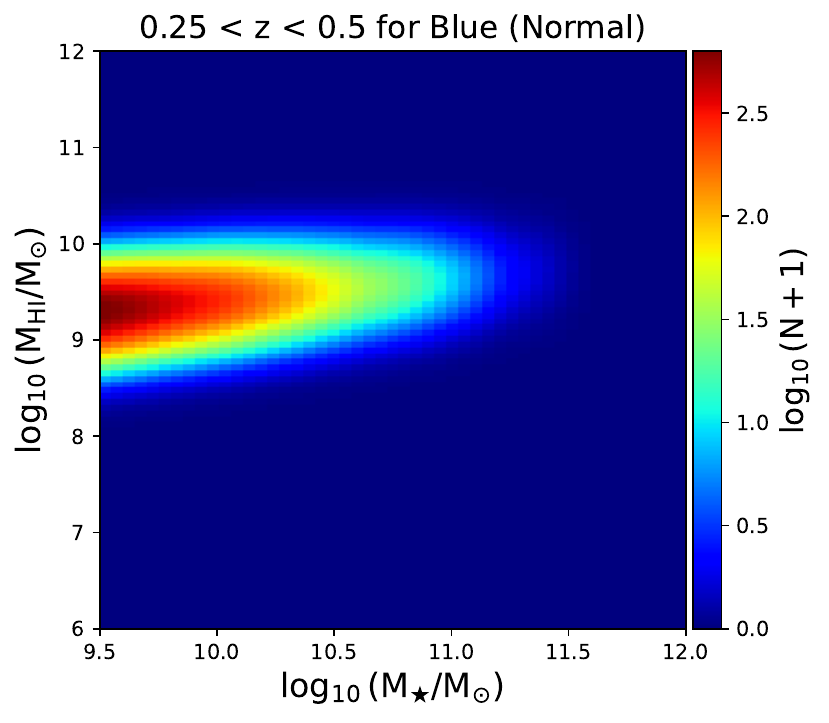}
    \includegraphics[height=0.94\columnwidth, width=1.\columnwidth]{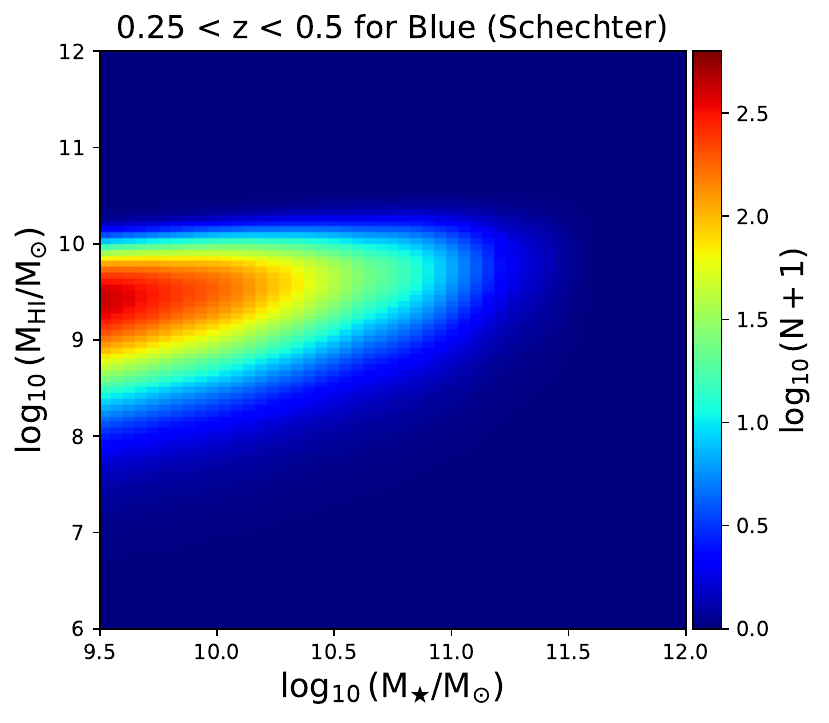}
    \includegraphics[height=0.94\columnwidth, width=1\columnwidth]{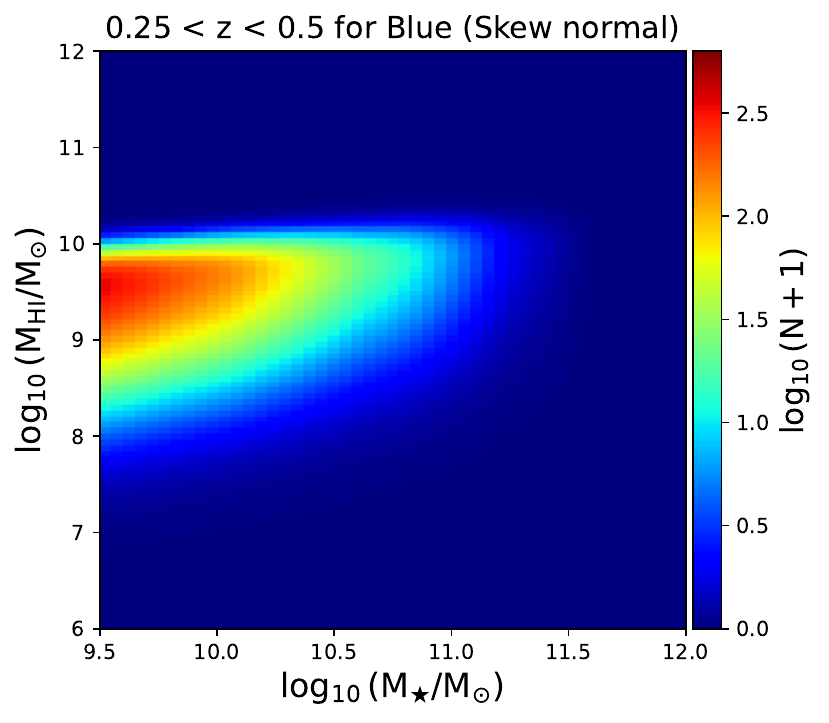}
  \end{subfigure}%    
  \begin{subfigure}[s]{0.45\textwidth}
    \includegraphics[height=0.94\columnwidth, width=1\columnwidth]{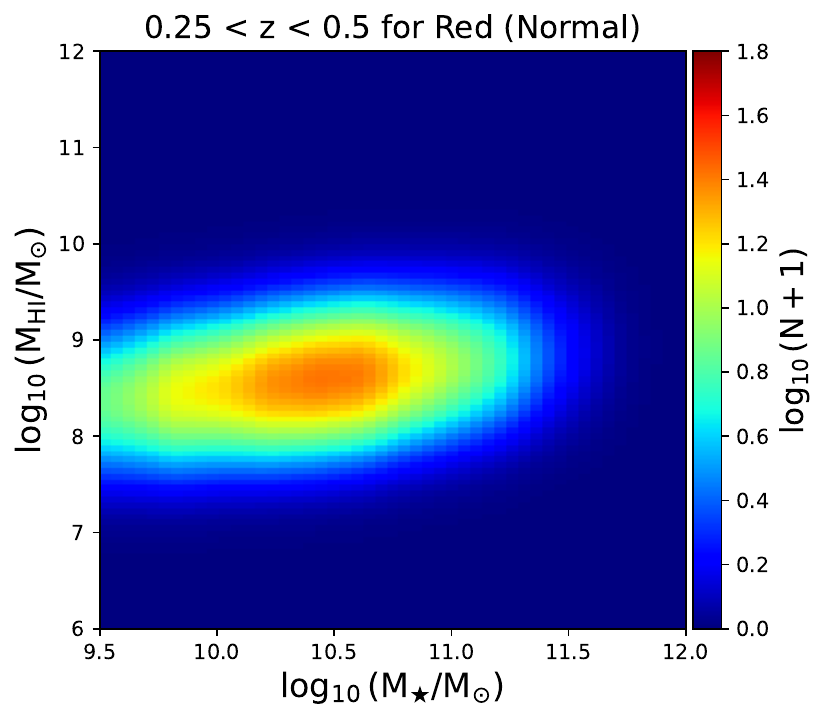}
    \includegraphics[height=0.94\columnwidth, width=1.\columnwidth]{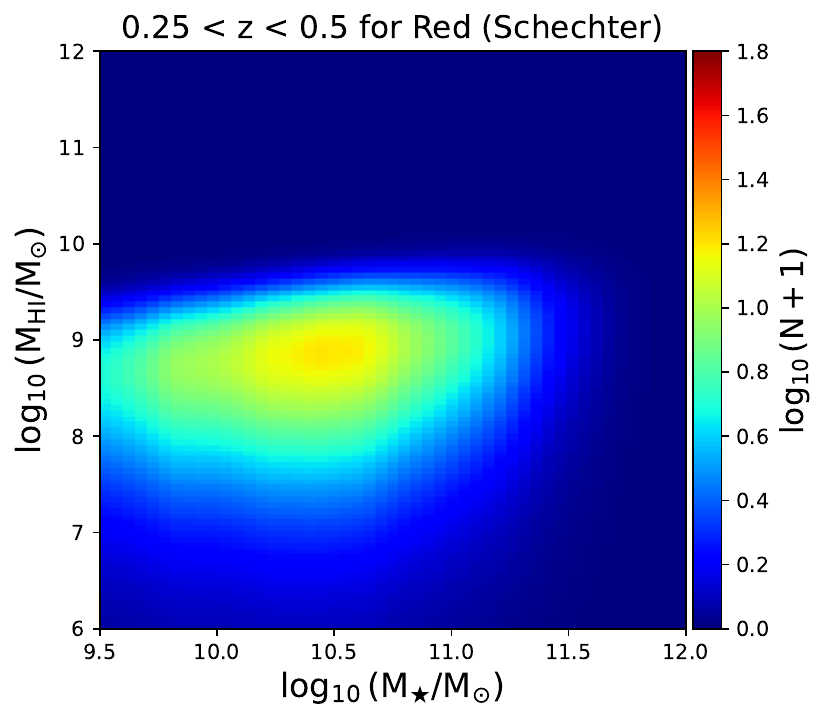}
    \includegraphics[height=0.94\columnwidth, width=1\columnwidth]{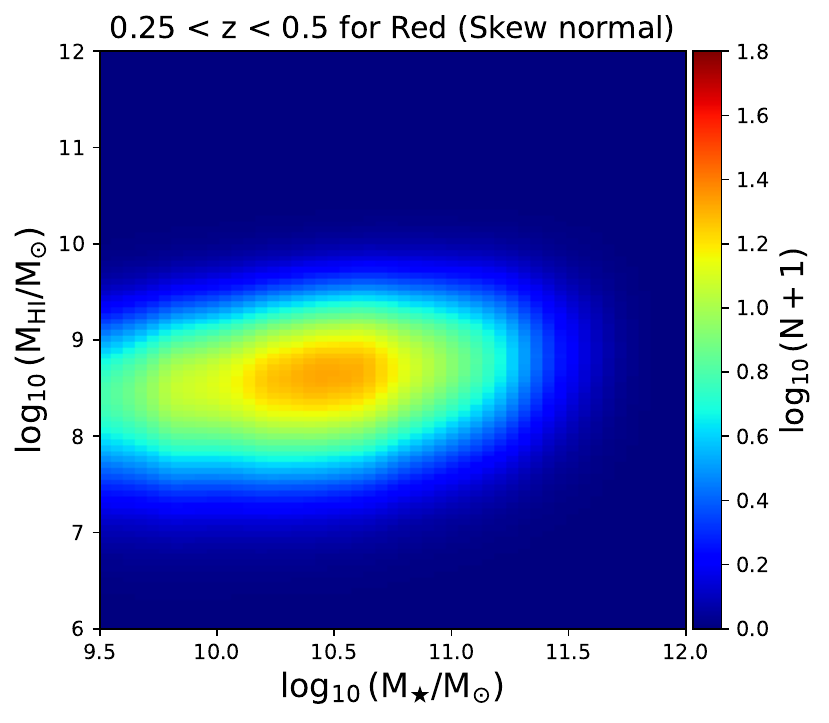}
  \end{subfigure}%
  \caption{Bivariate distribution of \hi mass and stellar mass for blue (left) and red (right) galaxies at $0.25 < z < 0.5$. Three models (Normal, Schechter, Skew normal) are fit to the data from top to bottom panels. The corresponding posteriors are appended in Figure~\ref{fig:color_post}.}
  \label{fig:color}
\end{figure*}

\begin{figure*}
  \centering
  \begin{subfigure}[s]{0.43\textwidth}
    \includegraphics[width=\columnwidth]{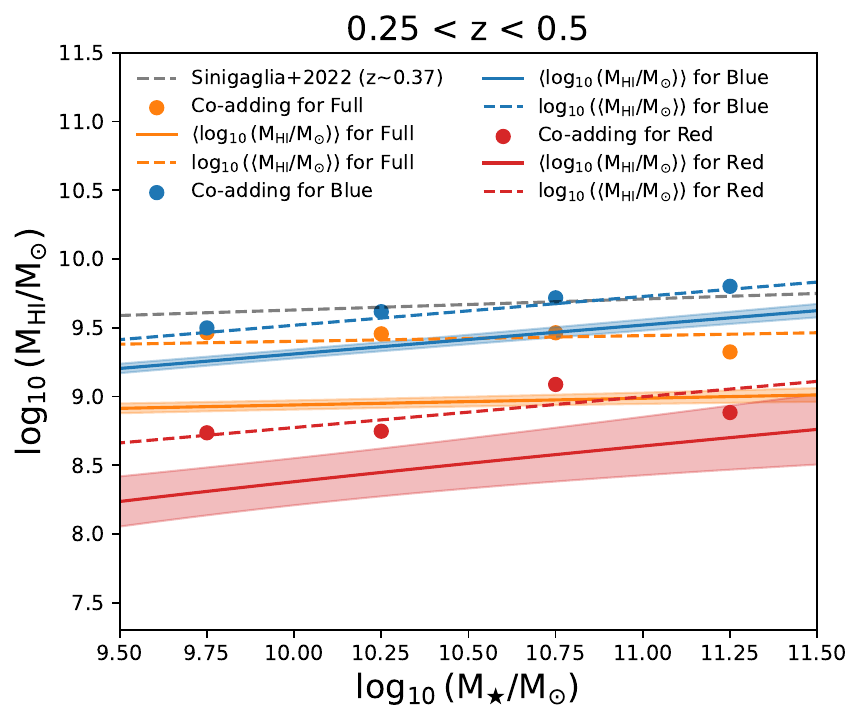}
  \end{subfigure}%    
  \hspace{3mm}
  \begin{subfigure}[s]{0.43\textwidth}
    \includegraphics[width=\columnwidth]{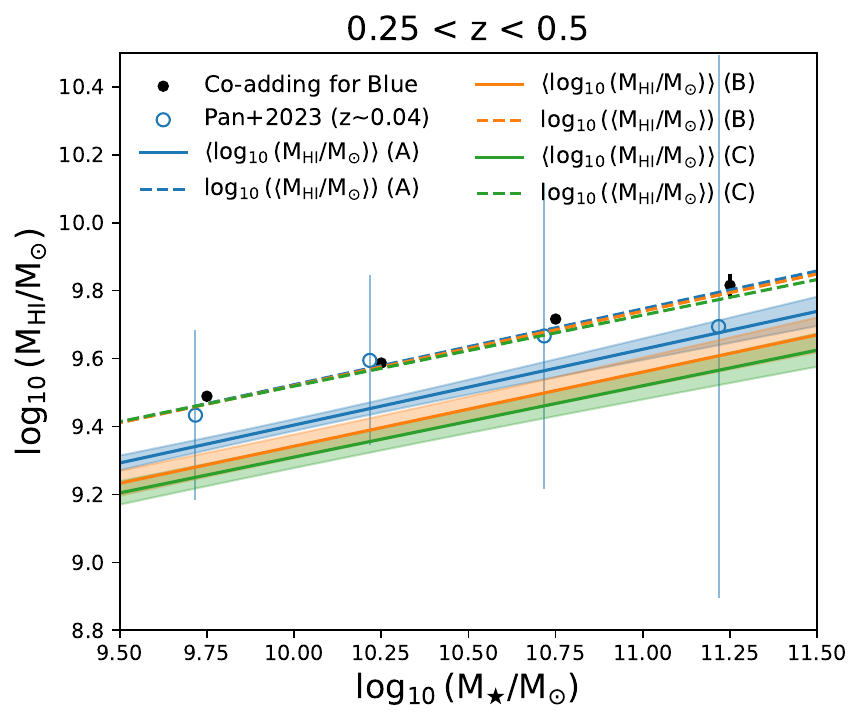}
  \end{subfigure}%
  \caption{Left panel: Best fitting $M_{\rm HI} - M_{\star}$ scaling relation for the full sample, blue and red galaxies with the Skew normal model, colour-coded by orange, blue and red, respectively. Right panel: Best fitting $M_{\rm HI} - M_{\star}$ scaling relation for blue galaxies with Normal, Schechter and Skew normal models labeled as Models A, B and C. The solid and dashed lines are the $\langle\log_{10}(M_{\rm HI})\rangle$ and $\log_{10}(\langle M_{\rm HI}\rangle)$, respectively. The solid dots are the logarithmic of directly averaging $M_{\rm HI}$, labelled as ``Co-adding". The dashed grey line is from \protect\cite{Sinigaglia_2022}, and the blue circles are from \protect\cite{Pan_2023}. The shaded areas are 1$\sigma$ statistical uncertainties on the $\langle\log_{10}(M_{\rm HI})\rangle$.}
  \label{fig:blue}
\end{figure*}

\begin{figure*}
    \centering
  \begin{subfigure}[b]{0.33\textwidth}
    \includegraphics[height=0.94\columnwidth, width=1\columnwidth]{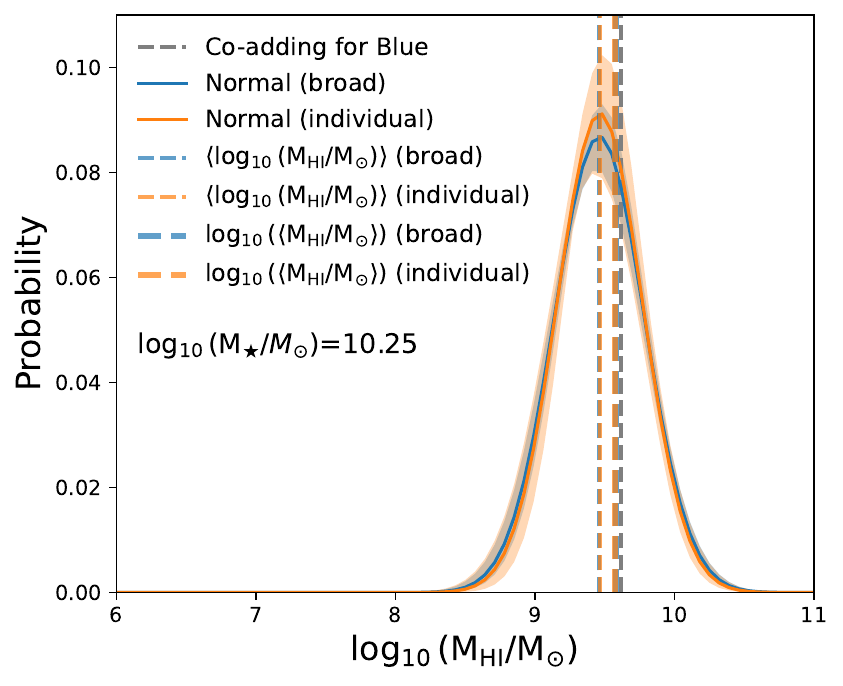}
  \end{subfigure}%
    \hfill
  \begin{subfigure}[b]{0.33\textwidth}
    \includegraphics[height=0.94\columnwidth, width=1.\columnwidth]{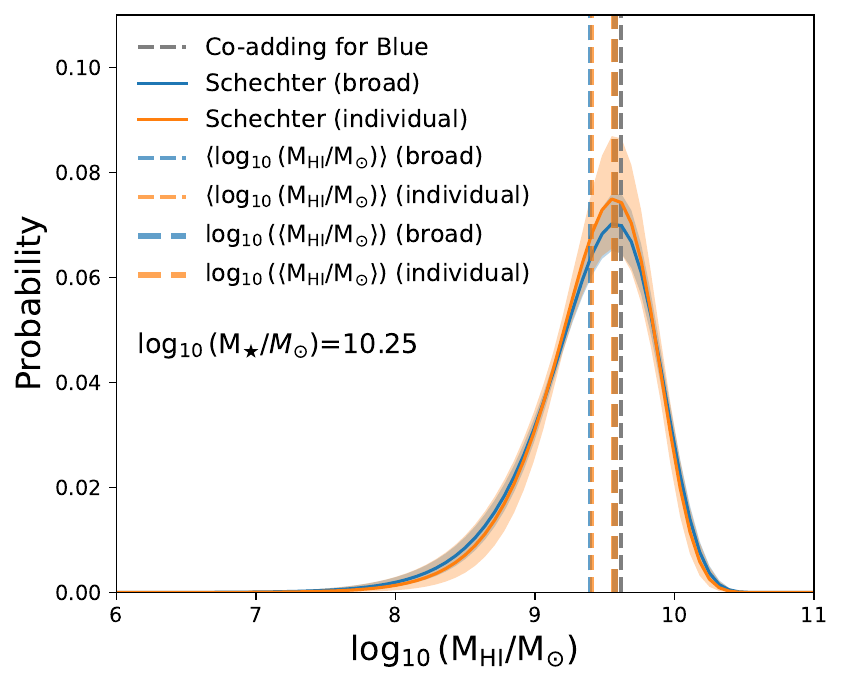}
  \end{subfigure}%
    \hfill
  \begin{subfigure}[b]{0.33\textwidth}
    \includegraphics[height=0.94\columnwidth, width=1\columnwidth]{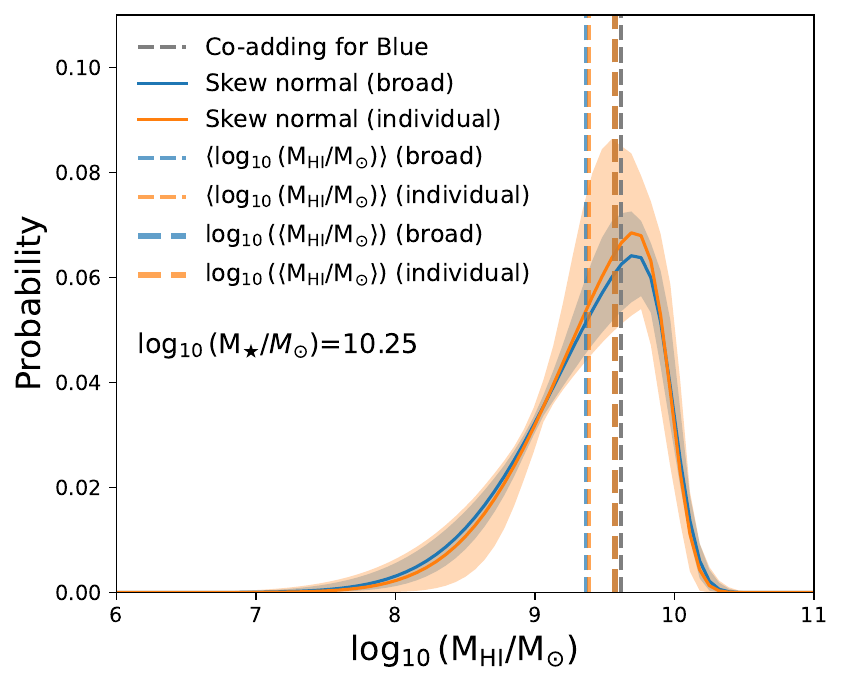}
  \end{subfigure}%
  \caption{Conditional probability distribution of the \hi mass on the stellar mass of $\log_{10}(M_{\star}/M_\odot)=10.25$ for blue galaxies at $0.25 < z < 0.5$. Three models (Normal, Schechter, Skew normal) are fit to the data from left to right panels.  The blue and orange are the probability distributions for the broad and individual mass ranges, respectively. The thin and thick dashed lines are the $\langle\log_{10}(M_{\rm HI})\rangle$ and $\log_{10}(\langle M_{\rm HI}\rangle)$, respectively. The grey dashed line is the logarithmic of directly averaging $M_{\rm HI}$, labelled as ``Co-adding".}
  \label{fig:color_BvsI}
\end{figure*}

\subsection{Colour dependence}
\label{sec:color}

The blue galaxies are generally \hi gas rich compared to the red population; therefore, they should occupy a different parameter space of \hi mass and stellar mass from red galaxies. We now investigate the color dependence of the $M_{\rm HI} - M_{\star}$ relation by dividing the full sample into blue and red galaxies as illustrated in Figure~\ref{fig:ur_vs_mstar}.

In Figure~\ref{fig:color}, we show the bivariate distribution of \hi mass and stellar mass for the blue and red samples in the left and right panels, with the three best fitting models from top to bottom panels. Indeed, the blue galaxies are tightly clustered in the lowest stellar mass bins of $9.5<\log_{10}(M_{\star}/M_\odot)<10.5$, and relatively high \hi mass range of $9<\log_{10}(M_{\rm HI}/M_\odot)<10$, especially with the preferred models of Schechter and Skew normal. The red galaxies instead sit broadly in the intermediate stellar mass range of $9.5<\log_{10}(M_{\star}/M_\odot)<11.5$, and the \hi mass range of $7.5<\log_{10}(M_{\rm HI}/M_\odot)<9.5$, therefore tend to have lower average \hi mass than the blue galaxies.
This trend can be noticed more clearly in the left panel of Figure~\ref{fig:blue}, where the $\langle\log_{10}(M_{\rm HI})\rangle$ and $\log_{10}(\langle M_{\rm HI}\rangle)$ of red galaxies in red solid and dashed lines are both lower than those of the full sample. The difference between the $\log_{10}(\langle M_{\rm HI}\rangle)$ derived from our modelling and dots measured from the ``Co-adding" experiment is due to the small sample size of red galaxies for their broad distribution in the $M_{\rm HI} - M_{\star}$ space.

The $\log_{10}(\langle M_{\rm HI}\rangle)$ of the blue galaxies (blue dashed line) is in good agreement with the stacking result (grey dashed line) from \cite{Sinigaglia_2022} where they use the MIGHTEE E.S. data to measure the logarithmic average of the \hi mass as a function of the stellar mass bin for selected star forming galaxies, with a minor differing slope. We note that \cite{Bianchetti_2025} use both MIGHTEE E.S. and COSMOS \hi Large Extra-galactic Survey \citep[CHILES;][]{chiles2016} data and measure a relatively steeper slope than that from \cite{Sinigaglia_2022} mostly due to the difference in the spectroscopic catalogue used.

In the right panel of Figure~\ref{fig:blue}, we show the difference for our three best fitting models to the blue galaxy sample. Although the logarithmic average of the \hi mass, $\log_{10}(\langle M_{\rm HI}\rangle)$, among the three models (dashed lines) is in excellent agreement, the average of the logarithmic \hi mass, $\langle\log_{10}(M_{\rm HI})\rangle$, demonstrates noticeable differences. This is fundamentally rooted in their different mathematical forms, where the logarithmic average is only sensitive to the the massive end of $\log_{10}(M_{\rm HI})$ distribution while the average of the logarithmic is sensitive to the whole mass space of $\log_{10}(M_{\rm HI})$ distribution. Therefore, the difference of these models manifests in the $\langle\log_{10}(M_{\rm HI})\rangle$ generally. Greater asymmetry in the fitting model results in a larger deviation of $\langle\log_{10}(M_{\rm HI})\rangle$ from the value predicted by our symmetrical Model A.

The maximum difference on the zero point of the $\langle\log_{10}(M_{\rm HI})\rangle$ as a function of the stellar mass between Models A, B and C is about 0.1 dex which is smaller than that for the full sample in Figure~\ref{fig:full_ave} due to a tighter $M_{\rm HI} - M_{\star}$ relation for the blue sample. Their slopes are rather close, and all are steep when compared to the relatively flat slopes for the full sample, indicating that the \hi mass of massive blue galaxies is positively correlated with the stellar mass. This trend is also true for the red galaxies due to the existence of a large fraction of red disk galaxies. The $\langle\log_{10}(M_{\rm HI})\rangle$ for our model A (i.e. the Normal model) is slightly lower than the double-power law fitting for the underlying average \hi mass for the late-type galaxies from \cite{Pan_2023} but still within 1$\sigma$ uncertainties. 

The above analysis is based on the modelling of galaxies across a broad mass range of $\log_{10}(M_{\star}/M_\odot)>9.5$, with only one set of free parameters. To investigate whether we can improve the modeling by splitting the total stellar masses into individual mass bins, we plot the probability distribution of the \hi mass in a typical mass bin of $10<\log_{10}(M_{\star}/M_\odot)<10.5$ for the blue sample in Figure~\ref{fig:color_BvsI}, and show the three models of Normal, Schechter and Skew normal from left to right panels. 

The probability distributions for the broad and individual mass ranges are represented in blue and orange, respectively. The $\langle\log_{10}(M_{\rm HI})\rangle$ and $\log_{10}(\langle M_{\rm HI}\rangle)$ are shown in thin and thick dashed lines. For the blue sample, the best fitting probability profiles from the broad mass range are in excellent agreement with those from the individual mass range for all three models, except for the regions surrounding the probability peaks. The Skew normal distribution has the largest difference between the $\langle\log_{10}(M_{\rm HI})\rangle$ and $\log_{10}(\langle M_{\rm HI}\rangle)$, as shown already in the right panel of Figure~\ref{fig:blue}. The minimal difference between the $\log_{10}(\langle M_{\rm HI}\rangle)$ and the result from the Co-adding measurement is also the same as we see at $\log_{10}(M_{\star}/M_\odot)=10.25$ in the right panel of Figure~\ref{fig:blue}. 
For the red galaxies, the Schechter and Skew normal models are not preferred by our data over the Normal distribution, likely due to the small sample size of the red subsamples, based on the negative Bayes factor of $\Delta\ln(\mathcal{Z})=-0.21$ for the Schechter model and the $\Delta\ln(\mathcal{Z})=-1.47$ for the Skew normal at $\log_{10}(M_{\star})>9.5$ in Table~\ref{tab:post}.

\begin{figure*}
  \centering
  \begin{subfigure}[s]{0.45\textwidth}
    \includegraphics[height=0.94\columnwidth, width=1\columnwidth]{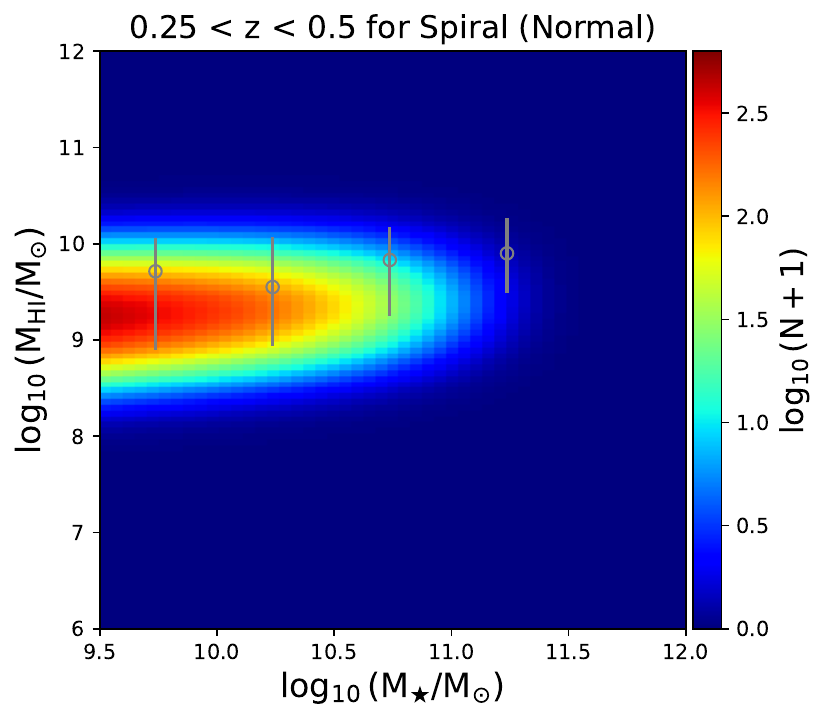}
    \includegraphics[height=0.94\columnwidth, width=1.\columnwidth]{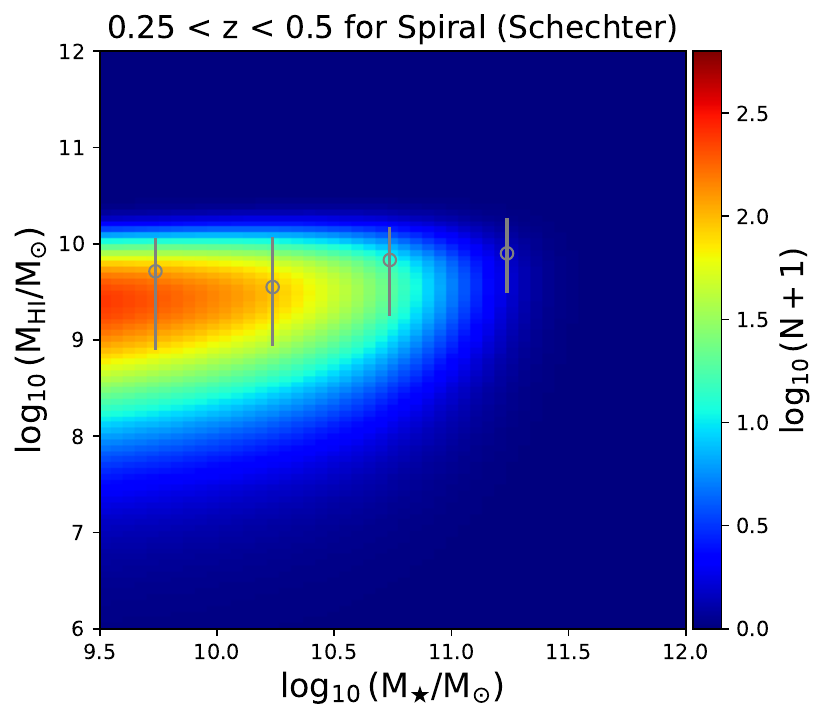}
    \includegraphics[height=0.94\columnwidth, width=1\columnwidth]{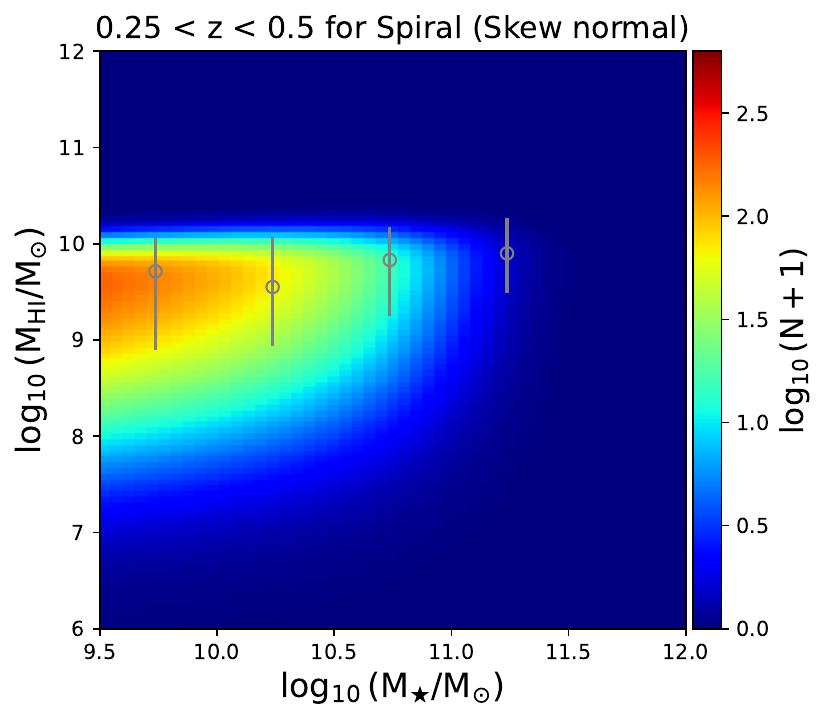}
  \end{subfigure}%    
  \begin{subfigure}[s]{0.45\textwidth}
    \includegraphics[height=0.94\columnwidth, width=1\columnwidth]{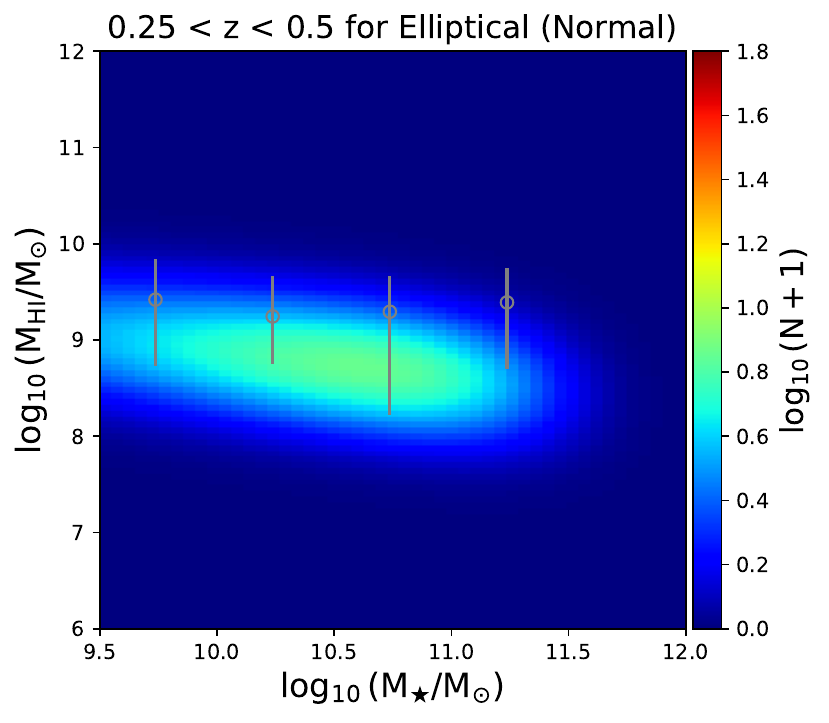}
    \includegraphics[height=0.94\columnwidth, width=1.\columnwidth]{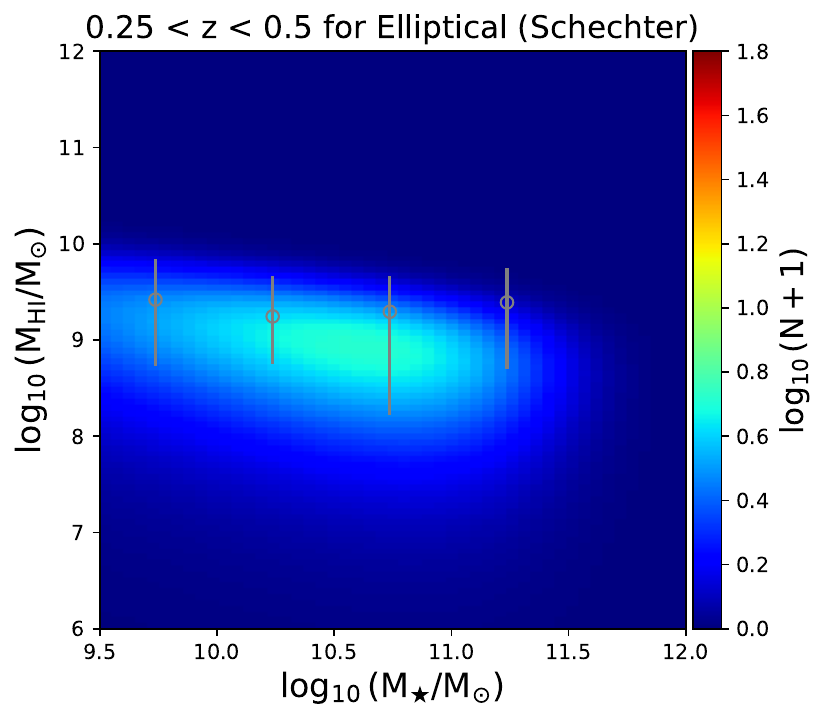}
    \includegraphics[height=0.94\columnwidth, width=1\columnwidth]{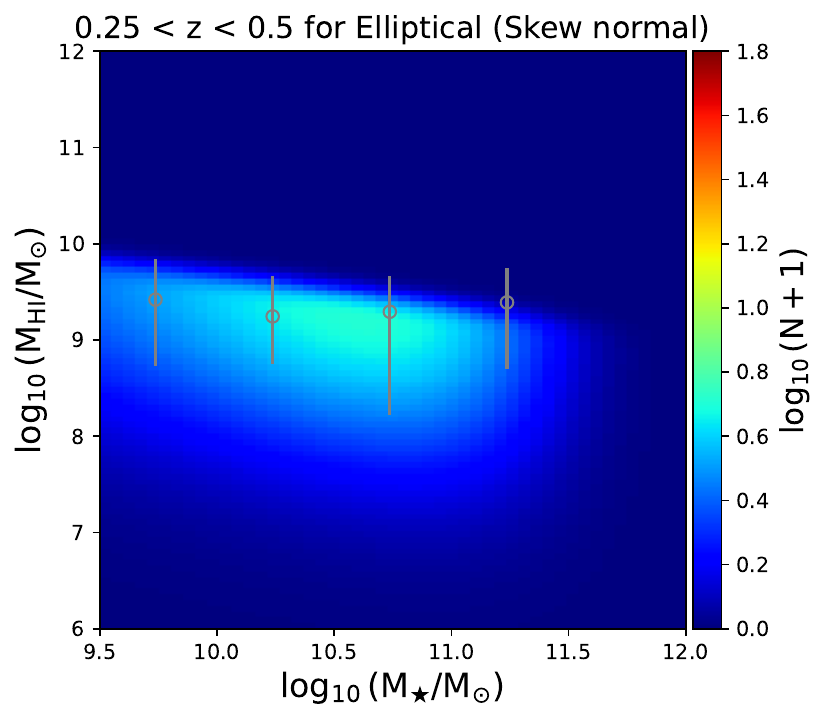}
  \end{subfigure}%
  \caption{Bivariate distribution of \hi mass and stellar mass for spiral (left) and elliptical (right) galaxies at $0.25 < z < 0.5$. Three models (Normal, Schechter, Skew normal) are used to fit the data from top to bottom panels. The grey circles with the error bars corresponds to the 16,50 and 84th percentiles for the \hi mass in four stellar mass bins from the SIMBA simulation \protect\citep{Dav__2019}. The corresponding posteriors are appended in Figure~\ref{fig:morph_post}.}
  \label{fig:morph}
\end{figure*}

\begin{figure}
    \centering
    \includegraphics[width=\columnwidth]{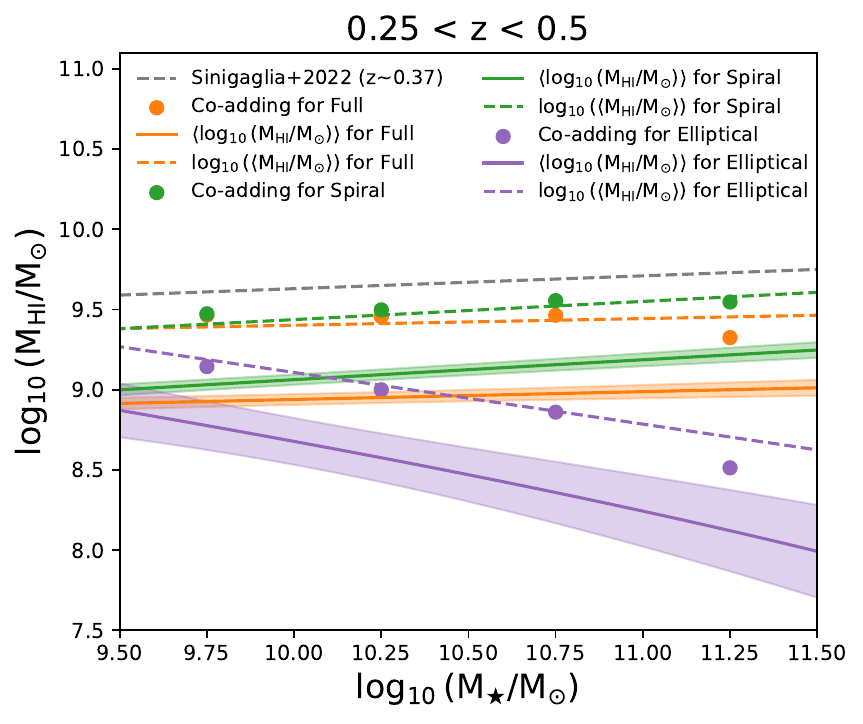}
  \caption{Best fitting $M_{\rm HI} - M_{\star}$ scaling relation for the full sample, spiral and elliptical galaxies with the Skew normal model, colour-coded by orange, green and purple, respectively. The solid and dashed lines are the $\langle\log_{10}(M_{\rm HI})\rangle$ and $\log_{10}(\langle M_{\rm HI}\rangle)$, respectively. The solid dots are the logarithmic of directly averaging $M_{\rm HI}$, labelled as ``Co-adding". The dashed grey line is from \protect\cite{Sinigaglia_2022}.}
  \label{fig:spiral}
\end{figure}

\subsection{Morphology dependence}
\label{sec:morph}

We also split the full sample into spirals and ellipticals to study the morphology dependence of the $M_{\rm HI} - M_{\star}$ relation, motivated by the morphology-colour correlation of galaxies as a function of the environment \citep[e.g.][]{Bamford_2009,Lopes_2016}. In Figure~\ref{fig:morph}, the bivariate distributions of \hi mass and stellar mass for the spirals and elllipticals largely follow the features for the blue and red galaxies in Figure~\ref{fig:color}. The Schechter and Skew normal models are favored by our data for the spirals, with Bayes factors of $\Delta\ln(\mathcal{Z})=18.15\pm0.03$ and $\Delta\ln(\mathcal{Z})=19.26\pm0.09$, respectively. A notable difference is that the number of ellipticals is apparently lower than that of red galaxies as expected from Figure~\ref{fig:gsmf}. We show the SIMBA \citep{Dav__2019} spirals and ellipticals with the grey circles along with the error bars indicating the 16, 50 and 84th percentiles of the \hi mass in four stellar mass bins over a similar redshift range to our MIGHTEE observations. The spirals and ellipticals in SIMBA are divided by the fraction of rotational kinetic energy ($\kappa_{\rm rot}$) = 0.7 \citep{Sales_2012,Elson_2023}, according to the importance of their rotationally-supported components. Overall, we find reasonably good agreement between MIGHTEE and SIMBA spirals for the bivariate distributions of \hi mass and stellar mass, especially when we model the conditional \hi mass distribution with the preferred Skew normal and Schechter models. However, for the ellipticals, the MIGHTEE median \hi masses tend to be systematically lower than those predicted from the SIMBA simulation by $\sim$0.4 dex, which suggests that SIMBA may overestimate the amount of \hi gas in the massive dead galaxies in line with the finding from \cite{Pan_2023} at $z<0.08$.

The averaged $M_{\rm HI} - M_{\star}$ relation for the spirals in Figure~\ref{fig:spiral} has a relatively smaller zero-point by $\sim$0.1 dex than for the blue galaxies, and deviates from \cite{Sinigaglia_2022} slightly further than for the blue population which presumably is more similar to the star forming population. We note that the \hi mass seems to be anti-correlated with the stellar mass for the ellipticals, indicating that more massive elliptical galaxies consume or lose their \hi gas more efficiently. In contrast, the red massive galaxies include a non-negligible amount of disk galaxies which can still retain a certain fraction of \hi gas across a wide stellar mass range \citep[e.g.][]{Wang_2022}

\begin{figure*}
    \centering
  \begin{subfigure}[b]{0.33\textwidth}
    \includegraphics[height=0.94\columnwidth, width=1\columnwidth]{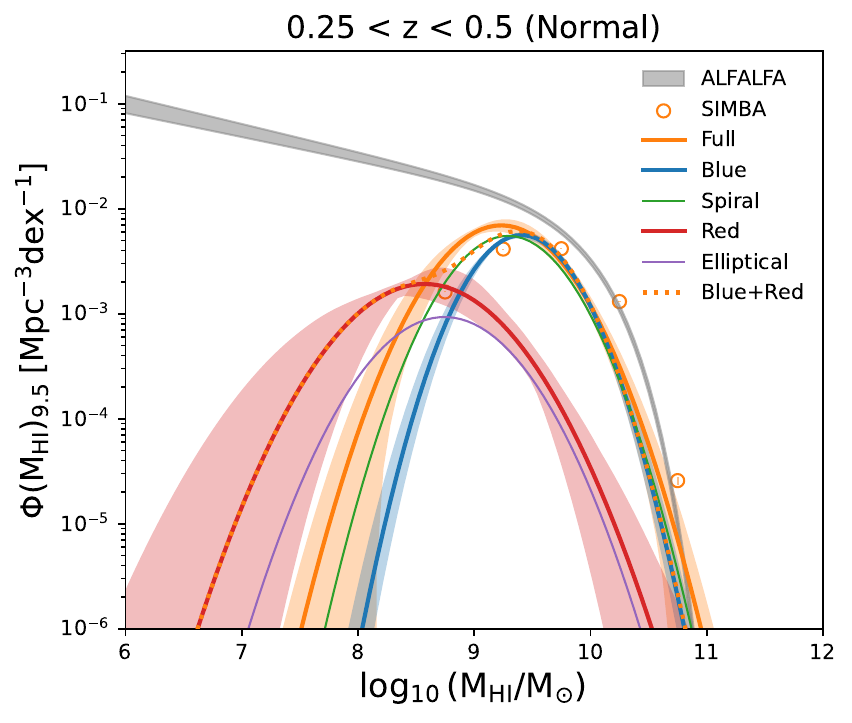}
  \end{subfigure}%
    \hfill
  \begin{subfigure}[b]{0.33\textwidth}
    \includegraphics[height=0.94\columnwidth, width=1.\columnwidth]{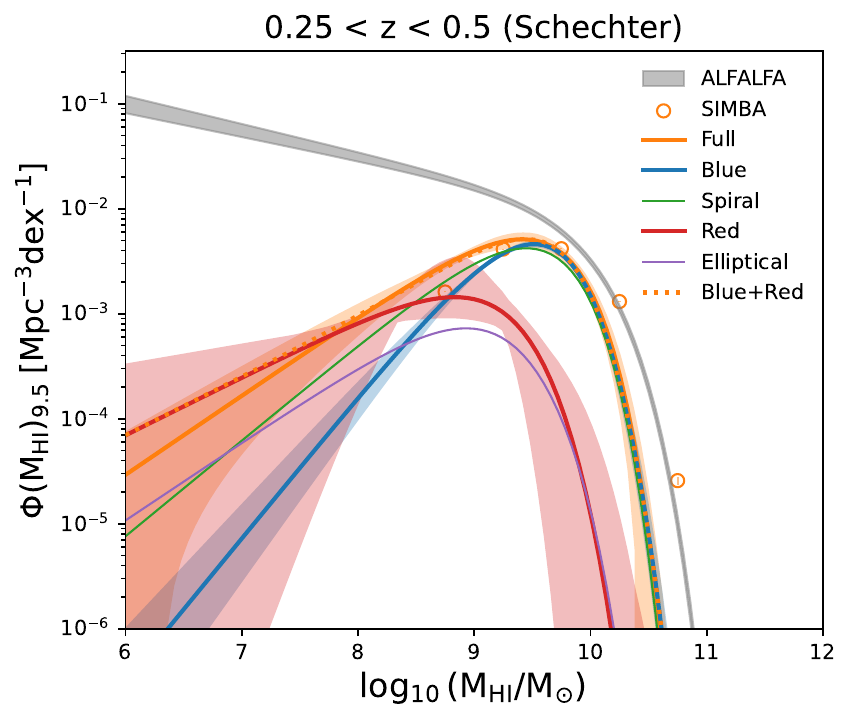}
  \end{subfigure}%
    \hfill
  \begin{subfigure}[b]{0.33\textwidth}
    \includegraphics[height=0.94\columnwidth, width=1\columnwidth]{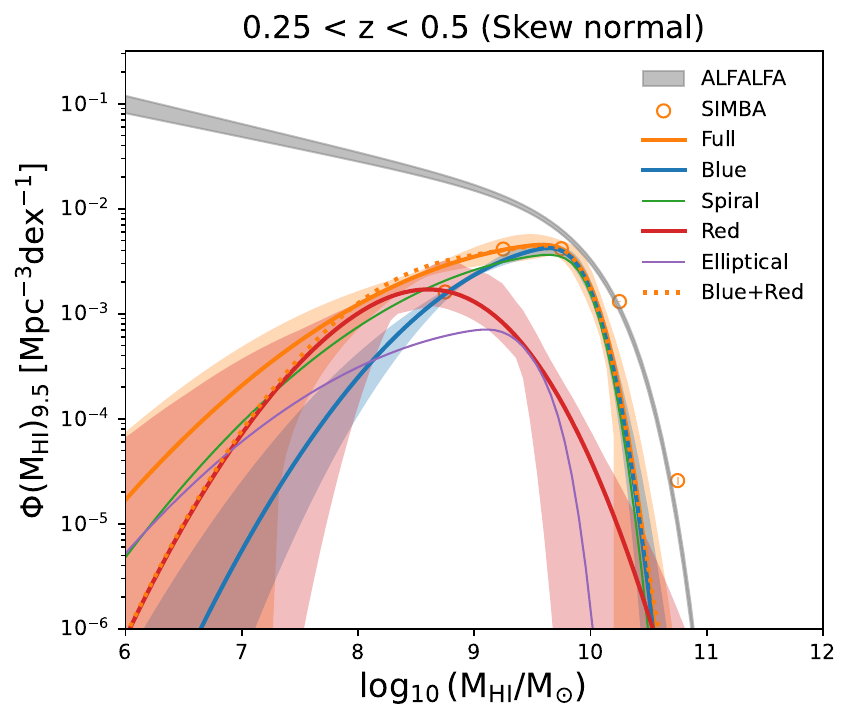}
  \end{subfigure}%
  \caption{Conditional \hi mass function for all samples at $0.25 < z < 0.5$. Three models (Normal, Schechter, Skew normal) are fit to the data from left to right panels. The blue, red, green and purple lines are for the blue, red, spiral and elliptical sub-samples while the corresponding shaded areas are 1$\sigma$ statistical uncertainties from our Bayesian modelling.  The orange lines are for the full sample with uncertainties derived from combining our modelling and error propagation from the GSMF \protect\citep{Hashemizadeh_2022}. Note that the shaded areas for spiral and elliptical sub-samples are omitted for simplicity. The grey shaded area is the 1$\sigma$ band of the \hi mass function at $z<0.06$ from the ALFALFA survey \protect\citep{Jones_2018}. The orange circles are the \hi mass function of $9.5<\log_{10}(M_{\star}/M_\odot)$ galaxies from the SIMBA simulation \protect\citep{Dav__2019}.}
  \label{fig:chimf}
\end{figure*}

\subsection{Conditional \hi mass function}
\label{sec:chimf}

We show the \hi mass function for galaxies with $\log_{10}(M_\star/M_\odot)>9.5$ at $0.25 < z < 0.5$ in Figure~\ref{fig:chimf}, determined using Eqn.~\ref{eq:himf}. The three best fitting models (Normal, Schechter, Skew normal) are shown from left to right panels. We also plot the \hi mass function from ALFALFA \citep{Jones_2018} for the \hi galaxy population in the local Universe, without any selection based on stellar mass, along with the SIMBA \hi mass function, applying the same selection criteria as imposed on our data at the same redshift. The difference between our conditional HIMF for the full sample and the ALFALFA general HIMF is significant at intermediate and low \hi masses, as expected as the DEVILS sample is limited to the $\log_{10}(M_\star/M_\odot)>9.5$ galaxies, which have the capability of hosting large amounts of \hi gas (i.e. most \hi gas in the local Universe resides in the high stellar mass galaxies based on \cite{Pan2024} for example). 

The deviation between the SIMBA HIMF and the MIGHTEE conditional \hi mass function (from our Schechter and Skew normal models) at the high \hi mass end is noticeable, again indicating that SIMBA is likely overestimating the amount of \hi gas in the massive galaxies. The overall (2-10 times) offset between the ALFALFA (and the MIGHTEE HIMF \citep{Ponomareva2023} at $z=0$) and the MIGHTEE HIMF at $0.25 < z < 0.5$ for the massive \hi galaxies at $10<\log_{10}(M_{\rm HI}/M_\odot)<10.5$ suggests a scenario of non-negligible evolution in their \hi gas content. This offset is in reasonably good agreement with the finding in \cite{Bera_2022} using the traditional stacking experiment at $z\sim0.35$, indicating that massive galaxies at $z\sim 0$ have acquired a significant amount of \hi over the past 4 Gyr, through either merger events or accretion from their surrounding CGM and IGM. This is also roughly consistent with the recent measurements of the \hi mass function from intensity mapping \citep{Paul2023} at $z\sim 0.32$ and the apparent mild evolution in the cosmic \hi mass density \citep[e.g.][]{PerouxHowk2020}.

We note that \cite{Sinigaglia2025} measured a noticeably higher HIMF than we do in this work, at a similar redshift. However, this is unsurprising considering that \cite{Sinigaglia2025} used the scaling relation between stellar mass and H{\sc i} mass for star-forming galaxies derived from \cite{Bianchetti_2025} for inferring the HIMF, as mentioned in Section~\ref{sec:color}. The \cite{Bianchetti_2025} scaling relation is much steeper than we find here, and also steeper than that found by \cite{Sinigaglia_2022} using the same dataset. \cite{Bianchetti_2025} discuss this in the appendix of their paper, and attribute the difference to the significantly more stringent definition of star-forming galaxies. This selection thus minimises the contamination from galaxies significantly below the star-formation main sequence. However, as noted in Section~2.1.2 in \cite{Bianchetti_2025}, by enforcing a symmetric scatter around the main sequence they remove around 25 per cent of the least star forming population, even if the colour selection criteria is satisfied.
On the other hand, the GSMF used to convert the scaling relation to the HIMF in \cite{Sinigaglia2025} is from \cite{Weaver2023}, who determine the star-forming GSMF from the purely colour-selected sample. As such the scaling relation from \cite{Bianchetti_2025} combined with the GSMF of {\em all} star-forming galaxies results in a HIMF that is skewed to high values, compared to the HIMF we find in this paper.
In this work, the sample used to determine the scaling relation is also that used to determine the GSMF that we use, as such they should result an HIMF that is less susceptible to possible biases in the mismatch of samples. We note that this is an important point for future studies, as a GSMF that is derived from a complete sample, and then used with an incomplete or mismatched sample used for determining the scaling relation, may lead to a bias in the predicted abundance of H{\sc i} at high stellar masses.

Although our sample is only complete at $\log_{10}(M_\star/M_\odot)>9.5$, it would require $>50$ times more mass in \hi than in stellar mass for galaxies below our stellar mass limit to contribute to the high \hi mass to bring our results down and in agreement with the $z\sim 0$ HIMF, which would appear to be unlikely given the upper envelope in the $M_\star - M_{\rm HI}$ relation discussed in \cite{Maddox2015} and \cite{Pan_2023}.

The \hi mass function for the massive galaxies with the preferred Schechter and Skew normal distributions peaks at $\log_{10}(M_{\rm HI}/M_\odot)=9.5\sim10$, which roughly corresponds to the knee of the \hi mass function from ALFALFA, further demonstrating that the $\log_{10}(M_\star/M_\odot)>9.5$ galaxies trace the galaxies containing the largest amount of \hi gas, albeit not necessarily the largest fraction of \hi. The existence of red spiral galaxies across a wide \hi mass range is also revealed by comparing the red and elliptical \hi mass functions measured from the sub-samples. The combined total HIMF from the red and blue sub-samples is shown in the orange dotted line, and in line with the HIMF measured from the full sample in the orange solid line across a wide \hi mass range. The discrepancy only increases towards the low-\hi mass end where the statistical uncertainties are large as shown by the shaded areas.

At the highest mass end of $\log_{10}(M_{\rm HI}/M_\odot)>10.5$, the \hi mass function measured using our symmetric Normal model shows the smallest deviation from the ALFALFA HIMF at $z=0$, and would imply a minimal amount of evolution in the HIMF if it was correct. Indeed, work assuming a symmetric underlying distribution around a mean value (as widely conducted using traditional stacking techniques), would potentially lead to an incorrect measurement of \hi distribution at the high mass end. The fact that our results disfavour the symmetric underlying normal distribution demonstrates the importance of properly modelling the underlying distribution. Specifically, \cite{Chowdhury2024} observed a strong evolution of the number density for high \hi mass galaxies at $z\sim1$, and they did not find a consistent evolution trend with \cite{Bera_2022} using the similar approach of combining the B-band luminosity function and the scaling relation between the \hi mass and B-band luminosity ($M_{\rm B}$) of star-forming galaxies at $z\sim0.35$. \cite{Chowdhury2024} suspected that this apparent contradiction is due to the differing net gas accretion rate for high star forming galaxies between $0.35<z<1$ and $0<z<0.35$ and the influence of the cosmic variance on the measurement of the $M_{\rm HI}-M_{\rm B}$ relation over a small cosmic volume. However, the assumption of using the symmetrical \hi scatter at $z\sim1$ may also be an important limiting factor for them to construct a consistent HIMF given the comparison between the Normal, Schechter and Skew normal for the \hi mass function on the high mass end in Figure~\ref{fig:chimf}. We append the best fitting conditional \hi mass function with Schechter and Skew normal models in Table~\ref{tab:chimf}.

\section{Conclusions}
\label{sec:conclusions}

We implement a Bayesian technique, developed from our previous work \citep{Pan_2021,Pan_2023}, to the MIGHTEE DR1 images for measuring the underlying $M_{\rm HI}-M_{\star}$ scaling relation by using the information from the DEVILS spectroscopic catalogue at $0.25 < z < 0.5$, while taking into account the intrinsic \hi distribution as a function of the stellar mass. Our results are highlighted as follows:

\begin{itemize}

\item We measure the bivariate distribution of \hi mass and stellar mass of massive galaxies down to $\log_{10}(M_\star/M_\odot$) = 9.5 at $0.25 < z < 0.5$ for the full galaxy sample, with three conditional probability distributions of \hi mass (Normal, Schechter, and Skew normal). The asymmetric \hi distributions are strongly preferred by our data.

\item We differentiate the concepts of the average of the logarithmic \hi mass, i.e. $\langle\log_{10}(M_{\rm HI})\rangle$, and the logarithmic average of the \hi mass, i.e. $\log_{10}(\langle M_{\rm HI}\rangle)$, to clarify the statistical confusions in the stacking approaches, and find that the difference between $\langle\log_{10}(M_{\rm HI})\rangle$ and $\log_{10}(\langle M_{\rm HI}\rangle)$ is non-trivial with 0.2$\sim$0.3 dex for normal and Schechter models, and $\sim$0.5 dex for the Skew normal.

\item We observe shallow slopes in the underlying 
$M_{\rm HI}-M_{\star}$ scaling relation, suggesting the presence of an upper \hi mass limit beyond which a galaxy can no longer accrete or retain further \hi gas, and find tentative evidence of mild negative evolution for the \hi distribution in massive galaxies at $z<0.5$ when compared to local $M_{\rm HI}-M_{\star}$ relation.

\item By studying the colour dependence of the $M_{\rm HI}-M_{\star}$ relation, we find that the Schechter and Skew normal profiles provide the best description of the conditional \hi mass distribution for blue galaxies, and the Normal profile is sufficient for describing the \hi destribution of red galaxies becasue of their small sample size in our present catalog. This trend is also true for the spiral and ellipical sub-samples due to the strong colour-morphology correlation of galaxies.

\item The massive blue (or spiral) galaxies are tightly clustered in the stellar mass bins of $9.5<\log_{10}(M_{\star}/M_\odot)<10.5$ and relatively high \hi mass range of $9<\log_{10}(M_{\rm HI}/M_\odot)<10$ while the massive red (or elliptical) galaxies instead sit broadly in the intermediate stellar mass range of $9.5<\log_{10}(M_{\star}/M_\odot)<11.5$ and the \hi mass range of $7.5<\log_{10}(M_{\rm HI}/M_\odot)<9.5$.

\item The \hi masses of massive blue and spiral galaxies are positively correlated with their stellar masses. Red galaxies follow a similar trend, largely due to the presence of a significant fraction of red disk galaxies. In contrast, the \hi mass appears to be anti-correlated with stellar mass for ellipticals, suggesting that more massive elliptical galaxies are more efficient at consuming or losing their \hi gas.

We also present the \hi mass function for $\log_{10}(M_{\star}/M_\odot)>9.5$ galaxies and find that this conditional HIMF is several times lower than the HIMF for the general galaxy population from ALFALFA at the highest \hi mass end, indicating a mild or moderate evolutionary picture for the extremely \hi rich galaxies observed at low redshift. Although we note that understanding the environmental effects on the measured HIMF will be important for future work with larger samples \citep[e.g.][]{Sinigaglia2024}.
\end{itemize}

In conclusion, we present the Bayesian stacking technique that is capable of probing the underlying \hi distribution at the distant Universe, and measure the $M_{\rm HI}-M_{\star}$ relation above or below the detection threshold in a unified way while modeling the scatter self-consistently within the same procedure without binning the datasets. This technique allows us to put tighter constraints on the galaxy formation and evolution models compared to the traditional stacking approach. The $\log_{10}(\langle M_{\rm HI}\rangle)$, i.e. traditional stacking, is more robust at picking out the peak of the distribution for an asymmetrical underlying \hi distribution skewed toward lower mass galaxies and is less susceptible to the form of the underlying distribution compared to $\langle\log_{10}(M_{\rm HI})\rangle$.

\section*{Data availability}
The MIGHTEE DR1 data are available in \cite{Heywood_2024}.

\section*{Acknowledgements}
The MeerKAT telescope is operated by the South African Radio Astronomy Observatory, which is a facility of the National Research Foundation, an agency of the Department of Science and Innovation. We acknowledge use of the IDIA data intensive research cloud for data processing. The IDIA is a South African university partnership involving the University of Cape Town, the University of Pretoria and the University of the Western Cape. The authors acknowledge the Centre for High Performance Computing (CHPC), South Africa, for providing computational resources to this research project.

We acknowledge the use of the ilifu cloud computing facility - \url{www.ilifu.ac.za}, a partnership between the University of Cape Town, the University of the Western Cape, the University of Stellenbosch, Sol Plaatje University, the Cape Peninsula University of Technology and the South African Radio Astronomy Observatory. The ilifu facility is supported by contributions from IDIA and the Computational Biology division at UCT and the Data Intensive Research Initiative of South Africa (DIRISA).

We thank the anonymous referee for their helpful comments that have improved this paper. HP, MJJ, IH and TY acknowledge support from a UKRI Frontiers Research Grant [EP/X026639/1], which was selected by the European Research Council. MJJ, IH and MGS acknowledge support from the South African Radio Astronomy Observatory (SARAO) towards this research (\url{www.sarao.ac.za}). 
MJJ acknowledges generous support from the Hintze
Family Charitable Foundation through the Oxford Hintze Centre for Astrophysical Surveys and the UK Science and Technology Facilities Council [ST/S000488/1]. HP acknowledges support from National Key R\&D Program of China No. 2022YFA1602904, Guizhou Provincial Science and Technology Projects (No.~QKHFQ[2023]003, No.~QKHFQ[2024]001, No.~QKHPTRC-ZDSYS[2023]003) and Guizhou Province High-level Talent Program (No.~QKHPTRC-GCC[2022]003-1).

For the purpose of Open Access, the author has applied a CC BY public copyright licence to any Author Accepted Manuscript version arising from this submission. 

%%%%%%%%%%%%%%%%%%%%%%%%%%%%%%%%%%%%%%%%%%%%%%%%%%



\bibliographystyle{mnras}
\bibliography{references} % if your bibtex file is called example.bib

\begin{thebibliography}{}
\makeatletter
\relax
\def\mn@urlcharsother{\let\do\@makeother \do\$\do\&\do\#\do\^\do\_\do\%\do\~}
\def\mn@doi{\begingroup\mn@urlcharsother \@ifnextchar [ {\mn@doi@} {\mn@doi@[]}}
\def\mn@doi@[#1]#2{\def\@tempa{#1}\ifx\@tempa\@empty \href {http://dx.doi.org/#2} {doi:#2}\else \href {http://dx.doi.org/#2} {#1}\fi \endgroup}
\def\mn@eprint#1#2{\mn@eprint@#1:#2::\@nil}
\def\mn@eprint@arXiv#1{\href {http://arxiv.org/abs/#1} {{\tt arXiv:#1}}}
\def\mn@eprint@dblp#1{\href {http://dblp.uni-trier.de/rec/bibtex/#1.xml} {dblp:#1}}
\def\mn@eprint@#1:#2:#3:#4\@nil{\def\@tempa {#1}\def\@tempb {#2}\def\@tempc {#3}\ifx \@tempc \@empty \let \@tempc \@tempb \let \@tempb \@tempa \fi \ifx \@tempb \@empty \def\@tempb {arXiv}\fi \@ifundefined {mn@eprint@\@tempb}{\@tempb:\@tempc}{\expandafter \expandafter \csname mn@eprint@\@tempb\endcsname \expandafter{\@tempc}}}

\bibitem[\protect\citeauthoryear{Aihara et~al.,}{Aihara et~al.}{2017}]{Aihara_2017}
Aihara H.,  et~al., 2017, \mn@doi [Publications of the Astronomical Society of Japan] {10.1093/pasj/psx066}, 70

\bibitem[\protect\citeauthoryear{{Aihara} et~al.,}{{Aihara} et~al.}{2018}]{Aihara2018}
{Aihara} H.,  et~al., 2018, \mn@doi [\pasj] {10.1093/pasj/psx066}, \href {https://ui.adsabs.harvard.edu/abs/2018PASJ...70S...4A} {70, S4}

\bibitem[\protect\citeauthoryear{{Aihara} et~al.,}{{Aihara} et~al.}{2019}]{Aihara2019}
{Aihara} H.,  et~al., 2019, \mn@doi [\pasj] {10.1093/pasj/psz103}, \href {https://ui.adsabs.harvard.edu/abs/2019PASJ...71..114A} {71, 114}

\bibitem[\protect\citeauthoryear{Bamford et~al.,}{Bamford et~al.}{2009}]{Bamford_2009}
Bamford S.~P.,  et~al., 2009, \mn@doi [MNRAS] {10.1111/j.1365-2966.2008.14252.x}, 393, 1324–1352

\bibitem[\protect\citeauthoryear{{Barnes} et~al.,}{{Barnes} et~al.}{2001}]{barnes2001h}
{Barnes} D.~G.,  et~al., 2001, \mn@doi [\mnras] {10.1046/j.1365-8711.2001.04102.x}, \href {https://ui.adsabs.harvard.edu/abs/2001MNRAS.322..486B} {322, 486}

\bibitem[\protect\citeauthoryear{Bera, Kanekar, Chengalur  \& Bagla}{Bera et~al.}{2022}]{Bera_2022}
Bera A.,  Kanekar N.,  Chengalur J.~N.,   Bagla J.~S.,  2022, \mn@doi [The Astrophysical Journal Letters] {10.3847/2041-8213/ac9d32}, 940, L10

\bibitem[\protect\citeauthoryear{Bianchetti et~al.,}{Bianchetti et~al.}{2025}]{Bianchetti_2025}
Bianchetti A.,  et~al., 2025, \mn@doi [The Astrophysical Journal] {10.3847/1538-4357/adb1b8}, 982, 82

\bibitem[\protect\citeauthoryear{{Blyth} et~al.,}{{Blyth} et~al.}{2016}]{Blyth2016}
{Blyth} S.,  et~al., 2016, in MeerKAT Science: On the Pathway to the SKA. p.~4, \mn@doi{10.22323/1.277.0004}

\bibitem[\protect\citeauthoryear{{Buchner} et~al.,}{{Buchner} et~al.}{2014}]{Buchner_2014}
{Buchner} J.,  et~al., 2014, \mn@doi [\aap] {10.1051/0004-6361/201322971}, \href {https://ui.adsabs.harvard.edu/abs/2014A&A...564A.125B} {564, A125}

\bibitem[\protect\citeauthoryear{Capak et~al.,}{Capak et~al.}{2007}]{Capak_2007}
Capak P.,  et~al., 2007, \mn@doi [The Astrophysical Journal Supplement Series] {10.1086/519081}, 172, 99–116

\bibitem[\protect\citeauthoryear{Catinella et~al.,}{Catinella et~al.}{2010}]{Catinella_2010}
Catinella B.,  et~al., 2010, \mn@doi [MNRAS] {10.1111/j.1365-2966.2009.16180.x}, 403, 683

\bibitem[\protect\citeauthoryear{{Chowdhury}, {Kanekar}  \& {Chengalur}}{{Chowdhury} et~al.}{2024}]{Chowdhury2024}
{Chowdhury} A.,  {Kanekar} N.,   {Chengalur} J.~N.,  2024, \mn@doi [\apjl] {10.3847/2041-8213/ad3dfe}, \href {https://ui.adsabs.harvard.edu/abs/2024ApJ...966L..39C} {966, L39}

\bibitem[\protect\citeauthoryear{{Cuillandre} et~al.,}{{Cuillandre} et~al.}{2012}]{Cuillandre2012}
{Cuillandre} J.-C.~J.,  et~al., 2012, in {Peck} A.~B.,  {Seaman} R.~L.,   {Comeron} F.,  eds,  Society of Photo-Optical Instrumentation Engineers (SPIE) Conference Series Vol. 8448, Observatory Operations: Strategies, Processes, and Systems IV. p. 84480M, \mn@doi{10.1117/12.925584}

\bibitem[\protect\citeauthoryear{Dav{\'{e} }, Angl{\'{e}}s-Alc{\'{a}}zar, Narayanan, Li, Rafieferantsoa  \& Appleby}{Dav{\'{e} } et~al.}{2019}]{Dav__2019}
Dav{\'{e} } R.,  Angl{\'{e}}s-Alc{\'{a}}zar D.,  Narayanan D.,  Li Q.,  Rafieferantsoa M.~H.,   Appleby S.,  2019, \mn@doi [MNRAS] {10.1093/mnras/stz937}, 486, 2827

\bibitem[\protect\citeauthoryear{Davies et~al.,}{Davies et~al.}{2018}]{Davies_2018}
Davies L. J.~M.,  et~al., 2018, \mn@doi [MNRAS] {10.1093/mnras/sty1553}, 480, 768–799

\bibitem[\protect\citeauthoryear{Davies et~al.,}{Davies et~al.}{2021}]{Davies_2021}
Davies L. J.~M.,  et~al., 2021, \mn@doi [Monthly Notices of the Royal Astronomical Society] {10.1093/mnras/stab1601}, 506, 256–287

\bibitem[\protect\citeauthoryear{Delhaize, Meyer, Staveley-Smith  \& Boyle}{Delhaize et~al.}{2013}]{Delhaize_2013}
Delhaize J.,  Meyer M.~J.,  Staveley-Smith L.,   Boyle B.~J.,  2013, \mn@doi [MNRAS] {10.1093/mnras/stt810}, 433, 1398

\bibitem[\protect\citeauthoryear{Elson, Glowacki  \& Davé}{Elson et~al.}{2023}]{Elson_2023}
Elson E.,  Glowacki M.,   Davé R.,  2023, \mn@doi [New Astronomy] {10.1016/j.newast.2022.101964}, 99, 101964

\bibitem[\protect\citeauthoryear{{Fern{\'a}ndez} et~al.,}{{Fern{\'a}ndez} et~al.}{2016}]{chiles2016}
{Fern{\'a}ndez} X.,  et~al., 2016, \mn@doi [\apjl] {10.3847/2041-8205/824/1/L1}, \href {https://ui.adsabs.harvard.edu/abs/2016ApJ...824L...1F} {824, L1}

\bibitem[\protect\citeauthoryear{Feroz, Hobson  \& Bridges}{Feroz et~al.}{2009}]{Feroz_2009}
Feroz F.,  Hobson M.~P.,   Bridges M.,  2009, \mn@doi [MNRAS] {10.1111/j.1365-2966.2009.14548.x}, 398, 1601

\bibitem[\protect\citeauthoryear{{Giovanelli} et~al.,}{{Giovanelli} et~al.}{2005}]{giovanelli2005arecibo}
{Giovanelli} R.,  et~al., 2005, \mn@doi [\aj] {10.1086/497431}, \href {https://ui.adsabs.harvard.edu/abs/2005AJ....130.2598G} {130, 2598}

\bibitem[\protect\citeauthoryear{{Hashemizadeh} et~al.,}{{Hashemizadeh} et~al.}{2021}]{Hashemizadeh2021}
{Hashemizadeh} A.,  et~al., 2021, \mn@doi [\mnras] {10.1093/mnras/stab600}, \href {https://ui.adsabs.harvard.edu/abs/2021MNRAS.505..136H} {505, 136}

\bibitem[\protect\citeauthoryear{Hashemizadeh et~al.,}{Hashemizadeh et~al.}{2022}]{Hashemizadeh_2022}
Hashemizadeh A.,  et~al., 2022, \mn@doi [MNRAS] {10.1093/mnras/stac1195}, 515, 1175–1198

\bibitem[\protect\citeauthoryear{Healy, Blyth, Elson, van Driel, Butcher, Schneider, Lehnert  \& Minchin}{Healy et~al.}{2019}]{Healy_2019}
Healy J.,  Blyth S.-L.,  Elson E.,  van Driel W.,  Butcher Z.,  Schneider S.,  Lehnert M.~D.,   Minchin R.,  2019, \mn@doi [MNRAS] {10.1093/mnras/stz1555}, 487, 4901

\bibitem[\protect\citeauthoryear{Heywood et~al.,}{Heywood et~al.}{2024}]{Heywood_2024}
Heywood I.,  et~al., 2024, \mn@doi [MNRAS] {10.1093/mnras/stae2081}, 534, 76–96

\bibitem[\protect\citeauthoryear{{Huang}, {Haynes}, {Giovanelli}  \& {Brinchmann}}{{Huang} et~al.}{2012}]{huang2012}
{Huang} S.,  {Haynes} M.~P.,  {Giovanelli} R.,   {Brinchmann} J.,  2012, \mn@doi [\apj] {10.1088/0004-637X/756/2/113}, \href {https://ui.adsabs.harvard.edu/abs/2012ApJ...756..113H} {756, 113}

\bibitem[\protect\citeauthoryear{{Jarvis} et~al.,}{{Jarvis} et~al.}{2016}]{Jarvis2016}
{Jarvis} M.,  et~al., 2016, in MeerKAT Science: On the Pathway to the SKA. p.~6 (\mn@eprint {arXiv} {1709.01901})

\bibitem[\protect\citeauthoryear{{Jarvis} et~al.,}{{Jarvis} et~al.}{2024}]{Jarvis2024}
{Jarvis} M.~J.,  et~al., 2024, \mn@doi [\mnras] {10.1093/mnras/stad3821}, \href {https://ui.adsabs.harvard.edu/abs/2024MNRAS.529.3484J} {529, 3484}

\bibitem[\protect\citeauthoryear{{Jarvis} et~al.,}{{Jarvis} et~al.}{2025}]{Jarvis2025}
{Jarvis} M.~J.,  et~al., 2025, \mn@doi [arXiv e-prints] {10.48550/arXiv.2506.11935}, \href {https://ui.adsabs.harvard.edu/abs/2025arXiv250611935J} {p. arXiv:2506.11935}

\bibitem[\protect\citeauthoryear{{Jonas} \& {MeerKAT Team}}{{Jonas} \& {MeerKAT Team}}{2016}]{Jonas2016mksJ}
{Jonas} J.,  {MeerKAT Team} 2016, in MeerKAT Science: On the Pathway to the SKA. p.~1

\bibitem[\protect\citeauthoryear{Jones, Haynes, Giovanelli  \& Moorman}{Jones et~al.}{2018}]{Jones_2018}
Jones M.~G.,  Haynes M.~P.,  Giovanelli R.,   Moorman C.,  2018, \mn@doi [MNRAS] {10.1093/mnras/sty521}, 477, 2

\bibitem[\protect\citeauthoryear{{Kazemi-Moridani} et~al.,}{{Kazemi-Moridani} et~al.}{2024}]{2024arXiv241211426K}
{Kazemi-Moridani} A.,  et~al., 2024, \mn@doi [arXiv e-prints] {10.48550/arXiv.2412.11426}, \href {https://ui.adsabs.harvard.edu/abs/2024arXiv241211426K} {p. arXiv:2412.11426}

\bibitem[\protect\citeauthoryear{Kleiner, Pimbblet, Heath~Jones, Koribalski  \& Serra}{Kleiner et~al.}{2016}]{Kleiner_2016}
Kleiner D.,  Pimbblet K.~A.,  Heath~Jones D.,  Koribalski B.~S.,   Serra P.,  2016, \mn@doi [MNRAS] {10.1093/mnras/stw3328}, p. stw3328

\bibitem[\protect\citeauthoryear{Lopes, Rembold, Ribeiro, Nascimento  \& Vajgel}{Lopes et~al.}{2016}]{Lopes_2016}
Lopes P. A.~A.,  Rembold S.~B.,  Ribeiro A. L.~B.,  Nascimento R.~S.,   Vajgel B.,  2016, \mn@doi [MNRAS] {10.1093/mnras/stw1497}, 461, 2559–2579

\bibitem[\protect\citeauthoryear{{Maddox}, {Hess}, {Obreschkow}, {Jarvis}  \& {Blyth}}{{Maddox} et~al.}{2015}]{Maddox2015}
{Maddox} N.,  {Hess} K.~M.,  {Obreschkow} D.,  {Jarvis} M.~J.,   {Blyth} S.~L.,  2015, \mn@doi [\mnras] {10.1093/mnras/stu2532}, \href {https://ui.adsabs.harvard.edu/abs/2015MNRAS.447.1610M} {447, 1610}

\bibitem[\protect\citeauthoryear{{Maddox} et~al.,}{{Maddox} et~al.}{2021}]{maddox2021}
{Maddox} N.,  et~al., 2021, \mn@doi [\aap] {10.1051/0004-6361/202039655}, \href {https://ui.adsabs.harvard.edu/abs/2021A&A...646A..35M} {646, A35}

\bibitem[\protect\citeauthoryear{Mahajan et~al.,}{Mahajan et~al.}{2019}]{Mahajan_2019}
Mahajan S.,  et~al., 2019, \mn@doi [MNRAS] {10.1093/mnras/stz2993}, 491, 398–408

\bibitem[\protect\citeauthoryear{Masters et~al.,}{Masters et~al.}{2010}]{Masters_2010}
Masters K.~L.,  et~al., 2010, \mn@doi [MNRAS] {10.1111/j.1365-2966.2010.16503.x}

\bibitem[\protect\citeauthoryear{{McCracken} et~al.,}{{McCracken} et~al.}{2012}]{McCracken2012}
{McCracken} H.~J.,  et~al., 2012, \mn@doi [\aap] {10.1051/0004-6361/201219507}, \href {https://ui.adsabs.harvard.edu/abs/2012A&A...544A.156M} {544, A156}

\bibitem[\protect\citeauthoryear{{Meyer}, {Robotham}, {Obreschkow}, {Westmeier}, {Duffy}  \& {Staveley-Smith}}{{Meyer} et~al.}{2017}]{meyer2017tracing}
{Meyer} M.,  {Robotham} A.,  {Obreschkow} D.,  {Westmeier} T.,  {Duffy} A.~R.,   {Staveley-Smith} L.,  2017, \mn@doi [\pasa] {10.1017/pasa.2017.31}, \href {https://ui.adsabs.harvard.edu/abs/2017PASA...34...52M} {34}

\bibitem[\protect\citeauthoryear{Nan et~al.,}{Nan et~al.}{2011}]{NAN_2011}
Nan R.,  et~al., 2011, \mn@doi [Int. J. Mod. Phys. D] {10.1142/s0218271811019335}, 20, 989–1024

\bibitem[\protect\citeauthoryear{Pan, Jarvis, Allison, Heywood, Santos, Maddox, Frank  \& Kang}{Pan et~al.}{2019}]{Pan_2019}
Pan H.,  Jarvis M.~J.,  Allison J.~R.,  Heywood I.,  Santos M.~G.,  Maddox N.,  Frank B.~S.,   Kang X.,  2019, \mn@doi [MNRAS] {10.1093/mnras/stz3030}

\bibitem[\protect\citeauthoryear{Pan, Jarvis, Ponomareva, Santos, Allison, Maddox  \& Frank}{Pan et~al.}{2021}]{Pan_2021}
Pan H.,  Jarvis M.~J.,  Ponomareva A.~A.,  Santos M.~G.,  Allison J.~R.,  Maddox N.,   Frank B.~S.,  2021, \mn@doi [MNRAS] {10.1093/mnras/stab2601}, 508, 1897

\bibitem[\protect\citeauthoryear{Pan et~al.,}{Pan et~al.}{2023}]{Pan_2023}
Pan H.,  et~al., 2023, \mn@doi [MNRAS] {10.1093/mnras/stad2343}, 525, 256–269

\bibitem[\protect\citeauthoryear{{Pan} et~al.,}{{Pan} et~al.}{2024}]{Pan2024}
{Pan} H.,  et~al., 2024, \mn@doi [\mnras] {10.1093/mnras/stae2054}, \href {https://ui.adsabs.harvard.edu/abs/2024MNRAS.534..202P} {534, 202}

\bibitem[\protect\citeauthoryear{Parkash, Brown, Jarrett  \& Bonne}{Parkash et~al.}{2018}]{Parkash2018}
Parkash V.,  Brown M. J.~I.,  Jarrett T.~H.,   Bonne N.~J.,  2018, \mn@doi [The Astrophysical Journal] {10.3847/1538-4357/aad3b9}, 864, 40

\bibitem[\protect\citeauthoryear{{Paul}, {Santos}, {Chen}  \& {Wolz}}{{Paul} et~al.}{2023}]{Paul2023}
{Paul} S.,  {Santos} M.~G.,  {Chen} Z.,   {Wolz} L.,  2023, \mn@doi [arXiv e-prints] {10.48550/arXiv.2301.11943}, \href {https://ui.adsabs.harvard.edu/abs/2023arXiv230111943P} {p. arXiv:2301.11943}

\bibitem[\protect\citeauthoryear{{P{\'e}roux} \& {Howk}}{{P{\'e}roux} \& {Howk}}{2020}]{PerouxHowk2020}
{P{\'e}roux} C.,  {Howk} J.~C.,  2020, \mn@doi [\araa] {10.1146/annurev-astro-021820-120014}, \href {https://ui.adsabs.harvard.edu/abs/2020ARA&A..58..363P} {58, 363}

\bibitem[\protect\citeauthoryear{Ponomareva et~al.,}{Ponomareva et~al.}{2021}]{Ponomareva_2021}
Ponomareva A.~A.,  et~al., 2021, \mn@doi [MNRAS] {10.1093/mnras/stab2654}, 508, 1195

\bibitem[\protect\citeauthoryear{{Ponomareva} et~al.,}{{Ponomareva} et~al.}{2023}]{Ponomareva2023}
{Ponomareva} A.~A.,  et~al., 2023, \mn@doi [\mnras] {10.1093/mnras/stad1249}, \href {https://ui.adsabs.harvard.edu/abs/2023MNRAS.522.5308P} {522, 5308}

\bibitem[\protect\citeauthoryear{Rodr{\'\i}guez-Puebla, Calette, Avila-Reese, Rodriguez-Gomez  et~al.}{Rodr{\'\i}guez-Puebla et~al.}{2020}]{Rodr_guez_Puebla_2020}
Rodr{\'\i}guez-Puebla A.,  Calette A.,  Avila-Reese V.,  Rodriguez-Gomez V.,   et~al., 2020, Publications of the Astronomical Society of Australia, 37, e024

\bibitem[\protect\citeauthoryear{{Saintonge} \& {Catinella}}{{Saintonge} \& {Catinella}}{2022}]{Saintonge2022}
{Saintonge} A.,  {Catinella} B.,  2022, arXiv e-prints, \href {https://ui.adsabs.harvard.edu/abs/2022arXiv220200690S} {p. arXiv:2202.00690}

\bibitem[\protect\citeauthoryear{Sales, Navarro, Theuns, Schaye, White, Frenk, Crain  \& Vecchia}{Sales et~al.}{2012}]{Sales_2012}
Sales L.~V.,  Navarro J.~F.,  Theuns T.,  Schaye J.,  White S. D.~M.,  Frenk C.~S.,  Crain R.~A.,   Vecchia C.~D.,  2012, \mn@doi [MNRAS] {10.1111/j.1365-2966.2012.20975.x}, 423, 1544

\bibitem[\protect\citeauthoryear{{Schawinski} et~al.,}{{Schawinski} et~al.}{2014}]{Schawinski2014}
{Schawinski} K.,  et~al., 2014, \mn@doi [\mnras] {10.1093/mnras/stu327}, \href {https://ui.adsabs.harvard.edu/abs/2014MNRAS.440..889S} {440, 889}

\bibitem[\protect\citeauthoryear{Sinigaglia et~al.,}{Sinigaglia et~al.}{2022}]{Sinigaglia_2022}
Sinigaglia F.,  et~al., 2022, \mn@doi [The Astrophysical Journal Letters] {10.3847/2041-8213/ac85ae}, 935, L13

\bibitem[\protect\citeauthoryear{{Sinigaglia} et~al.,}{{Sinigaglia} et~al.}{2024}]{Sinigaglia2024}
{Sinigaglia} F.,  et~al., 2024, \mn@doi [\mnras] {10.1093/mnras/stae713}, \href {https://ui.adsabs.harvard.edu/abs/2024MNRAS.529.4192S} {529, 4192}

\bibitem[\protect\citeauthoryear{{Sinigaglia}, {Bianchetti}, {Rodighiero}, {Mayer}, {Dessauges-Zavadsky}, {Elson}, {Vaccari}  \& {Jarvis}}{{Sinigaglia} et~al.}{2025}]{Sinigaglia2025}
{Sinigaglia} F.,  {Bianchetti} A.,  {Rodighiero} G.,  {Mayer} L.,  {Dessauges-Zavadsky} M.,  {Elson} E.,  {Vaccari} M.,   {Jarvis} M.~J.,  2025, \mn@doi [arXiv e-prints] {10.48550/arXiv.2506.11280}, \href {https://ui.adsabs.harvard.edu/abs/2025arXiv250611280S} {p. arXiv:2506.11280}

\bibitem[\protect\citeauthoryear{{Thorne} et~al.,}{{Thorne} et~al.}{2021}]{Thorne2021}
{Thorne} J.~E.,  et~al., 2021, \mn@doi [\mnras] {10.1093/mnras/stab1294}, \href {https://ui.adsabs.harvard.edu/abs/2021MNRAS.505..540T} {505, 540}

\bibitem[\protect\citeauthoryear{Wang et~al.,}{Wang et~al.}{2022}]{Wang_2022}
Wang L.,  et~al., 2022, \mn@doi [MNRAS] {10.1093/mnras/stac2292}, 516, 2337–2347

\bibitem[\protect\citeauthoryear{{Weaver} et~al.,}{{Weaver} et~al.}{2023}]{Weaver2023}
{Weaver} J.~R.,  et~al., 2023, \mn@doi [\aap] {10.1051/0004-6361/202245581}, \href {https://ui.adsabs.harvard.edu/abs/2023A&A...677A.184W} {677, A184}

\bibitem[\protect\citeauthoryear{{Xi}, {Peng}, {Staveley-Smith}, {For}, {Liu}  \& {Ding}}{{Xi} et~al.}{2024}]{Xi2024}
{Xi} H.,  {Peng} B.,  {Staveley-Smith} L.,  {For} B.-Q.,  {Liu} B.,   {Ding} D.,  2024, \mn@doi [\apjs] {10.3847/1538-4365/ad67d5}, \href {https://ui.adsabs.harvard.edu/abs/2024ApJS..274...18X} {274, 18}

\bibitem[\protect\citeauthoryear{{Zhang} et~al.,}{{Zhang} et~al.}{2024}]{zhang2024}
{Zhang} C.-P.,  et~al., 2024, \mn@doi [Sci. China-Phys. Mech. Astron.] {10.1007/s11433-023-2219-7}, \href {https://ui.adsabs.harvard.edu/abs/2024SCPMA..6719511Z} {67, 219511}

\bibitem[\protect\citeauthoryear{Zhou, Li, Hao, Guo, Mo  \& Xia}{Zhou et~al.}{2021}]{Zhou_2021}
Zhou S.,  Li C.,  Hao C.-N.,  Guo R.,  Mo H.,   Xia X.,  2021, \mn@doi [The Astrophysical Journal] {10.3847/1538-4357/ac06cc}, 916, 38

\makeatother
\end{thebibliography}
%%%%%%%%%%%%%%%%%%%%%%%%%%%%%%%%%%%%%%%%%%%%%%%%%%

%%%%%%%%%%%%%%%%%%%%%%%%%%%%%%%%%%%%%%%%%%%%%%%%%%
%%%%%%%%%%%%%%%%% APPENDICES %%%%%%%%%%%%%%%%%%%%%

\appendix
\section{Measured \hi mass}
\label{sec:mhi_m}

\begin{figure}
    \centering
    \includegraphics[height=0.85\columnwidth, width=\columnwidth]{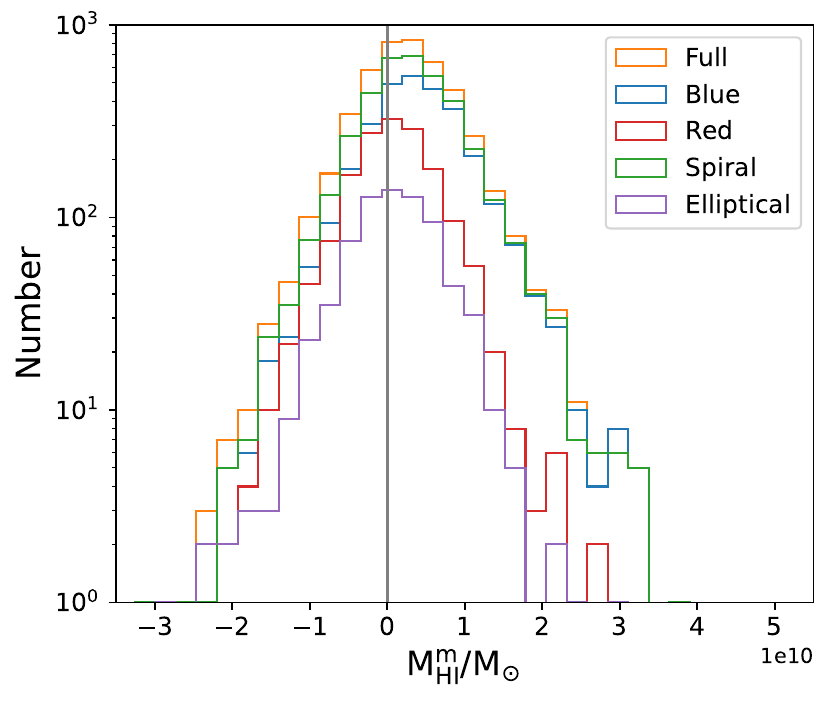}
  \caption{Measured \hi mass distribution from MIGHTEE DR1 data at $0.25 < z < 0.5$.}
  \label{fig:mhi_distribution}
\end{figure}

\begin{figure*}
    \centering
  \begin{subfigure}[b]{0.33\textwidth}
    \includegraphics[height=0.94\columnwidth, width=1\columnwidth]{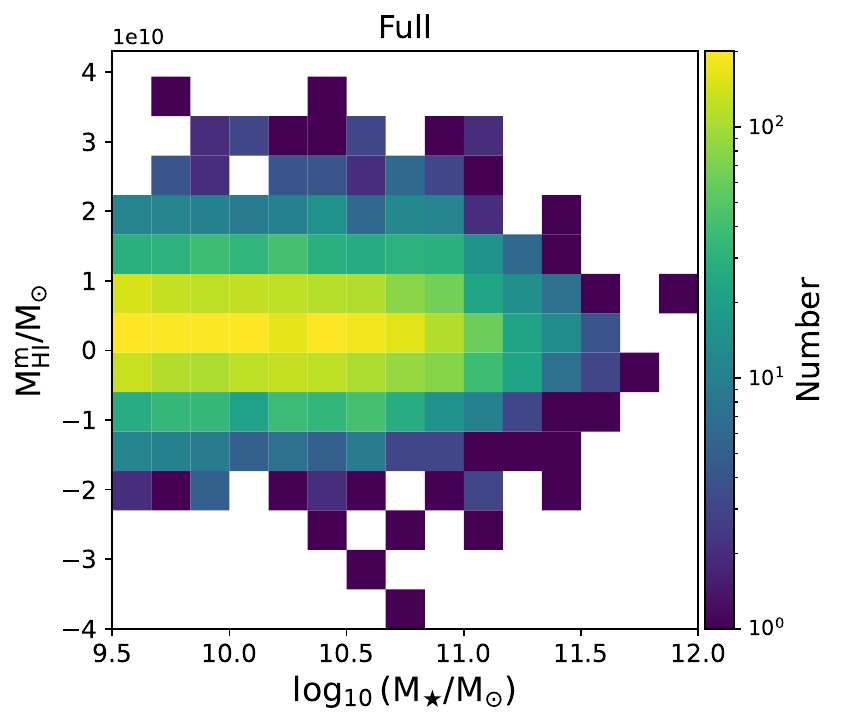}
  \end{subfigure}%
  % \hspace{6mm}
    \hfill
  \begin{subfigure}[b]{0.33\textwidth}
    \includegraphics[height=0.94\columnwidth, width=1.\columnwidth]{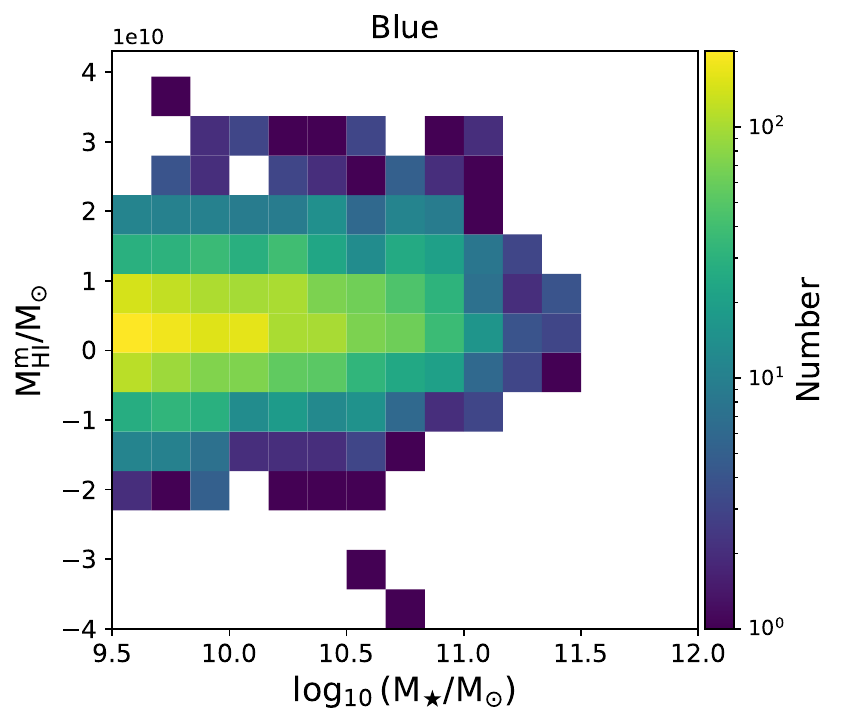}
  \end{subfigure}%
  % \hspace{-7mm}
    \hfill
  \begin{subfigure}[b]{0.33\textwidth}
    \includegraphics[height=0.94\columnwidth, width=1\columnwidth]{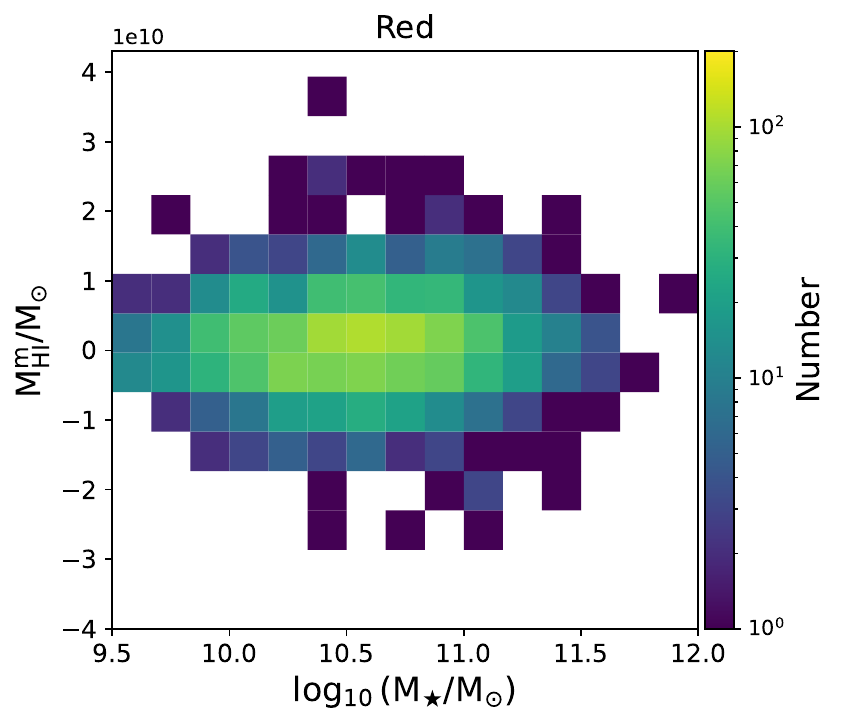}
  \end{subfigure}%
  \caption{Bivariate distribution of the measured \hi mass and stellar mass for the full (left), blue (middle) and red (right) galaxies at $0.25 < z < 0.5$.}
  \label{fig:mhi_mstar}
\end{figure*}

We used Eq.~\eqref{eq:factor} to calculate the intrinsic \hi fluxes with assumed \hi masses driven from the underlying \hi mass and stellar mass relation as the majority of our extracted fluxes are below the detection threshold at the redshift range of 0.25 $< z <$ 0.5. However, in this section, we use Eq.~\eqref{eq:factor} to measure the individual \hi masses directly from the extracted fluxes to intuitively understand the distribution of the measured \hi mass of $M_{\rm HI}^{\rm m}$, although we note that many of these measurements could be negative due to the presence of low SNR sources. We show the measured \hi mass ($M_{\rm HI}^{\rm m}$) distribution in Figure~\ref{fig:mhi_distribution} which can be turned into the SNR distribution in Figure~\ref{fig:snr_distribution} after the noise normalization. We also show the bivariate distribution of the $M_{\rm HI}^{\rm m}$ and $M_{\star}$ in Figure~\ref{fig:mhi_mstar} which corresponds to the best fitting bivariate distribution of the intrinsic \hi mass and stellar mass in Figure~\ref{fig:full_hist}.

\begin{figure}
    \centering
    \includegraphics[height=0.8\columnwidth, width=0.9\columnwidth]{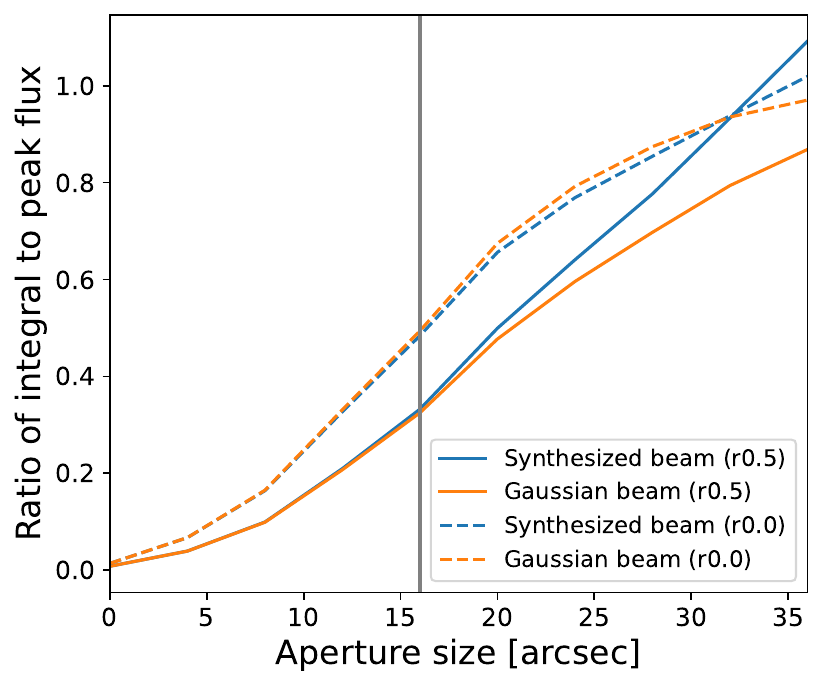}
  \caption{Ratio of the integral to peak flux for the discovered maser by \protect\cite{Jarvis2024}. The blue and orange lines are the synthesized beam and the corresponding Gaussian model. The solid and dashed lines are for the r0.5 and r0.0 \hi spectral cubes, respectively.}
  \label{fig:extraction}
\end{figure}

The aperture size for our flux extraction is 16 arcsec, which is slightly lower than the typical synthesized beam size for the robust$=$0.5 (r0.5) \hi spectral cube at the redshift range of 0.25 $< z <$ 0.5. This choice is to reduce the influence from the difference between the synthesized beam and the corresponding Gaussian model when we use the integral over the source to measure the \hi flux. We plot the ratio of the integral to peak flux as a function of the aperture size for the synthesized beam and the corresponding Gaussian model in Figure~\ref{fig:extraction} for the maser discovered by \cite{Jarvis2024}, and use the vertical line to mark the size of 16 arcsec. At this size, the fraction difference between the synthesized beam and the corresponding Gaussian model is less than $\sim$2\% for both r0.5 and r0.0 \hi cubes, and it increases gradually when the aperture size is larger. To also minimise the noise influence, we chose the r0.5 \hi spectral cube to extract the \hi fluxes, and measure the \hi masses based on them.

\section{Injection and recovery}
\label{sec:injection_and_recovery}

Our approach for measuring the \hi mass and stellar mass relation relies on assuming the Gaussianity of the noise distribution in the \hi spectral cube. However, in practice the real noise behavior deviates from the perfect Gaussian distribution.  To examine the influence of the real noise on our approach, we inject fake sources with known $M_{\rm HI} - M_{\star}$ relation and try to reconstruct this relation to verify if our approach is immune to the minor non-Gaussian noise behavior.

The input $M_{\rm HI} - M_{\star}$ relation follows our Model A with parameters $a, b$ and $\sigma_{\rm HI}$ of 0.21, 9.37 and 0.47, which are in the range of typical values for the $M_{\rm HI} - M_{\star}$ relation of blue or star forming galaxies. We use this relation to generate the \hi masses, and convert them into fluxes based on their stellar masses and redshifts. We then inject the flux for each fake source into a random position surrounding the real source (over a distance range of 5-10 beams), and we repeat this process for 100 times to get a high statistical significance on the agreement between the input parameters and reconstructed ones. This whole process is to further verify that our approach is not influenced by the random sampling, which in turn proves that the noise surrounding the source in our data is indeed well-behaved. We also simulate a data cube with Gaussian background noise for conducting the same injections as a reference.

\begin{figure*}
    \centering
  \begin{subfigure}[b]{0.33\textwidth}
    \includegraphics[height=0.94\columnwidth, width=1\columnwidth]{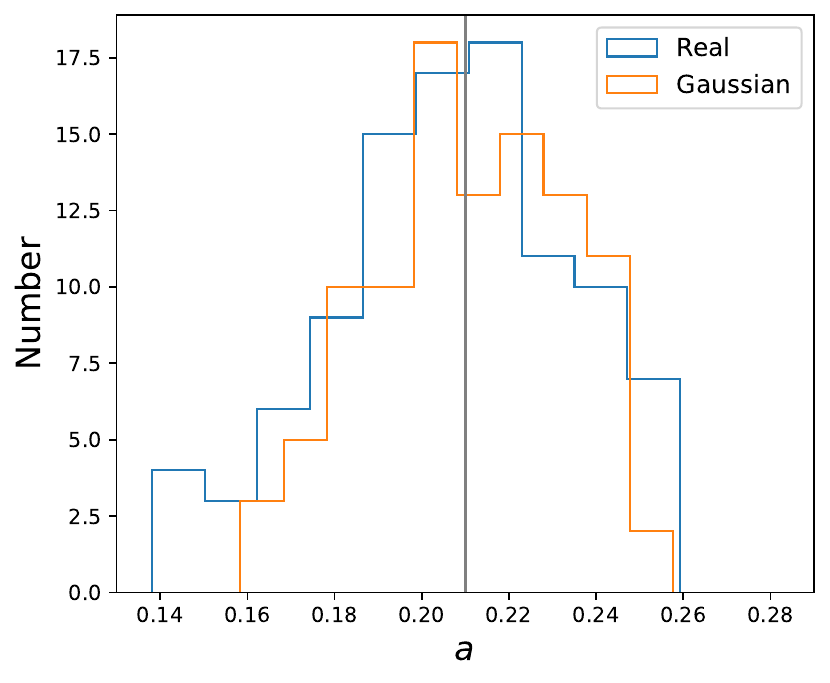}
  \end{subfigure}%
  % \hspace{6mm}
    \hfill
  \begin{subfigure}[b]{0.33\textwidth}
    \includegraphics[height=0.94\columnwidth, width=1.\columnwidth]{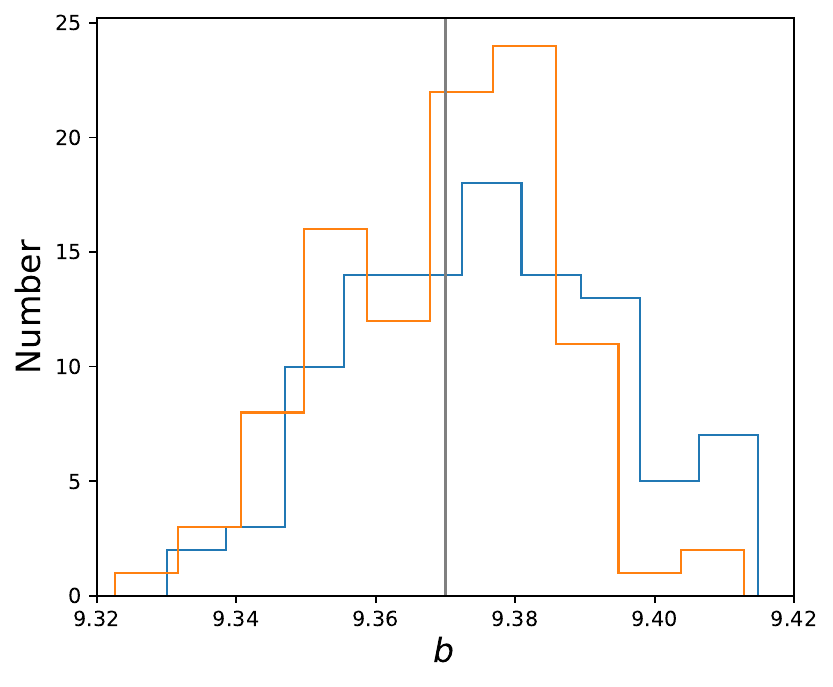}
  \end{subfigure}%
  % \hspace{-7mm}
    \hfill
  \begin{subfigure}[b]{0.33\textwidth}
    \includegraphics[height=0.94\columnwidth, width=1\columnwidth]{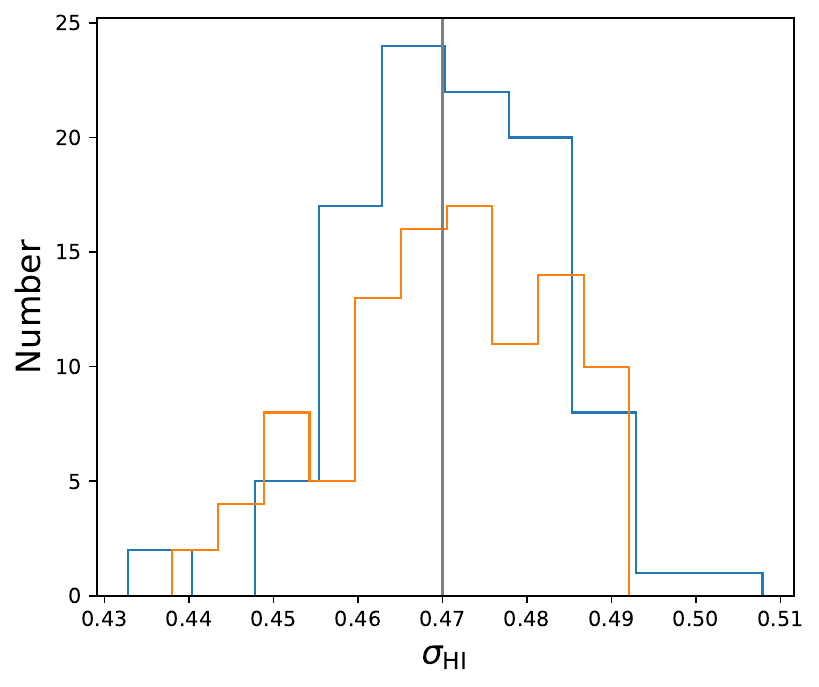}
  \end{subfigure}%
  \caption{Reconstructed parameters of $a, b, \sigma_{\rm HI}$ for the input \hi mass and stellar mass relation from left to right panels. The blue and orange lines correspond to the results derived from real and simulated Gaussian background noise, respectively. The grey vertical lines are the input values for an assumed $M_{\rm HI} - M_{\star}$ relation model.}
  \label{fig:input_recovery}
\end{figure*}

We show the histograms for the recovered parameters in Figure~\ref{fig:input_recovery}. The blue and orange lines correspond to the results reconstructed from the real background noise and fake Gaussian noise, respectively. The grey vertical lines are the input values for our assumed model. For all three parameters in both cases, the median reconstructed values are in great agreement with the input values, with a fraction difference of $<$1\% . The spread widths are on a scale of $\sim$0.02 as expected from our Bayesian modelling. The fraction difference between the Real noise and Gaussian noise on our approach is also less than 1\% for the median reconstructed values, demonstrating that the influence of the real noise distribution (slightly deviated from the perfect Gaussian) on the accuracy of our approach for measuring the $M_{\rm HI} - M_{\star}$ relation is minimal.

\section{Posteriors}
\label{sec:Posteriors}

\begin{figure*}
  \centering
  \begin{subfigure}[s]{0.43\textwidth}
    \includegraphics[height=0.94\columnwidth, width=1\columnwidth]{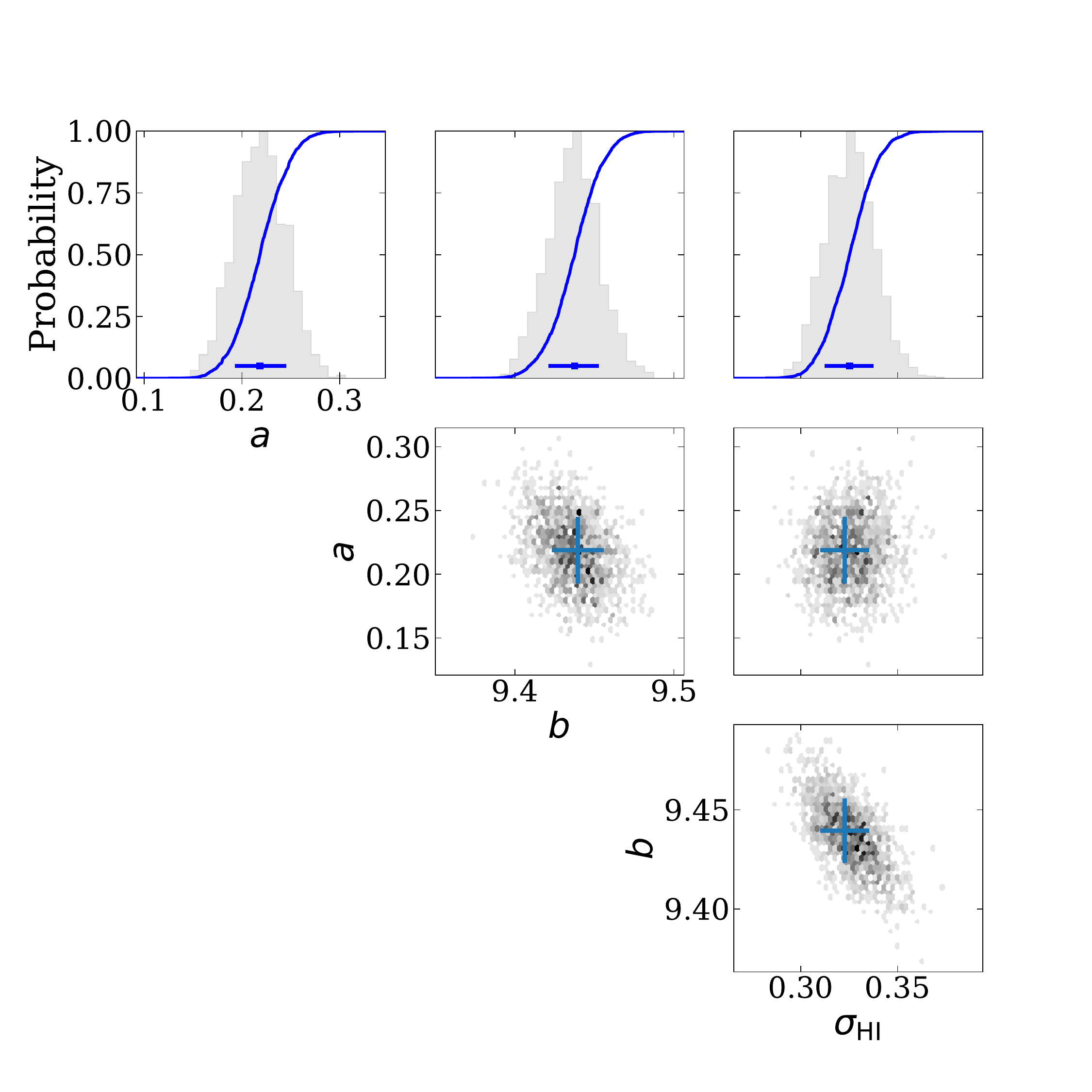}
    \includegraphics[height=0.94\columnwidth, width=1.\columnwidth]{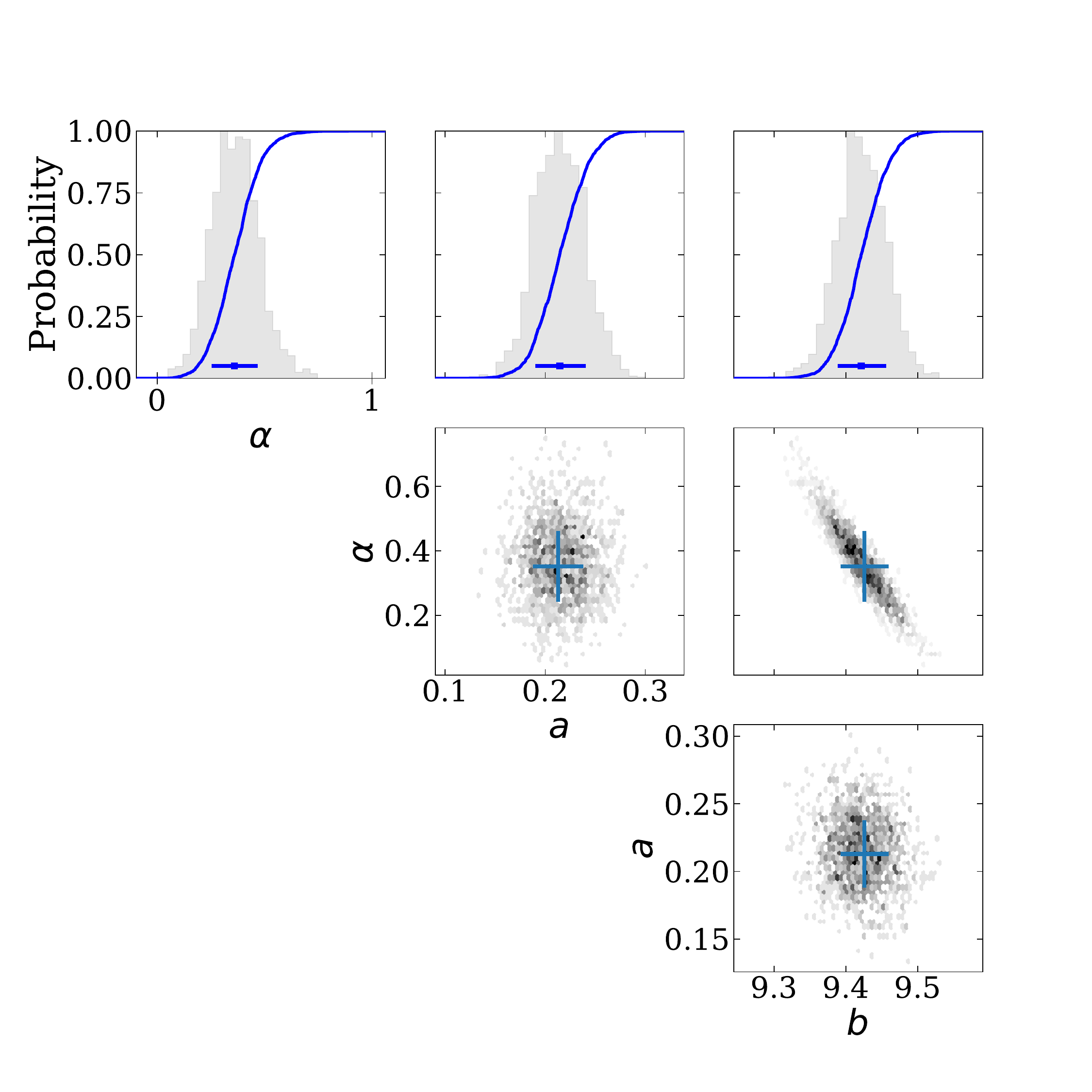}
    \includegraphics[height=0.94\columnwidth, width=1\columnwidth]{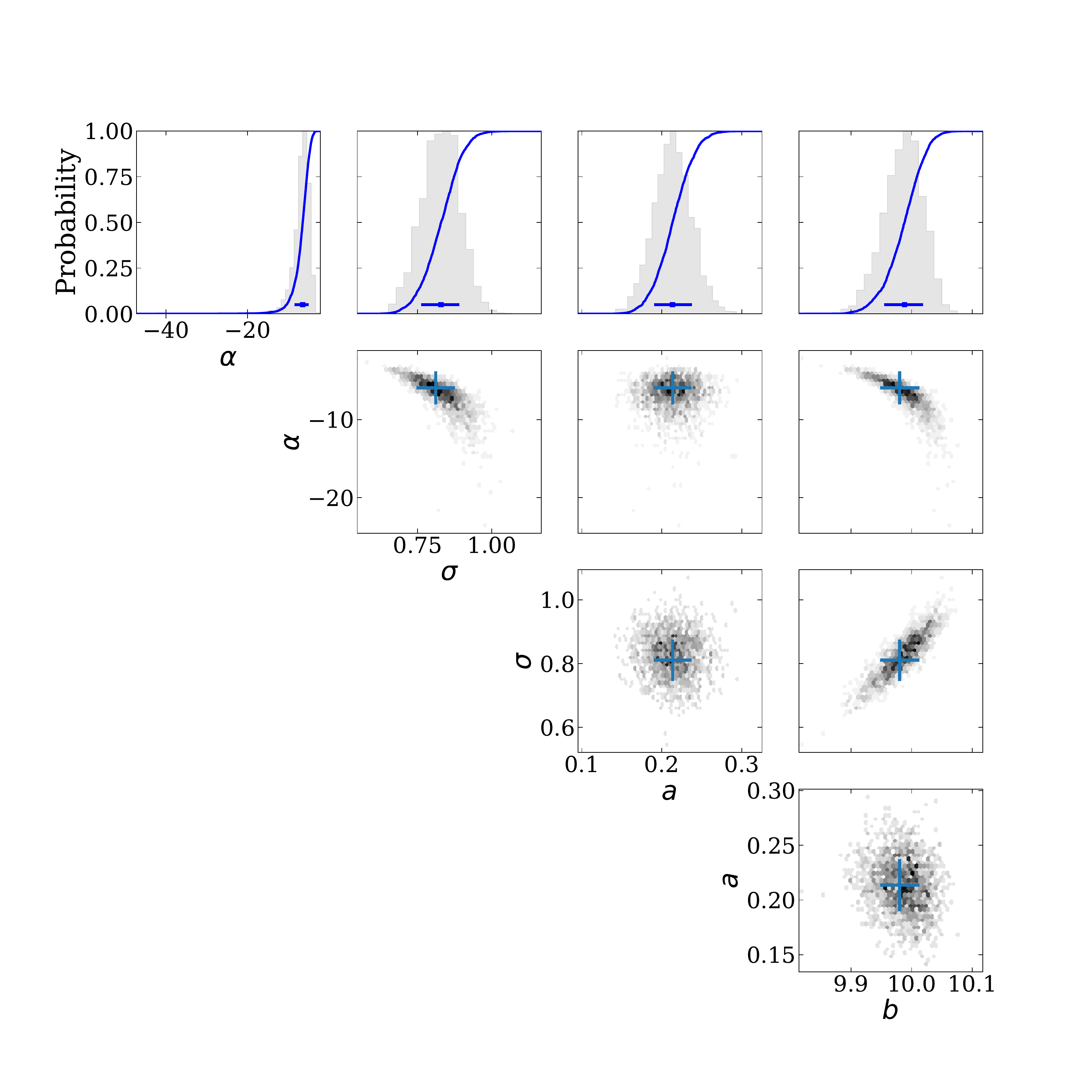}
    %\label{fig:sub2}
  \end{subfigure}%    
  % \hfill
  % \hspace{-10mm}
  \begin{subfigure}[s]{0.43\textwidth}
    \includegraphics[height=0.94\columnwidth, width=1\columnwidth]{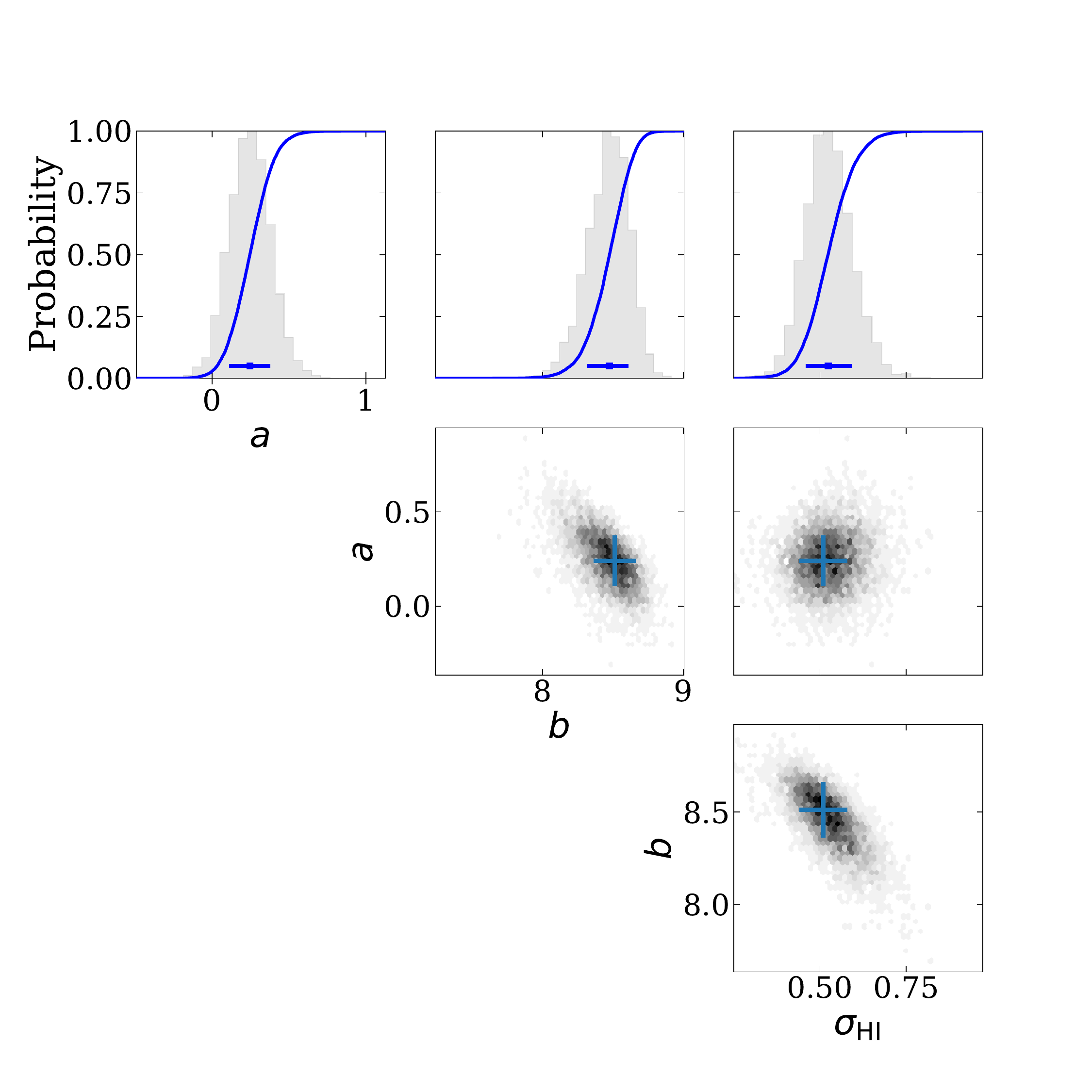}
    \includegraphics[height=0.94\columnwidth, width=1.\columnwidth]{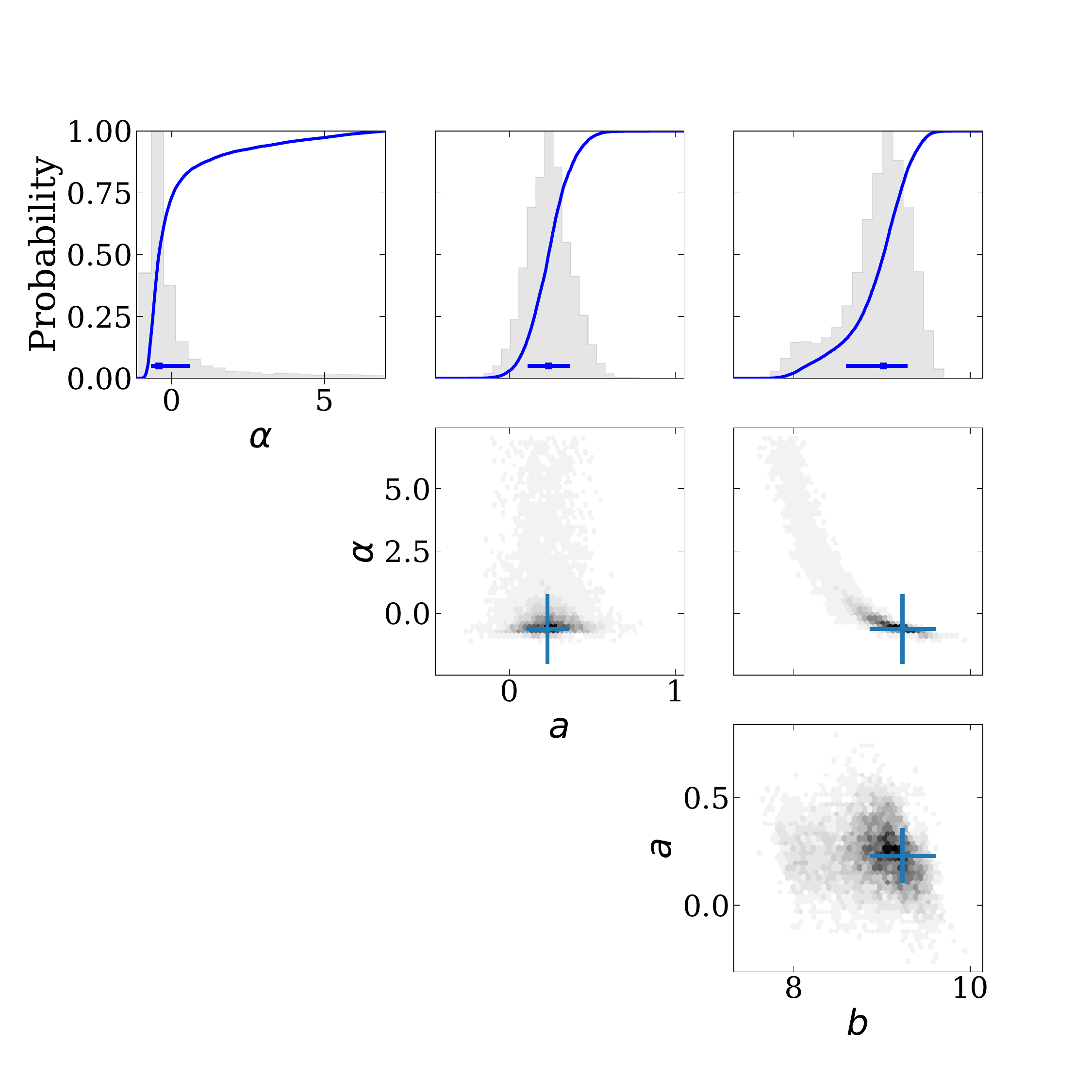}
    \includegraphics[height=0.94\columnwidth, width=1\columnwidth]{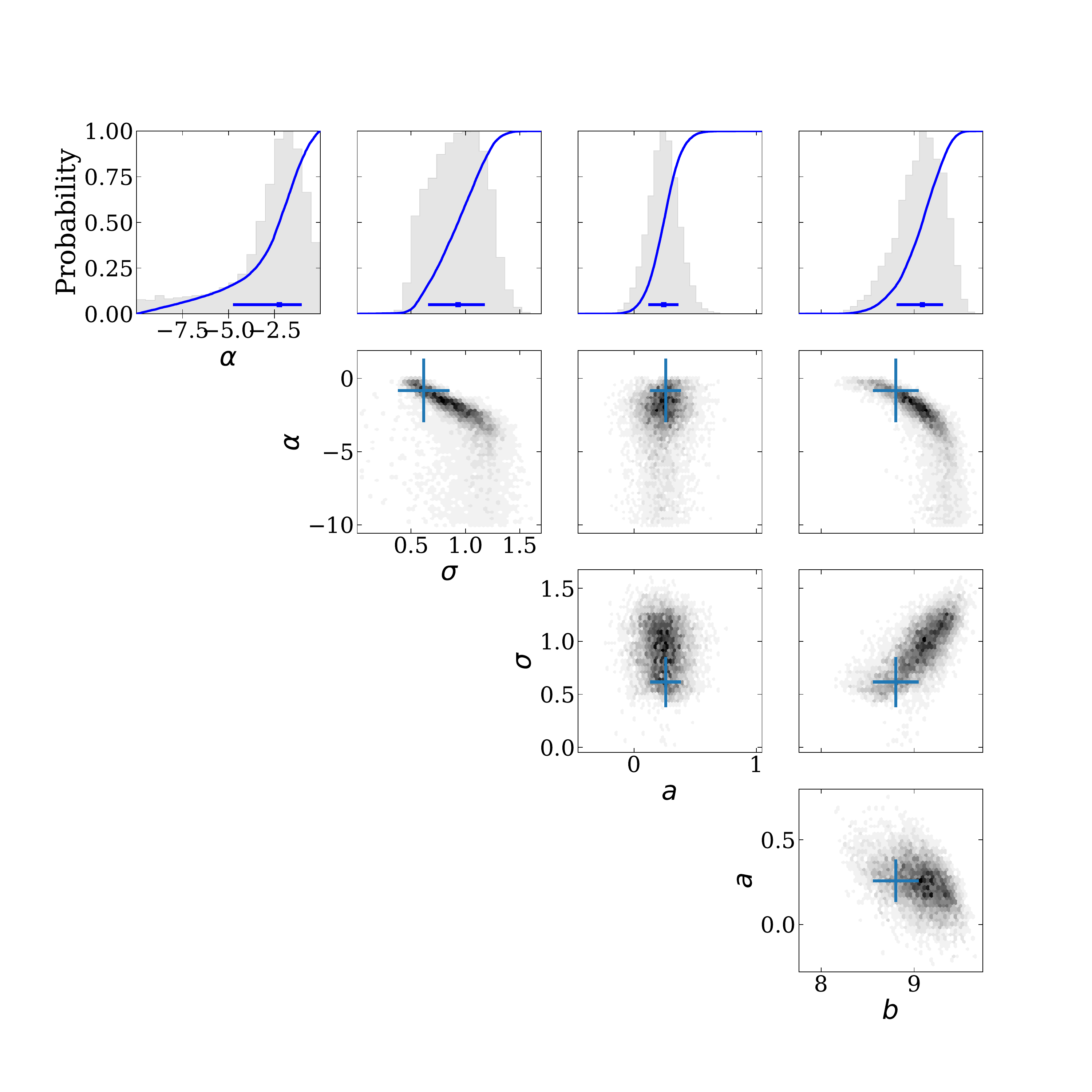}
    %\label{fig:sub2}
  \end{subfigure}%
  \caption{Posterior distributions for three models (Normal, Schechter, Skew normal) of the \hi mass and stellar mass relation at $0.25 < z < 0.5$ from top to bottom panels. The left and right panels are for the blue and red galaxies, respectively. The gray histograms represent the marginal posterior probabilities (1D or 2D), while the blue curves indicate the cumulative distributions. In the 2D posterior plots, the blue crosses mark the parameter set with the maximum likelihood, and the 1$\sigma$ error bars are estimated from the 1D marginal posterior distributions.}
  \label{fig:color_post}
\end{figure*}

\begin{figure*}
  \centering
  \begin{subfigure}[s]{0.43\textwidth}
    \includegraphics[height=0.94\columnwidth, width=1\columnwidth]{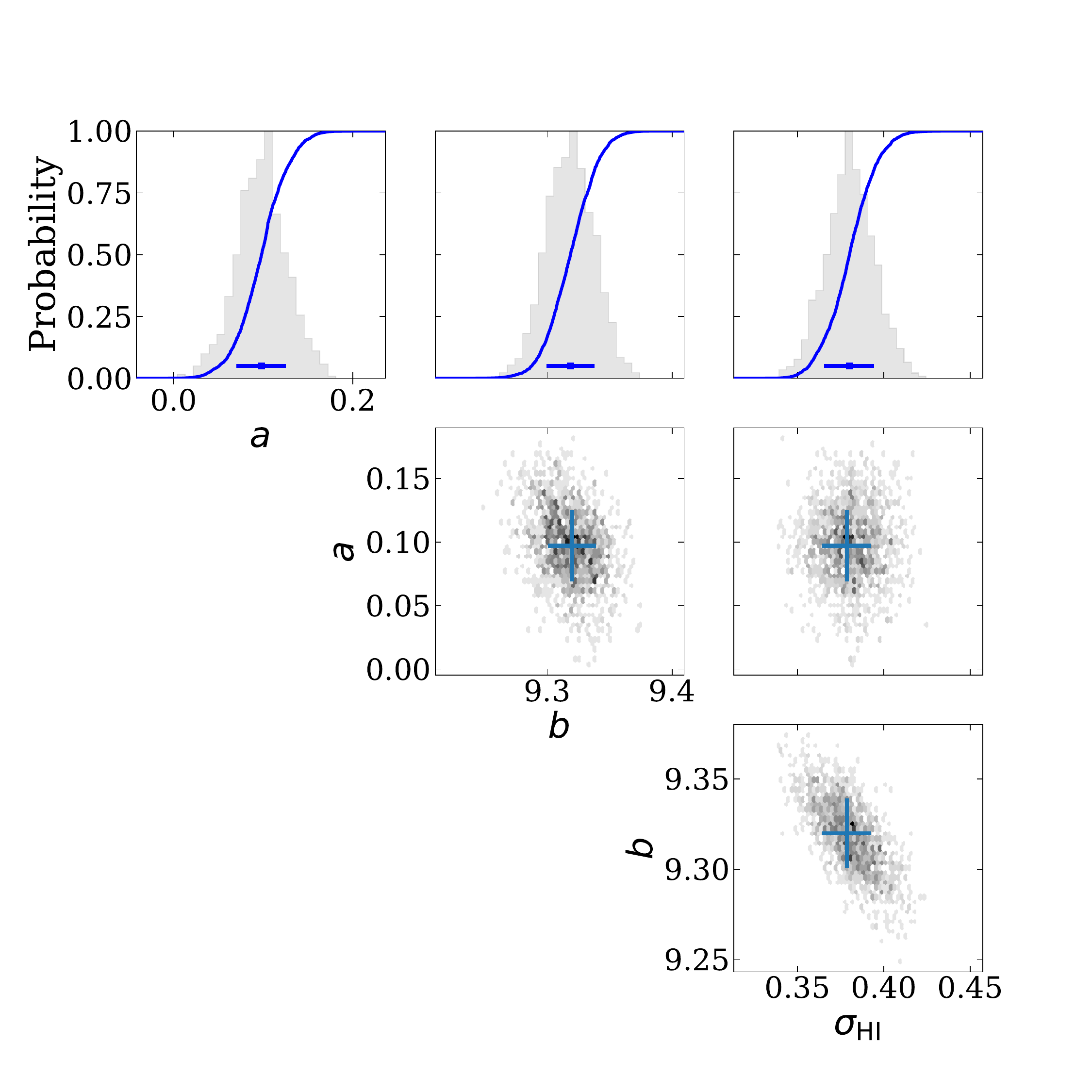}
    \includegraphics[height=0.94\columnwidth, width=1.\columnwidth]{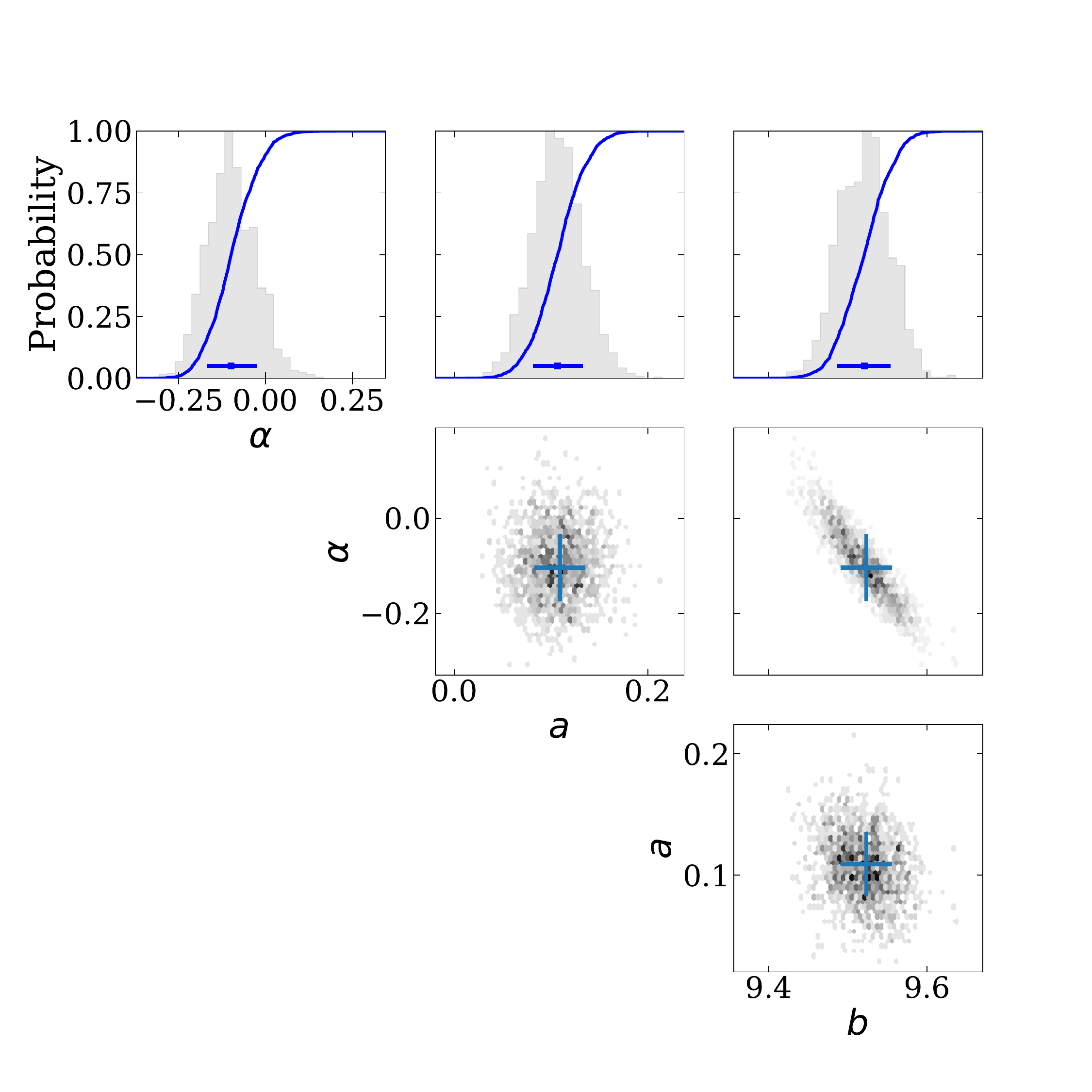}
    \includegraphics[height=0.94\columnwidth, width=1\columnwidth]{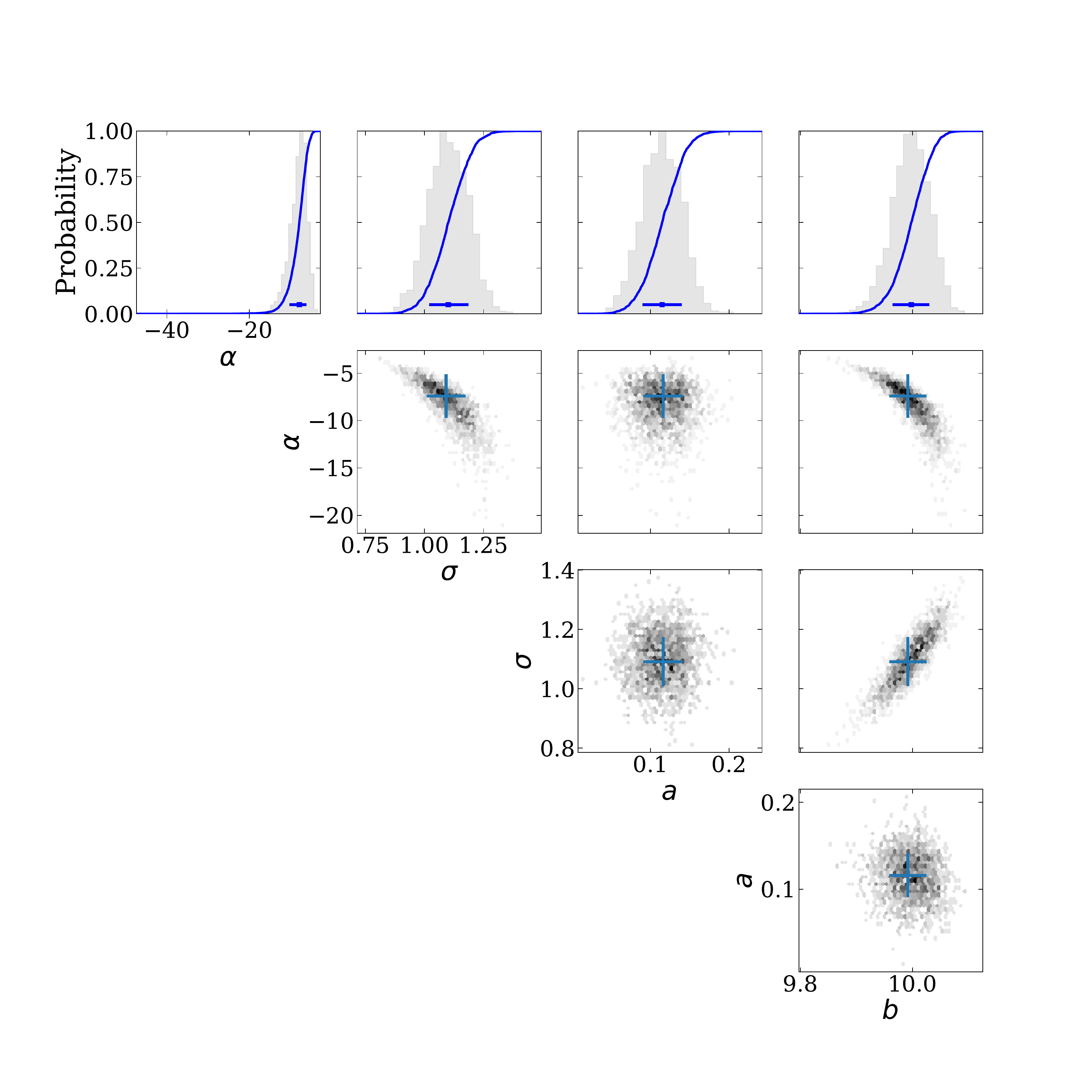}
    %\label{fig:sub2}
  \end{subfigure}%    
  % \hfill
  % \hspace{-10mm}
  \begin{subfigure}[s]{0.43\textwidth}
    \includegraphics[height=0.94\columnwidth, width=1\columnwidth]{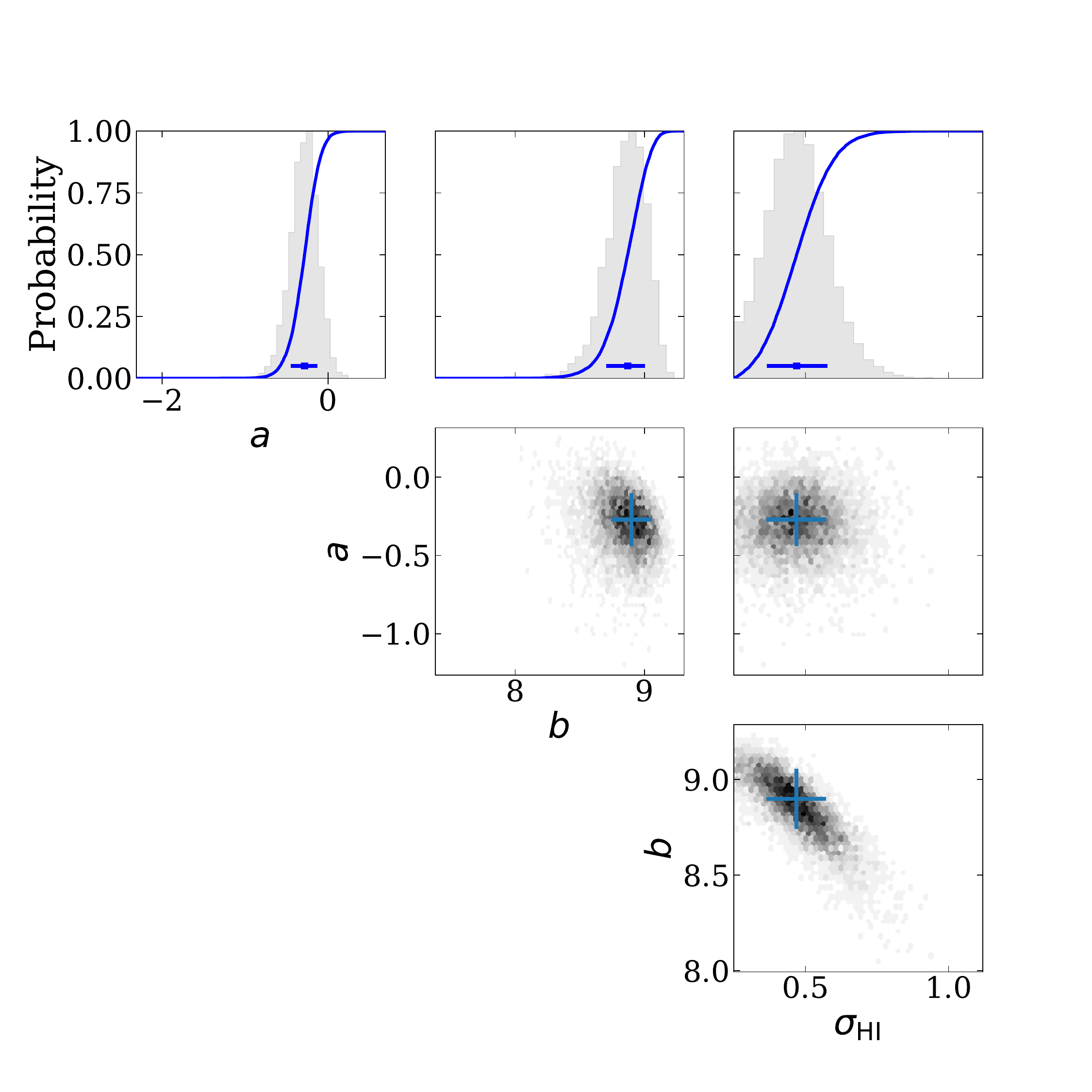}
    \includegraphics[height=0.94\columnwidth, width=1.\columnwidth]{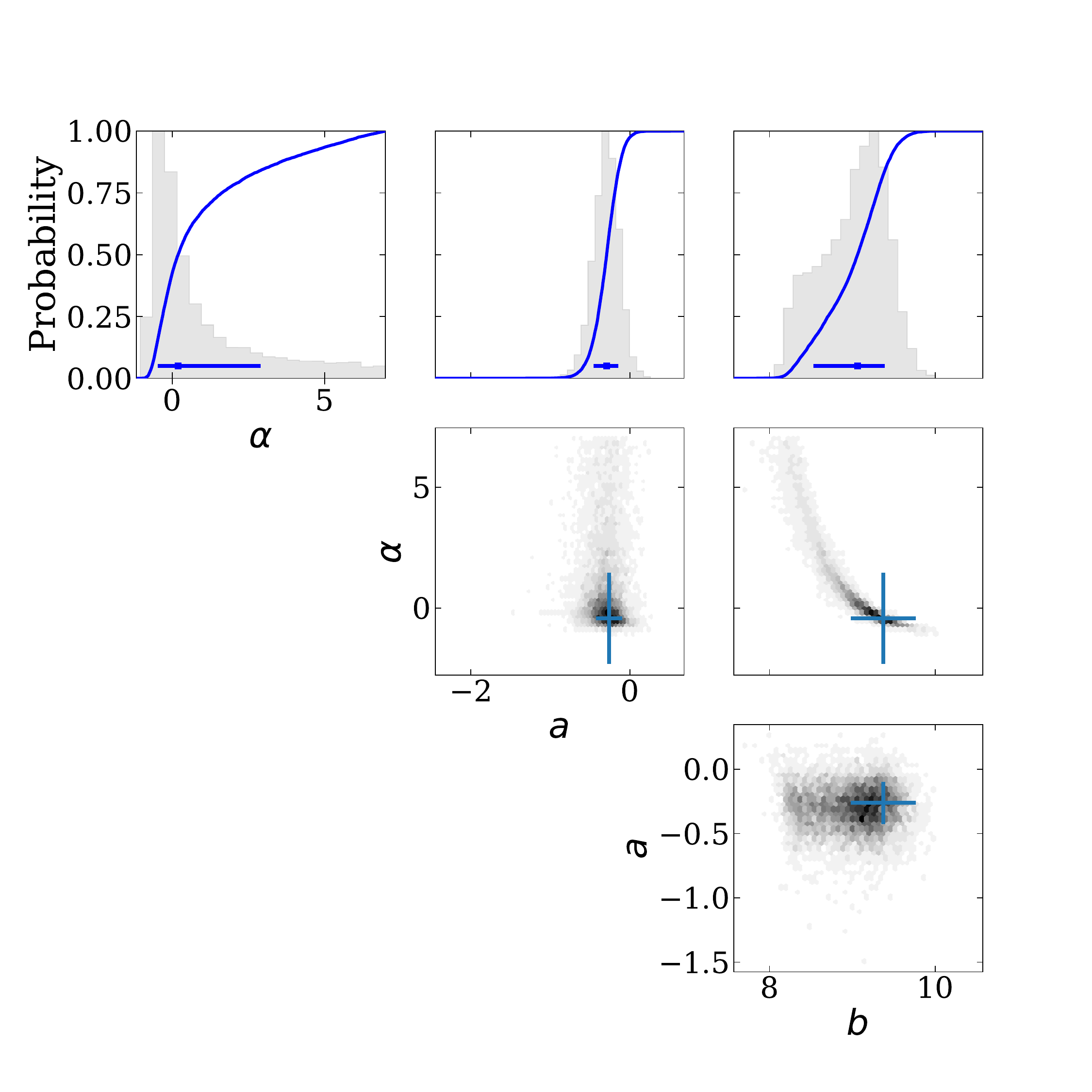}
    \includegraphics[height=0.94\columnwidth, width=1\columnwidth]{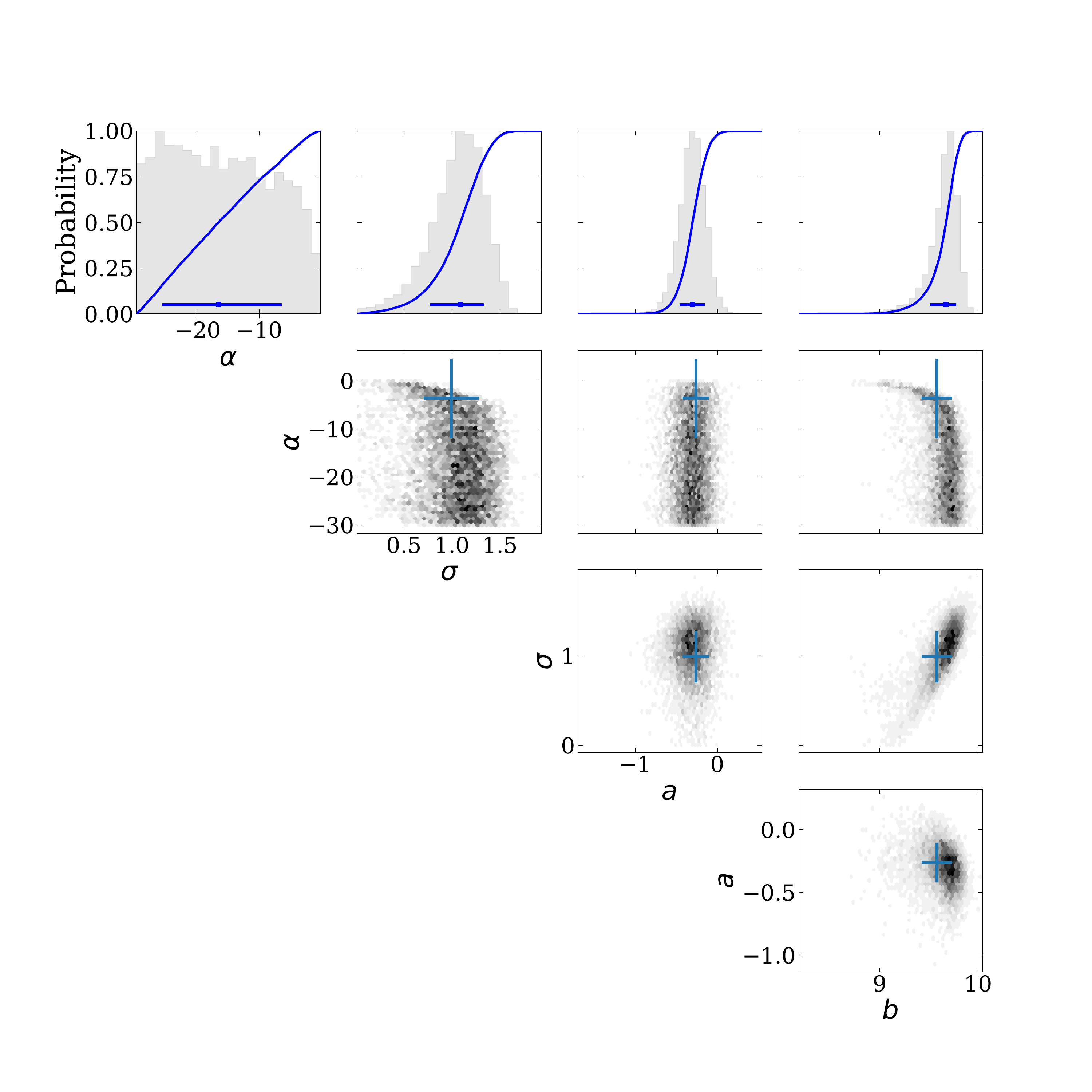}
    %\label{fig:sub2}
  \end{subfigure}%
  \caption{The same as the caption of Figure~\ref{fig:color_post}, except that the left and right panels are for the spiral and elliptical galaxies, respectively.}
  \label{fig:morph_post}
\end{figure*}

The posterior distributions of modelling the \hi mass and stellar mass relation at $0.25 < z < 0.5$ for the blue and red galaxies are shown in Figure~\ref{fig:color_post}, and for the spiral and elliptical galaxies in Figure~\ref{fig:morph_post}, respectively. 

\section{Best fitting conditional HIMF}
\label{sec:chimf_table}

We list the best fitting conditional \hi mass functions with Schechter and Skew normal models in Table~\ref{tab:chimf} for reference and ease of comparison for future work.

\begin{table*}
\centering

\caption{Best fitting conditional \hi mass functions with Schechter and Skew normal models for the full galaxy sample, blue, red, spiral, elliptical sub-samples at $0.25<z<0.5$. The conditional HIMF for the Full (Blue+Red) is the sum of the conditional HIMFs for the Blue and Red sub-samples, which has better performance than for the Full sample, particularly at the low-mass end, due to more accurate modelling.}
\label{tab:chimf}
\begin{tabular}{ c c c c c c c}
 \hline
 \hline
$\log_{10}(M_{\rm HI}/M_{\odot})$ & \multicolumn{6}{c}{$\mathrm{\Phi(M_{\rm HI})_{9.5}}h_{70}^{-3}\; \mathrm{[Mpc^{-3} dex^{-1}]}$}   \\
 & Full & Blue & Spiral & Red & Elliptical & Full (Blue+Red) \\
 \hline
  & \multicolumn{6}{c}{Schechter}   \\
 \hline
6.0 & $2.90 \times 10^{-5}$ & $-$ & $8.00 \times 10^{-6}$ & $6.90 \times 10^{-5}$ & $1.10 \times 10^{-5}$ & $7.00 \times 10^{-5}$ \\
6.2 & $4.10 \times 10^{-5}$ & $1.00 \times 10^{-6}$ & $1.20 \times 10^{-5}$ & $8.90 \times 10^{-5}$ & $1.50 \times 10^{-5}$ & $9.00 \times 10^{-5}$ \\
6.4 & $5.80 \times 10^{-5}$ & $1.00 \times 10^{-6}$ & $1.80 \times 10^{-5}$ & $1.15 \times 10^{-4}$ & $2.10 \times 10^{-5}$ & $1.16 \times 10^{-4}$ \\
6.6 & $8.30 \times 10^{-5}$ & $2.00 \times 10^{-6}$ & $2.70 \times 10^{-5}$ & $1.48 \times 10^{-4}$ & $3.00 \times 10^{-5}$ & $1.50 \times 10^{-4}$ \\
6.8 & $1.17 \times 10^{-4}$ & $4.00 \times 10^{-6}$ & $4.10 \times 10^{-5}$ & $1.91 \times 10^{-4}$ & $4.20 \times 10^{-5}$ & $1.95 \times 10^{-4}$ \\
7.0 & $1.65 \times 10^{-4}$ & $7.00 \times 10^{-6}$ & $6.20 \times 10^{-5}$ & $2.45 \times 10^{-4}$ & $5.80 \times 10^{-5}$ & $2.52 \times 10^{-4}$ \\
7.2 & $2.33 \times 10^{-4}$ & $1.30 \times 10^{-5}$ & $9.40 \times 10^{-5}$ & $3.14 \times 10^{-4}$ & $8.20 \times 10^{-5}$ & $3.28 \times 10^{-4}$ \\
7.4 & $3.28 \times 10^{-4}$ & $2.50 \times 10^{-5}$ & $1.42 \times 10^{-4}$ & $4.02 \times 10^{-4}$ & $1.14 \times 10^{-4}$ & $4.27 \times 10^{-4}$ \\
7.6 & $4.62 \times 10^{-4}$ & $4.60 \times 10^{-5}$ & $2.15 \times 10^{-4}$ & $5.11 \times 10^{-4}$ & $1.58 \times 10^{-4}$ & $5.57 \times 10^{-4}$ \\
7.8 & $6.50 \times 10^{-4}$ & $8.50 \times 10^{-5}$ & $3.25 \times 10^{-4}$ & $6.46 \times 10^{-4}$ & $2.17 \times 10^{-4}$ & $7.31 \times 10^{-4}$ \\
8.0 & $9.09 \times 10^{-4}$ & $1.56 \times 10^{-4}$ & $4.88 \times 10^{-4}$ & $8.07 \times 10^{-4}$ & $2.95 \times 10^{-4}$ & $9.63 \times 10^{-4}$ \\
8.2 & $1.27 \times 10^{-3}$ & $2.84 \times 10^{-4}$ & $7.28 \times 10^{-4}$ & $9.91 \times 10^{-4}$ & $3.92 \times 10^{-4}$ & $1.27 \times 10^{-3}$ \\
8.4 & $1.74 \times 10^{-3}$ & $5.08 \times 10^{-4}$ & $1.08 \times 10^{-3}$ & $1.18 \times 10^{-3}$ & $5.05 \times 10^{-4}$ & $1.69 \times 10^{-3}$ \\
8.6 & $2.36 \times 10^{-3}$ & $8.90 \times 10^{-4}$ & $1.56 \times 10^{-3}$ & $1.35 \times 10^{-3}$ & $6.20 \times 10^{-4}$ & $2.24 \times 10^{-3}$ \\
8.8 & $3.12 \times 10^{-3}$ & $1.50 \times 10^{-3}$ & $2.20 \times 10^{-3}$ & $1.44 \times 10^{-3}$ & $7.07 \times 10^{-4}$ & $2.94 \times 10^{-3}$ \\
9.0 & $3.96 \times 10^{-3}$ & $2.39 \times 10^{-3}$ & $2.96 \times 10^{-3}$ & $1.38 \times 10^{-3}$ & $7.20 \times 10^{-4}$ & $3.77 \times 10^{-3}$ \\
9.2 & $4.73 \times 10^{-3}$ & $3.49 \times 10^{-3}$ & $3.72 \times 10^{-3}$ & $1.12 \times 10^{-3}$ & $6.22 \times 10^{-4}$ & $4.61 \times 10^{-3}$ \\
9.4 & $5.11 \times 10^{-3}$ & $4.41 \times 10^{-3}$ & $4.18 \times 10^{-3}$ & $7.15 \times 10^{-4}$ & $4.28 \times 10^{-4}$ & $5.13 \times 10^{-3}$ \\
9.6 & $4.72 \times 10^{-3}$ & $4.49 \times 10^{-3}$ & $3.94 \times 10^{-3}$ & $3.22 \times 10^{-4}$ & $2.17 \times 10^{-4}$ & $4.81 \times 10^{-3}$ \\
9.8 & $3.40 \times 10^{-3}$ & $3.29 \times 10^{-3}$ & $2.82 \times 10^{-3}$ & $8.90 \times 10^{-5}$ & $7.20 \times 10^{-5}$ & $3.38 \times 10^{-3}$ \\
10.0 & $1.65 \times 10^{-3}$ & $1.50 \times 10^{-3}$ & $1.31 \times 10^{-3}$ & $1.30 \times 10^{-5}$ & $1.30 \times 10^{-5}$ & $1.51 \times 10^{-3}$ \\
10.2 & $4.29 \times 10^{-4}$ & $3.56 \times 10^{-4}$ & $3.15 \times 10^{-4}$ & $1.00 \times 10^{-6}$ & $1.00 \times 10^{-6}$ & $3.57 \times 10^{-4}$ \\
10.4 & $4.20 \times 10^{-5}$ & $3.70 \times 10^{-5}$ & $2.80 \times 10^{-5}$ & $-$ & $-$ & $3.70 \times 10^{-5}$ \\
10.6 & $1.00 \times 10^{-6}$ & $1.00 \times 10^{-6}$ & $1.00 \times 10^{-6}$ & $-$ & $-$ & $1.00 \times 10^{-6}$ \\
\hline
  &\multicolumn{6}{c}{Skew normal}   \\
\hline
6.0 & $1.70 \times 10^{-5}$ & $-$ & $5.00 \times 10^{-6}$ & $1.00 \times 10^{-6}$ & $5.00 \times 10^{-6}$ & $1.00 \times 10^{-6}$ \\
6.2 & $3.00 \times 10^{-5}$ & $-$ & $9.00 \times 10^{-6}$ & $2.00 \times 10^{-6}$ & $9.00 \times 10^{-6}$ & $2.00 \times 10^{-6}$ \\
6.4 & $5.00 \times 10^{-5}$ & $-$ & $1.70 \times 10^{-5}$ & $6.00 \times 10^{-6}$ & $1.50 \times 10^{-5}$ & $6.00 \times 10^{-6}$ \\
6.6 & $8.30 \times 10^{-5}$ & $1.00 \times 10^{-6}$ & $3.10 \times 10^{-5}$ & $1.50 \times 10^{-5}$ & $2.50 \times 10^{-5}$ & $1.60 \times 10^{-5}$ \\
6.8 & $1.33 \times 10^{-4}$ & $2.00 \times 10^{-6}$ & $5.40 \times 10^{-5}$ & $3.40 \times 10^{-5}$ & $3.90 \times 10^{-5}$ & $3.60 \times 10^{-5}$ \\
7.0 & $2.06 \times 10^{-4}$ & $6.00 \times 10^{-6}$ & $9.20 \times 10^{-5}$ & $7.10 \times 10^{-5}$ & $6.00 \times 10^{-5}$ & $7.70 \times 10^{-5}$ \\
7.2 & $3.12 \times 10^{-4}$ & $1.30 \times 10^{-5}$ & $1.49 \times 10^{-4}$ & $1.41 \times 10^{-4}$ & $8.90 \times 10^{-5}$ & $1.54 \times 10^{-4}$ \\
7.4 & $4.59 \times 10^{-4}$ & $3.00 \times 10^{-5}$ & $2.35 \times 10^{-4}$ & $2.58 \times 10^{-4}$ & $1.28 \times 10^{-4}$ & $2.88 \times 10^{-4}$ \\
7.6 & $6.55 \times 10^{-4}$ & $6.50 \times 10^{-5}$ & $3.58 \times 10^{-4}$ & $4.39 \times 10^{-4}$ & $1.77 \times 10^{-4}$ & $5.04 \times 10^{-4}$ \\
7.8 & $9.07 \times 10^{-4}$ & $1.31 \times 10^{-4}$ & $5.26 \times 10^{-4}$ & $6.93 \times 10^{-4}$ & $2.38 \times 10^{-4}$ & $8.24 \times 10^{-4}$ \\
8.0 & $1.22 \times 10^{-3}$ & $2.47 \times 10^{-4}$ & $7.48 \times 10^{-4}$ & $1.01 \times 10^{-3}$ & $3.09 \times 10^{-4}$ & $1.25 \times 10^{-3}$ \\
8.2 & $1.60 \times 10^{-3}$ & $4.39 \times 10^{-4}$ & $1.03 \times 10^{-3}$ & $1.34 \times 10^{-3}$ & $3.88 \times 10^{-4}$ & $1.77 \times 10^{-3}$ \\
8.4 & $2.03 \times 10^{-3}$ & $7.32 \times 10^{-4}$ & $1.37 \times 10^{-3}$ & $1.60 \times 10^{-3}$ & $4.72 \times 10^{-4}$ & $2.33 \times 10^{-3}$ \\
8.6 & $2.51 \times 10^{-3}$ & $1.15 \times 10^{-3}$ & $1.75 \times 10^{-3}$ & $1.71 \times 10^{-3}$ & $5.55 \times 10^{-4}$ & $2.85 \times 10^{-3}$ \\
8.8 & $3.00 \times 10^{-3}$ & $1.69 \times 10^{-3}$ & $2.17 \times 10^{-3}$ & $1.60 \times 10^{-3}$ & $6.32 \times 10^{-4}$ & $3.29 \times 10^{-3}$ \\
9.0 & $3.50 \times 10^{-3}$ & $2.34 \times 10^{-3}$ & $2.61 \times 10^{-3}$ & $1.30 \times 10^{-3}$ & $6.94 \times 10^{-4}$ & $3.64 \times 10^{-3}$ \\
9.2 & $3.95 \times 10^{-3}$ & $3.04 \times 10^{-3}$ & $3.02 \times 10^{-3}$ & $9.11 \times 10^{-4}$ & $6.90 \times 10^{-4}$ & $3.95 \times 10^{-3}$ \\
9.4 & $4.34 \times 10^{-3}$ & $3.71 \times 10^{-3}$ & $3.38 \times 10^{-3}$ & $5.43 \times 10^{-4}$ & $5.15 \times 10^{-4}$ & $4.25 \times 10^{-3}$ \\
9.6 & $4.53 \times 10^{-3}$ & $4.20 \times 10^{-3}$ & $3.63 \times 10^{-3}$ & $2.73 \times 10^{-4}$ & $2.41 \times 10^{-4}$ & $4.47 \times 10^{-3}$ \\
9.8 & $3.85 \times 10^{-3}$ & $3.72 \times 10^{-3}$ & $3.28 \times 10^{-3}$ & $1.15 \times 10^{-4}$ & $4.80 \times 10^{-5}$ & $3.83 \times 10^{-3}$ \\
10.0 & $1.84 \times 10^{-3}$ & $1.69 \times 10^{-3}$ & $1.57 \times 10^{-3}$ & $4.10 \times 10^{-5}$ & $2.00 \times 10^{-6}$ & $1.73 \times 10^{-3}$ \\
10.2 & $3.43 \times 10^{-4}$ & $2.89 \times 10^{-4}$ & $2.36 \times 10^{-4}$ & $1.20 \times 10^{-5}$ & $-$ & $3.01 \times 10^{-4}$ \\
10.4 & $2.00 \times 10^{-5}$ & $1.60 \times 10^{-5}$ & $8.00 \times 10^{-6}$ & $3.00 \times 10^{-6}$ & $-$ & $1.90 \times 10^{-5}$ \\
10.6 & $-$ & $-$ & $-$ & $1.00 \times 10^{-6}$ & $-$ & $1.00 \times 10^{-6}$ \\
\hline
% \bottomrule
\end{tabular}
\end{table*}


%%%%%%%%%%%%%%%%%%%%%%%%%%%%%%%%%%%%%%%%%%%%%%%%%%

% Don't change these lines
\bsp	% typesetting comment
\label{lastpage}
\end{document}